\documentclass[floatfix,aps,showpacs,amsmath,amssymb,twocolumn,eqsecnum,superscriptaddress]{revtex4-1}
\let\Twocolumn

\newif\ifTwocolumn
\ifx\Twocolumn\undefined
  \Twocolumnfalse
\else
  \Twocolumntrue
\fi
\ifTwocolumn
\fi
\usepackage{graphicx}
\usepackage{gensymb}
\usepackage{subfigure}
\usepackage{bm}
\usepackage{upgreek}
\usepackage{wasysym}
\usepackage{bibentry}
\usepackage[T1]{fontenc}
\usepackage{textcomp}
\usepackage{mathptmx}
\usepackage[scaled=0.92]{helvet}
\usepackage{courier}
\usepackage{setspace}
\usepackage{color}
\usepackage{epstopdf}
\usepackage{epsfig}
\usepackage{hyperref}
\definecolor{red_n}{rgb}{1.0, 0.0, 0.0}
\definecolor{brown_n}{rgb}{0.6, 0.4, 0.2}
\definecolor{cyan_n}{RGB}{0.0, 255.0, 255.0}
\definecolor{blue_n}{rgb}{0.0, 0.0, 1.0}
\definecolor{green_n}{rgb}{0.0, 0.5, 0.0}
\definecolor{orange_n}{RGB}{255.0, 127.0, 0.0}
\definecolor{magenta_n}{RGB}{255.0, 0.0, 255.0}
\definecolor{purple_n}{RGB}{128.0, 0.0, 128.0}
\definecolor{gray_n}{RGB}{128.0, 128.0, 128.0}
\definecolor{dark_green}{rgb}{0.0, 0.5, 0.0}
\newcommand{\eref}[1]{Eq.~\eqref{#1}}
\newcommand{\oncite}[1]{Ref.~[\onlinecite{#1}]}

\ifTwocolumn
  
\else
  
\fi
\begin{document}
\title{Sign change in the net force in sphere-plate and sphere-sphere systems immersed in nonpolar
critical fluid due to the interplay between the critical Casimir and dispersion van der Waals forces}
\author{Galin~Valchev}
\email[Electronic address: ]{gvalchev@imbm.bas.bg}
\affiliation{Institute of Mechanics -- Bulgarian Academy of Sciences, Academic Georgy Bonchev St. building 4, 1113 Sofia, Bulgaria}
\author{Daniel~Dantchev}
\email[Electronic address: ]{daniel@imbm.bas.bg}
\affiliation{Institute of Mechanics -- Bulgarian Academy of Sciences, Academic Georgy Bonchev St. building 4, 1113 Sofia, Bulgaria}
\affiliation{Max-Planck-Institut f\"{u}r Intelligente Systeme, Heisenbergstrasse 3, D-70569 Stuttgart, Germany and
IV. Institut f\"{u}r Theoretische Physik, Universit\"{a}t Stuttgart, Pfaffenwaldring 57, D-70569 Stuttgart, Germany}

\date{\today}
%
\begin{abstract}
We study systems in which both long-ranged van der Waals and critical Casimir interactions are present. The last arise as an effective force between bodies when immersed in a near-critical medium, say a nonpolar one-component fluid or a binary liquid mixture. They are due to the fact that the presence of the bodies modifies the order parameter profile of the medium between them as well as the spectrum of its allowed fluctuations. We study the interplay between these forces, as well as the {\it total} force (TF) between a spherical colloid particle and a thick planar slab, and between two spherical colloid particles. We do that using general scaling arguments and mean-field type calculations utilizing the Derjaguin and the surface integration approaches. They both are based on data of the forces between two parallel slabs separated at a distance $L$ from each other, confining the fluctuating fluid medium characterized by its temperature $T$ and chemical potential $\mu$.  The surfaces of the colloid particles and the slab are coated by thin layers exerting strong preference to the liquid phase of the fluid, or one of the components of the mixture, modeled by strong adsorbing local surface potentials, ensuring the so-called  $(+,+)$ boundary conditions. On the other hand, the core region of the slab and the particles, influence the fluid by long-ranged competing dispersion potentials. We demonstrate that for a suitable set of colloids-fluid, slab-fluid, and fluid-fluid coupling parameters the competition between the effects due to the coatings and the core regions of the objects involved result, when one changes $T$, $\mu$ or $L$, in {\it sign change} of the Casimir force (CF) {\it and} the TF acting between the colloid and the slab, as well as between the colloids.  This can be used for governing the behavior of objects, say colloidal particles, at small distances, say in colloid suspensions for preventing flocculation. It can also  provide a strategy for solving problems with handling, feeding, trapping and fixing of microparts in nanotechnology. Data for specific substances in support of the experimental feasibility of the theoretically predicted behavior of the CF and TF have been also presented.
\end{abstract}
\maketitle
%
\section{Introduction}\label{sec:Inrtoduction}
%

In a recent article \cite{VaDa2015}, we have demonstrated that the critical Casimir force (CCF) between two plates, placed at a distance $L$ from each other and immersed in a critical nonpolar fluid governed by dispersion van der Waals forces (vdWF) can change sign below a given threshold thickness of the system $L_{\rm crit}$ when one changes the temperature $T$, the chemical potential of the fluid $\mu$, or $L$. Because of the high relevance to, say, physics of colloids, as well as nanotechnology, it is important to clarify if one can change the sign of the {\it total} force between two object when one changes a parameter that can be externally controlled, like $T$ and $L$, i.e., if one can govern, on wish, this force to be attractive, or repulsive. This, of course, will be of especial interest, if at least one of the object involved in the interaction is small in size. The simplest nonplanar geometrical shape which one can study is then, of course, the spherical one.

\begin{figure*}[t!]
	\centering
	\includegraphics[width=17.8 cm]{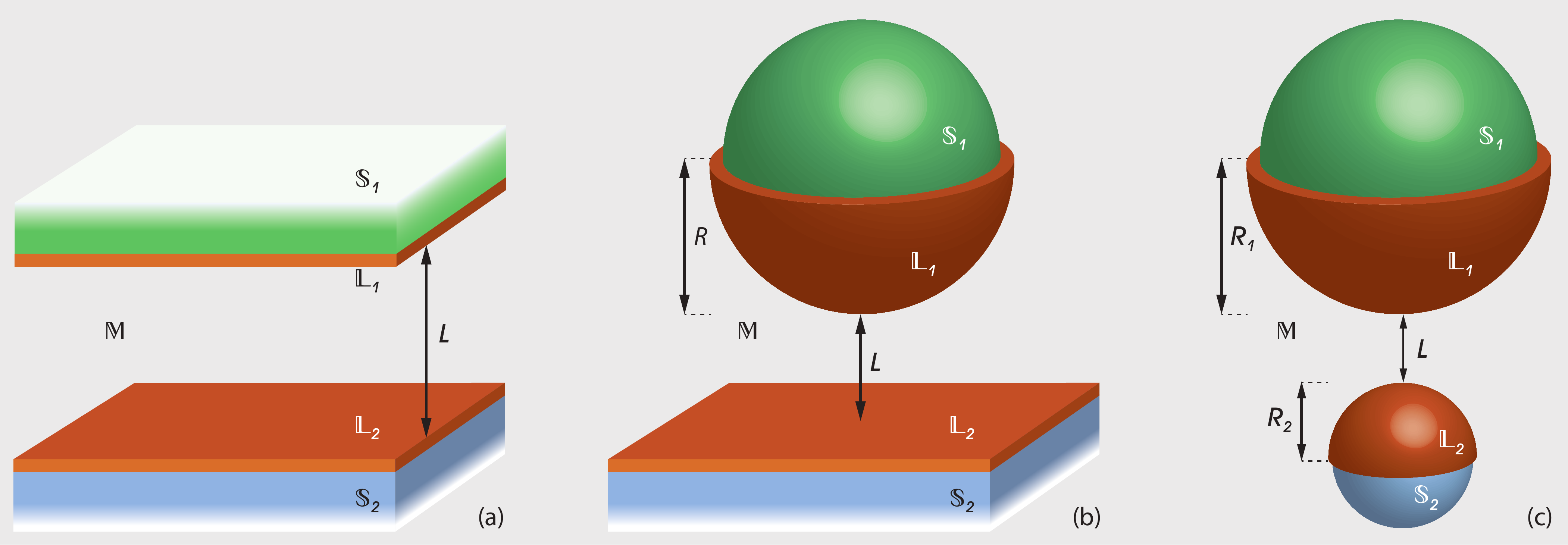}
	\caption{(Color online) Schematic depiction of the considered fluid systems: $\mathbf{(a)}$ pair of parallel plates, $\mathbf{(b)}$ sphere of radius $R$ above a plate and $\mathbf{(c)}$ pair of spherical particles with radii $R_{1}$ and $R_{2}$ (dissimilar in the most general case). In all three cases the interacting objects are assumed immersed in some fluid medium $\mathbb{M}$ -- a nonpolar one-component fluid or a binary mixture composed out of the molecules of some nonpolar liquids $A$ and $B$, which is close to its critical/demixing point. The minimal separation between the interacting objects in every of the described systems is denoted by $L$. The substances composing the objects are denoted by $\mathbb{S}_{1}$ and $\mathbb{S}_{2}$, coated by thin layers of some other substances $\mathbb{L}_{1}$ and $\mathbb{L}_{2}$, respectively. The fluid medium is considered embedded on a lattice in which, some nodes are occupied by a particle and others are not (for a simple one-component fluid) -- thus depicting the "liquid" and "gas" states respectively at some values of the temperature $T$ and chemical potential $\mu$ of the fluid, or some of the nodes are occupied by a molecule from the substance $A$ (the "liquid" state) and the rest are occupied by the molecules belonging to the species $B$ ("gas" state). The surfaces of the interacting objects impose on the fluid medium boundary conditions of strong adsorption, the so-called $(+,+)$ boundary conditions, i.e., the nearest to the coating substances layers are entirely occupied by the particles of the one-component fluid or if the medium is a binary liquid mixture -- by the particles of one of its components. The bulk phase (core) of the objects, on the other hand, influence the fluid by long-range {\it competing} dispersion potentials.}
	\label{fig:fluid_systems}
\end{figure*}

In the current article, we study the interplay between the CCF and the dispersion vdWF, as well as the resultant {\it total} force $F_{\rm tot}$ between a sphere and a plate and between two spheres -- see Fig. \ref{fig:fluid_systems}(\textbf{b}) and (\textbf{c}). To be more specific, we consider a spherical in shape colloidal particle with mesoscopic radius $R$, or two such particles with, in general, different mesoscopic radii $R_1$ and $R_2$ immersed in a medium that is either simple fluid or a binary liquid mixture. We envisage the case in which the particle is placed at a distance $L$ of closest surface-to-surface approach from a planar boundary wall (plate), or the situation in which the  surfaces of the two spheres are at such a distance. The above mentioned fluid mediated forces between two surfaces or large particles are usually referred to as solvation forces \cite{E90,E90book,ES94} in colloid science. Here we are going to consider the special case when the temperature of the medium is close to the critical temperature $T_c$ of  either the liquid-vapor critical point of the simple fluid or the critical demixing point of a binary liquid mixture. In this case the solvation force acquires, as pointed out by Fisher and de Gennes  \cite{FG78},  a contribution due to the critical fluctuation of the medium.  This contribution is of a long-ranged character. It is characterized, to a great extend, by the gross features of the medium \cite{K94,BDT2000,G2009} depending also on the boundary conditions which the bodies immersed in the medium impose on it at their surfaces. This fluctuation induced force, which is due to the critical fluctuations of the order parameter of the medium is, thus, {\it universal} in nature. It has a lot of similarities with the force between neutral bodies due to the quantum and temperature fluctuations of the charge distributions in them, i.e., of the electromagnetic field, which force is known today under the general name of a Casimir force \cite{C48,CP48,C53}, or, more specifically, quantum electrodynamic (QED) CF. That is why the fluctuation part of the solvation force $F_{{\rm Cas}}$ near $T_c$ is termed {\it critical Casimir force} \cite{K94,BDT2000,G2009}.

The force $F_{\rm tot}$ separates into a regular background contribution $F_{{\rm tot}}^{\rm (reg)}$, which depends on the parameters characterizing the medium in an analytic way, and a singular contribution $F_{{\rm tot}}^{{\rm (sing)}}$, which is due to the critical fluctuations of the medium
\begin{equation}
\label{eq:decomp}
F_{{\rm tot}}=F_{{\rm tot}}^{\rm (reg)}+F_{{\rm tot}}^{{\rm (sing)}}.
\end{equation}
Obviously, the excess \cite{note} grand canonical potential $\Omega_{\rm ex}(T,\mu|L,\cdots)$, or the excess free energy, of the system depends on the geometrical characteristics like $L$, $R$, $R_1$ and $R_2$. Then one has
\begin{equation}\label{eq:total_force}
F_{{\rm tot}}=- \dfrac{\partial}{\partial L} \Omega_{\rm ex}(L|T,\mu,\cdots).
\end{equation}
One normally defines
\begin{equation}
\label{eq:def_Casimir}
F_{{\rm Cas}}(L|T,\mu)\equiv F_{{\rm tot}}^{{\rm (sing)}}(L|T,\mu).
\end{equation}
Within the systems considered, we suppose that the colloid particles, the solid wall and the medium are all governed by dispersion van der Waals interactions. Thus, one shall have
\begin{equation}
\label{eq:def_vdW}
F_{{\rm vdW}} \equiv F_{{\rm tot}}^{\rm (reg)}(L|T,\mu).
\end{equation}

It shall be stressed that both the CCF and vdWF are fluctuation induced ones but due to the fluctuations of different entities -- the first is due to massless excitations of the order parameter, while the second -- of the electromagnetic field. In the current article we are going to consider how the interplay of these two types of interactions govern the behavior of colloidal particles. Currently there is no general theory available to scope with the problem of quantitative description of the mutual influence of the fluctuation of the electromagnetic field and the order parameter fluctuations of a medium when it is close to its critical point. The Lifshitz theory \cite{L56,DLP61} -- which is the basic one for studying the Casimir effect due to the fluctuations of the electromagnetic field has never been meant to nor can deal with the problem of a critical medium between two other substances. For practical application of this theory the main quantity, which has to be known for any material involved, is the dielectric permittivity $\varepsilon(\omega,T)$. Under the normal approach it is usually tabulated at room  temperature for specific values of the angular frequency $\omega$ of the electromagnetic field. Then, in order to perform the specific calculations needed one supposes the analytic validity of a given dependence on $\omega$, say, the validity of the Drude or plasma models for the considered material. It shall be emphasized, however, that in a critical fluid $\varepsilon(\omega,T)$ is itself a {\it singular} function of the temperature \cite{SBMG80,MRDJ2003}. We are not aware of a theory that quantitatively predicts  how $\varepsilon(\omega,T)$ depends on the temperature and $\omega$ near the critical point of the medium for a specific material characterized by some characteristic spectrum.

Due to the hypothesis of universality and scaling when studying critical phenomena one normally is interested only in the gross features of the system and utilizes some effective Hamiltonian where only few basic features of the critical medium are reflected. Therefore, one observes that both basic theories in the filed, approach very differently the problem -- the Lifshitz theory that is based on the detailed knowledge of $\varepsilon(\omega,T)$, which is material specific, while the approach to  critical phenomena is based on universality, i.e., on few basic features of the system.

In \cite{DSD2007,DRB2007,DRB2009,VaDa2015} an approach has been suggested for such a situation in the case of a film geometry, based on the equation of state of the critical medium, which provides an uniform treatment of the contributions due to the vdWF and the CCF.  The specifics of the materials are reflected by the long-ranged tails of the potentials but, otherwise, the theory uses the standard approach to criticality. It utilizes general renormalization group arguments and makes use of field-theoretical methods within, in its simplest realization, a mean-field approximation. This relatively simple approach allows all the calculations for the forces involved to be performed on equal footing. Let us note that the expression derived within the suggested in \cite{DSD2007,DRB2007,DRB2009,VaDa2015}  approach for the Hamaker term is in full agreement with the Dzyaloshinskii-Lifshitz-Pitaevskii theory \cite{L56,DLP61}. Let us also remind that the mean-field theory is considered as a reliable theoretical horse for the qualitative description of critical phenomena.

The interest in the fluctuation-induced phenomena in the last years blossomed due to their importance in the rapidly developing field of nanotechnology where, below a micrometer distances, the vdWF and QED CF play a dominant role between neutral nonmagnetic objects. The last implies that these forces play a key role in micro- and nano-electromechanical systems (MEMS/NEMS) \cite{CAKBC2001a,DBKRCN2005,BEBDPS2012} operating at such distances. In vacuum, or gas medium, they lead to  irreversible, usually undesirable phenomena, such as stiction (i.e., irreversible adhesion) or pull-in due to mechanical instabilities \cite{BR2001,BR2001a,CAKBC2001}. Closely related to that is another troubling effect: when a particle's characteristic size is scaled down below a micrometre the role of its weight becomes negligible. As a result, when one tries to release such a neutral particle from, say, the surface of whatever handling device in air or vacuum, the particle will not drop down under
the gravity but, instead, will stick to the surface due to the effect of the omnipresent vdWF. If, in an attempt to release the particle, one charges the particle, forces vibration of the surface in question, etc., the released particle might move in an uncontrollable way leaving the observation field of the apparatus controlling the performance of the operation. That is the main reason why the handling, feeding, trapping and fixing of micro- and/or nano- particles is still the main bottleneck in micro manufacturing and is far from being solved in a satisfactory fashion \cite{CVP2005}. Thus, formalizing the above, one of the main problems in the micro- and nano-assembly is the precise and reliable manipulation of a micro- or nano-particles that includes moving it from a given starting point, where it is to be taken from, to some end point, when it is to be placed on. In that respect it seems ideal, if one can modify the net force between the manipulated particle and the operating device, sometimes called gripper, in such a way that it is repulsive at short distances between the handling surfaces and the particle and attractive at larger ones. It is clear that the ability to modify the Casimir interaction can strongly influence the development of MEMS/NEMS. Several theorems, however, seriously limit the possible search of repulsive QED CF \cite{KK2006,S2010,RKE2010}. Currently, apart from some suggestions for achieving QED Casimir repulsion in systems out of equilibrium the only experimentally well verified way to obtain such repulsive force is to have interaction between two different materials characterized by dielectric permittivities $\varepsilon_1$ and $\varepsilon_2$ such that \cite{L56,DLP61,LP80}
\begin{equation}
\label{permit}
\varepsilon_1<\varepsilon_M<\varepsilon_2
\end{equation}
along the imaginary frequency axis, with $\varepsilon_{M}$ being the dielectric permittivity of the medium between them. In Refs. \cite{MML96,MLB1997,LS2001,LS2002,MCP2009,IIIM2011} QED Casimir repulsion has indeed been observed experimentally for the sphere-plate geometry. In order to minimize the potential negative effects of all possible circuitry at such a small distances and the complications with the isolation, as well as possible problems involving chemical reactions it seems that one promising strategy for overcoming the obstacles mentioned above is to choose such a fluid as a medium that possesses no free changes dissolved in it and that is inert and do not interact chemically with the materials. That leads us to choose as a fluid a nonpolar liquefied noble gas that has critical parameters as close as possible to the normal ones.

Due to both scientific and technological reasons currently the Casimir effect is object of intense studies. The last is true both for the QED Casimir effect, as well as in its thermodynamic manifestation via the CCF. Let us stress out that the CCF has already been measured \cite{HHGDB2008}, utilizing light scattering technique, in the interaction between a single colloidal sphere and a flat silica surface immersed in a binary mixture near its critical point. The theoretical background of the obtained results is discussed in details in \oncite{GMHNHBD2009}. In \oncite{PCTBDGV2016}, using a system of three optically trapped \cite{T2014} spherical colloidal particles, immersed in a critical binary mixture, the authors demonstrated experimentally the theoretically predicted nonadditivity of the fluctuation-induced interactions. Other experimental setups which exploit a sphere-plate or sphere-sphere configurations immersed in a critical fluid include the interaction of spherical colloids with chemically patterned substrates \cite{SZHHB2008,TZGVHBD2011} which theoretical description is  discussed in Refs. \cite{TKGHD2009,TKGHD2010}, formation of critical colloid aggregates \cite{BOSGWS2009,GamDit2010,PMVWMSW2014}, phase behaviour studies \cite{ZAB2011,NFHWS2013,TaDi2017}, various techniques for fine-tuning of the CCF \cite{NHB2009,NDHCNVB2011}. As far as the theoretical side is concerned, it was de Gennes who first obtained the CCF between spherical particles \cite{PGdeGennes1981} considering a local free-energy functional. Among the other techniques used to study the CCF in sphere-plate and sphere-sphere geometries are the Ornstein-Zernike theory \cite{AUP1992}, conformal invariance methods \cite{BE1995,BuEi1995,BiEmKa2013}, Monte Carlo calculations \cite{GZTS2012,SNLS2013,DVNBS2013,GZS2014,H2013,V2014,HH2014,ETBERD2015,VM2015,TEvanRED2016}, fluid-particle dynamics simulations \cite{FuGaDiTa2013,YOO2015}, mean-field type \cite{KHD2009,MHD2013,MKMD2014,LHTD2014,MHD2015} and density-functional \cite{AnPa2016} theory calculations combined with the Derjaguin approximation \cite{HSED1998,SHD2003,TKGHD2010,LaDi2016}. Several review articles and works \cite{G2009,GD2011,N2016,NDNS2016,LZH2016} summarize both the experimental and theoretical results presented there.

In the current article we will demonstrate that by proper choice of the materials (cores) of the colloid particles and the handling surface of a gripper it is indeed possible to achieve control over the net interaction (TF) as well as the CCF between the surface and the particle by simply changing $T$, $\mu$ and $L$. We also present results for the forces between two colloidal particles. Let us stress that due to its unique temperature dependence, the CCF  allows \textit{in situ} control of reversible assembly in soft matter and nanoscience. A further advantage of the force is that both its magnitude and range of action depends on the separation between the objects and the thermodynamic parameters of the fluctuating medium. The last facts can potentially be used in controlling the properties of colloidal suspensions and for governing the behavior of objects at small, below micrometer, distances.

The content of the article is arranged as follows. In Sec. \ref{sec:general} we present some general predictions of the finite-size scaling theory for the interaction between sphere and a plate and between two spheres.  In Subsec. \ref{sec:ThermalCas} we recall and comment on the finite-size behaviour of systems with dispersion forces extending the known facts to the expected behaviour of the CCF, vdWF and TF when they act between pair of parallel plates, sphere and a plate and couple of spherical particles. By doing so, we introduce and compare the two general techniques, namely the Derjaguin approximation (DA) (Subsec. \ref{sec:DA}) and the surface integration approach (SIA) (Subsec. \ref{sec:SIA}), within which we study the commented spectrum of forces. Section \ref{sec:Model} briefly presents the corresponding lattice-gas model suitable for the investigation of fluid media with account of the long-ranged van der Waals interactions. Here we present the equation for the equilibrium profile of the finite-size order parameter, identify the main coupling parameters characterizing the interactions in the systems, which enter in it, and give the general expressions used to calculate the CCF and TF. Last but not least in Sec. \ref{sec:ResultsAndExpFeas} we present the exact equations (Subsec. \ref{subsec:SpPlSpSpDASIAd3}) used to obtain the numerical results for the behaviour of the investigated forces, and comment them in details in Subsec. \ref{subsec:Discussion}. Finally, we provide arguments in support of the experimental feasibility of the predicted effects -- Subsec. \ref{subsec:ExperimentalFeas}. The article ends with a summary and discussion section -- Sec. \ref{sec:DisandConcRem}. Important technical details concerning the derivation of the expressions for the interaction forces between spherical particles within the SIA are presented in Appendix \ref{sec:AppSpSpSIA}.

%
\section{Theoretical background}
\label{sec:general}

\subsection{Some general predictions of the finite-size scaling theory}

Since the CCF depends on the properties of the solvent near the bulk critical point, it is governed by universality and scaling \cite{K94,BDT2000}. The last implies that, in first approximation, this force depends only on the gross features of the system -- its dimensionality $d$ and the symmetry of the ordered state (both defining the so-called bulk universality class of the system) and on the boundary conditions (determined by the surface universality classes) imposed on the fluid by the bodies immersed in it. Therefore, to a great extend the CCF is \textit{universal}. The quantitative effects of the presence of a surfaces of the bodies on the thermodynamic behavior of the system depends on the penetration depth of their symmetry breaking effect into the volume. Obviously, the range to which these effects are felt within the system depends on two phenomena: on how long-ranged the interactions are and on how  long-ranged the fluid correlations, which mediate the interactions between the bodies, are. The long-ranginess of the correlations is set by the correlation length $\xi$ of the order parameter of the solvent; $\xi$ becomes large, and theoretically diverges, in the vicinity of the bulk critical point $(T_c,\mu_c)$: $\xi(T\to T_c^{+},\mu=\mu_c)\simeq \xi_0^{+}t^{-\nu}$, where $t=(T-T_c)/T_c$, and $\xi(T=T_c,\mu\to\mu_c)\simeq \xi_{0,\mu} |\Delta\mu/(k_B T_c)|^{-\nu/\Delta}$, where $\Delta \mu=\mu-\mu_c$. Here $\nu$ and $\Delta$ are the usual critical exponents which, for classical fluids, are those of the three-dimensional Ising model, and $\xi_0^{+}$ and $\xi_{0,\mu}$ are the corresponding nonuniversal amplitudes of the correlation length along the $t$ and $\mu$ axes. When $\xi$ becomes comparable to the characteristic dimension of the system, say the separation $L$ between the objects, the size dependence of the thermodynamic functions enters into the thermodynamic potentials through the ratio $L/\xi$, and takes a scaling form given by the finite-size scaling theory \cite{BDT2000,P90,Ba83}. Then

{\it i) } for the Casimir force between a spherical colloidal particle of radius $R$ and a plate in the case of a system governed only by short-ranged interactions the theory \cite{C88,BDT2000,Ba83,P90,Ped90,K94,HSED98,SHD2003} predicts:
\begin{equation}\label{eq:sphere_plane}
\beta F_{\rm Cas}^{R,|}(L|T,\mu)= L^{-1}X_{\rm Cas}^{R,|}\left(\Xi, x_t, x_\mu \right),
\end{equation}
where
\begin{equation}\label{eq:def_scaling_var}
 x_{t}=t\left(L/\xi_{0}^+\right)^{1/\nu}, \qquad  x_{\mu}=\beta\Delta\mu\left(L/\xi_{0,\mu}\right)^{\Delta/\nu}
\end{equation}
are the temperature and field relevant in renormalization group sense scaling variables, $\Xi\equiv R/L$ and $\beta=1/(k_{B}T)$. Note that this prediction for the $L^{-1}$ dependence in front of the scaling function of \eref{eq:sphere_plane}  shall be valid for any dimensionality since it simply takes into account that the dimension of a hypersphere in a $d$ dimensional space is one dimension less than that of the space itself. With respect to the scaling, \eref{eq:sphere_plane}  shall be valid when the hyperscaling holds, i.e., for $2<d<4$ in systems with short-ranged and subleading long-ranged interactions, with the scaling function $X_{\rm Cas}^{R,|}$ being universal. Within the mean-field theory one formally sets there the critical exponents pertinent to the $d=4$ case but shall keep in mind that a nonuniversal system dependent prefactor is expected to be present in the scaling function $X_{\rm Cas}^{R,|}$ that shall be taken into account. When $d>4$  the hyperscaling is violated and the scaling variables in \eref{eq:sphere_plane} change. This question has been discussed, say, in Ref. \cite{BDT2000}, see Section 6.3. there, Ref. \cite{CGGP2001}, etc. Here we will not be going in any details in it.

{\it ii)} for the interaction mediated by the critical fluid  \cite{HSED98,SHD2003} in the case of two spherical colloidal particles with radii $R_1$ and $R_2$ one has
\begin{eqnarray}
\label{eq:sphere_sphere_basic}
\beta F_{\rm Cas}^{R_1, R_2}(L|T,\mu) =L^{-1} X_{\rm Cas}^{R_1, R_2}\left(\Xi_{1},\Xi_{2}, x_t, x_\mu\right),
\end{eqnarray}
with $\Xi_{i}\equiv R_{i}/L,\ i=1,2$.

In the remainder we are going to study $F_{\rm Cas}^{R,|}(L|T,\mu)$ and $F_{\rm Cas}^{R_1, R_2}(L|T,\mu)$ in a system governed not by short-ranged, but by long-ranged dispersion interactions. That will require some modifications of \eref{eq:sphere_plane} and \eref{eq:sphere_sphere_basic}. Furthermore, we will obtain $F_{\rm Cas}^{R,|}(L|T,\mu)$ and $F_{\rm Cas}^{R_1, R_2}(L|T,\mu)$ utilizing the Derjaguin \cite{D34} and the recently introduced surface integration approach \cite{DV2012}. For this purpose we will need the corresponding results for a system with a film geometry that is governed by the same dispersion interactions as those occurring in the sphere-plate and sphere-sphere systems. That is why we are next going to concisely recall these topics.  We start with the behavior of the thermodynamic Casimir force in nonpolar fluid systems with dispersion forces.

\subsection{The thermodynamic Casimir force in nonpolar fluid systems with dispersion forces}
 \label{sec:ThermalCas}

The specifics of the scaling theory for systems with dispersion forces are described in details in Refs. \cite{DR2001,D2001,DRB2007,DRB2009,DDG2006,DSD2007,VaDa2015}. That is why we will just briefly remind here some very basic facts that will also serve us for introducing notations needed further in the main text.

We consider a fluid system consisting of a nonpolar medium $\mathbb{M}$ comprising two thick parallel plates of materials $\mathbb{S}_{1}$ and $\mathbb{S}_{2}$ which are coated by thin solid films of substances $\mathbb{L}_{1}$ and $\mathbb{L}_{2}$, respectively -- see Fig. \ref{fig:fluid_systems}({\bf a}). If the fluid medium is in contact with a particle reservoir with a chemical potential $\mu$, the grand canonical potential $\Omega_{\mathrm{ex}}(L|T,\mu)$ per unite area $\mathcal{A}$ of this medium in excess to its bulk value $\mathcal{A}L\omega_{\mathrm{bulk}}(T,\mu)$ depends on the film thickness $L$ and, thus, one can define the total effective force $F_{\mathrm{tot}}(L|T,\mu)$ [in a full accord with \eref{eq:total_force}], which is due to fluctuations of the medium and dispersion interactions in it. Here $\omega_{\mathrm{bulk}}(T,\mu)$ is the density of the bulk grand canonical potential, and $\mathcal{A}$ is the surface area of the plates.

We suppose that the dispersion forces governing all the parts of the considered system depend on the distance $r$ between the constituents of the bodies in the system $\propto r^{-d-\sigma}$, where $d$ is the dimensionality of the space and $\sigma$ is a parameter that controls the decay of the dispersion interactions. The last implies  interactions between the fluid particles $\propto J^{l}r^{-d-\sigma}$ and substrate potentials $\propto J^{s_{i},l}z^{-\sigma},\ i=1,2$ acting on the fluid particles at a distance $z$ from the surface of the colloid particles. When $\sigma>2$ such long-range interactions are termed subleading long-ranged interactions \cite{DR2001,D2001,DDG2006,DSD2007,VaDa2015}. The systems governed by them do also belong to the Ising universality class characterized by short-ranged forces \cite{PT77}, i.e., the critical exponents, e.g., do not depend on $\sigma$ for such type of interactions. For $d=3$ and $\sigma=3$ one has the usual van der Waals interactions, while $d=3$ and $\sigma=4$ corresponds to the retarded  Casimir-Polder one. These two interactions are two prominent  representative of the class of  subleading long-ranged interactions.

Clearly, by varying the ratio between the strengths of the long-ranged -- $J^{l}$ and the short-ranged -- $J_{{\rm sr}}^{l}$ contributions in the fluid interaction one can quantitatively probe the importance of the long-ranged parts of the interactions within the fluid medium. One can also in this way study potential experiments in colloidal systems which allow for a dedicated tailoring of the form of the effective interactions between colloidal particles.

The contribution of the dispersion forces to the total effective force \eref{eq:total_force} can be distinguished from that of the CCF by their temperature dependence, because the leading such of the former does not exhibit a singularity. Thus, one can perform the decompositions and identifications given in \eref{eq:decomp}, \eref{eq:def_Casimir} and \eref{eq:def_vdW}.

If the system is away from its bulk critical point for the occurring force $f_{\rm tot}^{\parallel}(L|T,\mu)$ per cross section area $\mathcal{A}$ {\it \underline {and}} $k_{B}T$ is customary to write the following expression
\begin{equation}\label{way}
f_{\mathrm{tot}}^{\parallel}(L|T,\mu) \simeq (\sigma-1)\beta H_A(T,\mu) L^{-\sigma}\vartheta^{\sigma-d},
\end{equation}
where, for dimensional reasons, the microscopic length scale $\vartheta$ is introduced. Let us note that one normally considers the case $d=\sigma$ and, thus,  omits the apparent dependence of this length, that can be taken to be, e.g., the so-called {\it retardation length} \cite{GC99,DV2012} $\xi_{\mathrm{ret}}$. In Eq. \eqref{way} $H_{A}$ is the Hamaker term, whose dependence from the temperature and chemical potential is given by the so-called Hamaker constant \cite{P2006,I2011} (for details see the Appendix in \oncite{VaDa2015}).

As already explained above, upon approaching the bulk critical point of the system the fluctuations of the order parameter of the confined fluid medium exhibit strong correlations which gives rise to new contribution to the TF, the CCF (see below). In the vicinity of this point (critical region) \eref{way} is no longer valid. Following \oncite{VaDa2015}, \eref{way} becomes
\begin{eqnarray}\label{CasimirF_l}
f_{\mathrm{tot}}^{\parallel}(L|T,\mu) &&\simeq L^{-d}X_{\rm crit}^{\parallel}\left[x_{t},x_{\mu},
x_{l},\left\{x_{s_{i}},i=1,2\right\},x_{g}\right] \nonumber\\
&&+(\sigma-1)\beta H_A(T,\mu) L^{-\sigma}\vartheta^{\sigma-d}.
\end{eqnarray}
Here $X_{\rm crit}^{\parallel}$ is dimensionless, universal scaling function, $x_{l}$, $x_{s_{i}},\ i=1,2$ and $x_{g}$ are the irrelevant in renormalization group sense scaling variables associated with the interactions in the system. Explicitly these variables are defined in the text below Eq. (2.6) in \oncite{VaDa2015}.

According to the scaling hypothesis of the CCF one expects that near the bulk critical point
\begin{equation}\label{scaling_function}
f_{\rm Cas}^{\parallel}(L|T,\mu)=L^{-d}X_{\mathrm{Cas}}^{\parallel}(x_t,x_\mu,\cdots),
\end{equation}
where $X_{\mathrm{Cas}}^{\parallel}$ is a scaling function, that for large enough $L$ with fixed $x_t={\cal O}(1)$ and  $x_\mu={\cal O}(1)$ it approaches the scaling function of the short-ranged system $X_{\mathrm{Cas}}^{\mathrm{sr},\parallel}(x_t,x_\mu)$ (for details see Eqs. (2.12) and (4.10) in \oncite{VaDa2015}). From Eqs. (\ref{eq:decomp}), (\ref{eq:def_Casimir}) and (\ref{eq:def_vdW}), together with \eref{CasimirF_l} it follows that the scaling function of the CCF $X_{\rm Cas}^{\parallel}$ is proportional to the sum of $X_{\rm crit}^{\parallel}$ and the singular part of the Hamaker term $H_{A}^{({\rm sing})}(T,\mu)$. The last implies that in order to determine the CF in systems with dispersion interaction one has to decompose the contribution captured through the Hamaker term in a singular and a regular parts, i.e.
\begin{equation}
\label{eq:Hamaker_decomposition}
H_{A}(T,\mu)=H_{A}^{({\rm reg})}(T,\mu)+H_{A}^{({\rm sing})}(T,\mu).
\end{equation}
Thus, with $d=\sigma$ one has
\begin{equation}\label{scaling_function_Casimir}
f_{\rm Cas}^{\parallel}=L^{-d}\left[X_{\mathrm{crit}}^{\parallel}+(d-1)\beta H_A^{(\rm sing)}\right],
\end{equation}
while \eref{eq:def_vdW}, normalized per unit area, in the general case coincides with \eref{way}, where $H_{A}\equiv H_{A}^{\rm (reg)}$.

We will often compare the behavior of the system with subleading long-ranged dispersion interactions present with this one of a system with purely short-ranged interactions which will serve as a reference system. In such a purely short-range system one has $H_{A}=0$. Then, at the bulk critical point $(T=T_{c},\ \mu=\mu_{c})$ the leading term of the CCF $f_{{{\cal A},\rm Cas}}^{\parallel}$ per unit area between the plates bounding the fluid has the form
\begin{equation}\label{ForceCasbulkcritpoint}
f_{{{\cal A},\rm Cas}}^{\parallel}(L|T_{c},\mu_{c})=(d-1)\Delta(d)\dfrac{k_{B}T_{c}}{L^{d}},
\end{equation}
where $X_{\rm crit}^{\rm sr,\parallel}(0)=(d-1)\Delta(d)$. Here $\Delta(d)={\cal O}(1)$ is an universal dimensionless quantity, called Casimir amplitude, which depends on the bulk and surface universality classes (and the geometry). Since the CCF is proportional to $k_{B}T_{c}$ the interaction between the plates can become rather strong in a system with high critical temperature such as, e.g., in classical binary liquid mixtures. Note that the {\it sign} of the force  depends on the sign of the Casimir amplitude $\Delta(d)$ which, on its turn, depends on the boundary conditions imposed by the bounding surfaces on the fluid. According to the usual convention negative sign corresponds to attraction, while positive sign means repulsion of the surfaces bounding the system.

In what follows, we are going to present results for the CCF, vdWF and TF in the cases of sphere-sphere and sphere-plate systems, utilizing the knowledge gained from studies of the corresponding  interactions between parallel plates.

%
\subsection{Sphere-plate and sphere-sphere geometries}\label{sec:SpPlandSpSpGeomForces}
%
%
\subsubsection{The studied forces within the DA}\label{sec:DA}
%
\begin{figure*}[t!]
\centering
\includegraphics[width=17.4 cm]{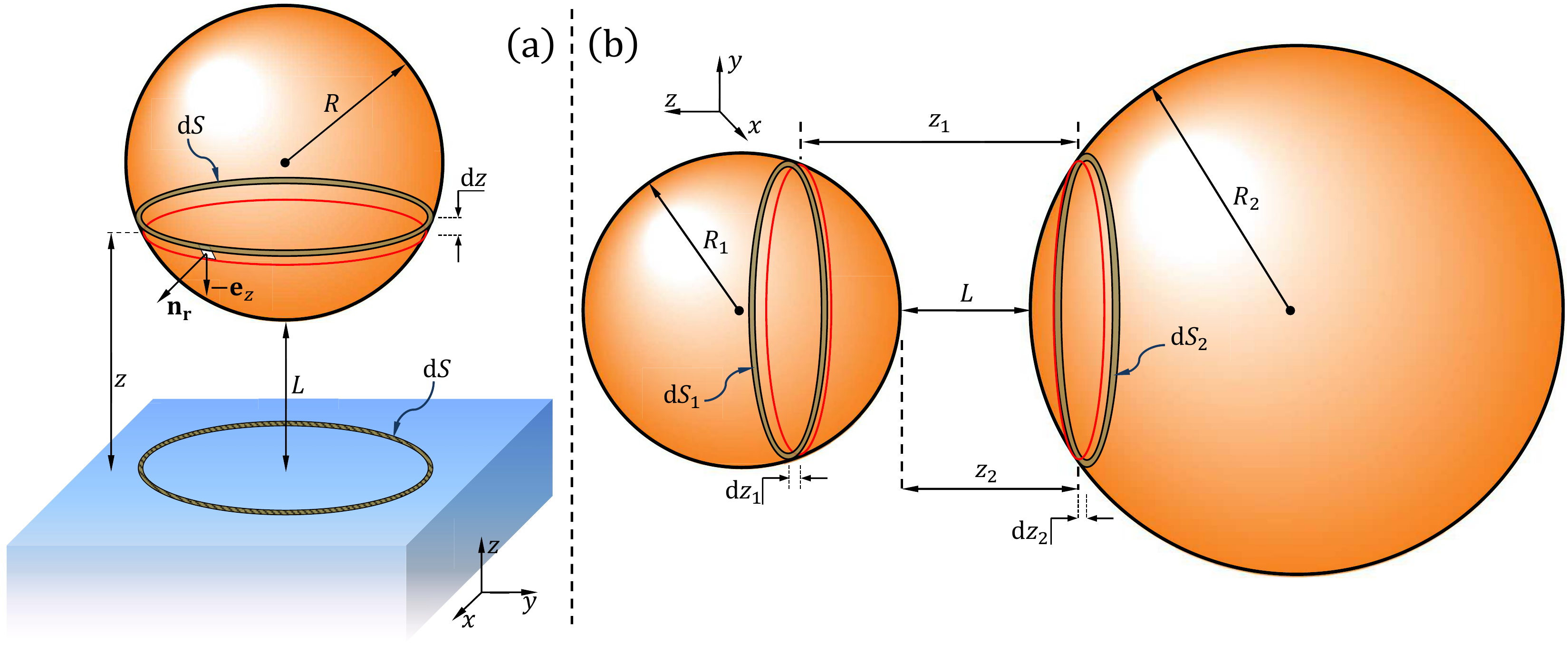}
  \caption{(Color online) Geometry of the {\it{surface integration approach}} for the interaction between $\mathbf{(a)}$ a spherical particle of radius $R$ with a planar substrate (plate) as well as between $\mathbf{(b)}$ pair of spherical particles of radii $R_{1}$ and $R_{2}$ (dissimilar in the most general case). On both figures the minimal distance at which the objects are situated is denoted by $L$. In $\mathbf{(a)}$ we consider the plate infinite in $-z$ direction. Here $\mathbf{n}_{\mathbf{r}}$ is the unit vector, normal to the sphere's surface, with $\mathbf{e}_{z}$ being its $z-$component, taken with a minus sign in the lower half of the sphere, which faces towards the plate. The red (full line) circle situated at a distance $z$ from the plate indicates the border of the cross section formed from the intersection between the sphere and a plane parallel to the $(x,y)-$plane. Then the infinitesimal projection area ${\rm{d}}S$ results from the cross section area difference at separations $z$ and $z+{\rm{d}}z$. In $\mathbf{(b)}$ the cross section of the sphere with radius $R_{2}$ with a plane parallel to the $(x,y)-$plane is situated at a distance $z_{2}$ from the surface of the second sphere, as the cross sections in both spheres are spaced apart $z_{1}$ from one another. The infinitesimal areas ${\rm{d}}S_{i}$ results from the cross section area difference at separations $z_{i}$ and $z_{i}+{\rm{d}}z_{i}$, with $i=1,2$.}
  \label{fig:SIAfig}
\end{figure*}
When it comes to calculating some geometry dependent interaction energy or force in systems where at least one of the objects has a nonplanar geometry, the most common approach used is the one first proposed by B. Derjaguin \cite{D34}.  It is know as Derjaguin approximation in colloidal science (see e.g. \oncite{SHD2003} and p. 34 in \oncite{BK2010}), and proximity force approximation in studies of QED Casimir effect (see e.g. p. 97 in \oncite{BKMM2009}). The main idea behind the DA is that one can relate the knowledge for the interaction force/potential between two parallel plates with the one between two gently curved colloidal particles, when the separation between them is much smaller than the geometrical characteristics of the particles in question. More specifically, the DA states that in $d=3$ the interaction force $F^{R_{1},R_{2}}(L)$ between two spherical particles with radii $R_{1}$ and $R_{2}$ placed at a distance $L\ll R_{1},R_{2}$  is given by
\begin{equation}\label{DASpSp}
F_{\rm DA}^{R_{1},R_{2}}(L)=2\pi R_{\rm eff}\int_{L}^{\infty}f_{{\cal A}}^{\parallel}(z){\rm d}z,
\end{equation}
where $R_{\rm eff}^{-1}=R_{1}^{-1}+R_{2}^{-1}$ is an effective radius and $f_{{\cal A}}^{\parallel}$ -- force per unit area between parallel plates. When the sphere with radius $R_{1}\equiv R$ interacts with a plate one has $R_{2}=\infty$ and then \eref{DASpSp} is still valid with $R_{\rm eff}=R$.

Now if one takes the integrand in \eref{DASpSp} in the form given by \eref{way} [see also the text below \eref{scaling_function_Casimir}] with $d=3$, and performs the integration there, the result is
\begin{equation}\label{intDAvdW}
\int_{L}^{\infty}f_{\rm vdW}^{\parallel}(z){\rm d}z=\dfrac{\beta H_{A}^{({\rm reg})}\vartheta^{\sigma-3}}{L^{\sigma-1}}.
\end{equation}
Here we recall that the Hamaker term $H_{A}$, depends both on the dimensionality $d$ of the system and the decay parameter $\sigma$ characterizing the strength of the van der Waals interactions [see Eqs. (\ref{HA_ns}) and (\ref{HA_sing}) below]. As far as the CCF is concerned, the DA for $d=3$, after substituting the force per unit area in the form \eref{scaling_function}, the integration of \eref{DASpSp}  delivers
\begin{eqnarray}\label{intDACas}
\int_{L}^{\infty}f_{\rm Cas}^{\parallel}(z){\rm d}z&&=\int_{L}^{\infty}z^{-3}X_{\mathrm{Cas}}^{\parallel}[x_t(z),x_\mu(z),\cdots]{\rm d}z\nonumber\\ &&\equiv X_{\mathrm{Cas,DA}}(x_t,x_\mu,\cdots),
\end{eqnarray}
where, due to the rapid decay of the interaction, the upper limit of integration has been set to infinity. Of course, this can be justified only if $L$ is much smaller than the characteristic sizes of the interacting objects involved.

%
\subsubsection{The studied forces within the SIA}\label{sec:SIA}
%
\begin{figure*}[t!]
\centering
\mbox{\subfigure{\includegraphics[width=8.7 cm]{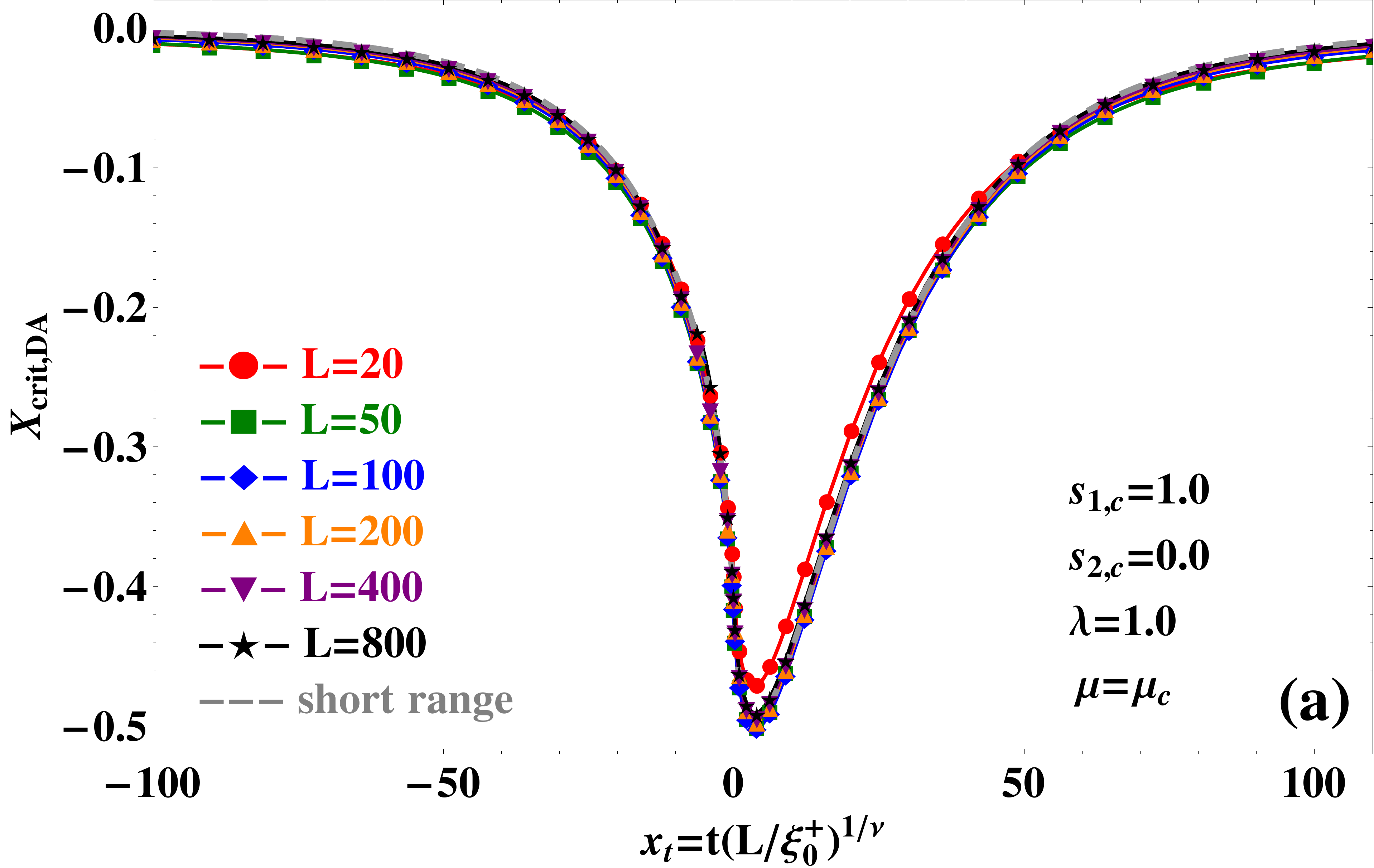}}\quad
      \subfigure{\includegraphics[width=8.7 cm]{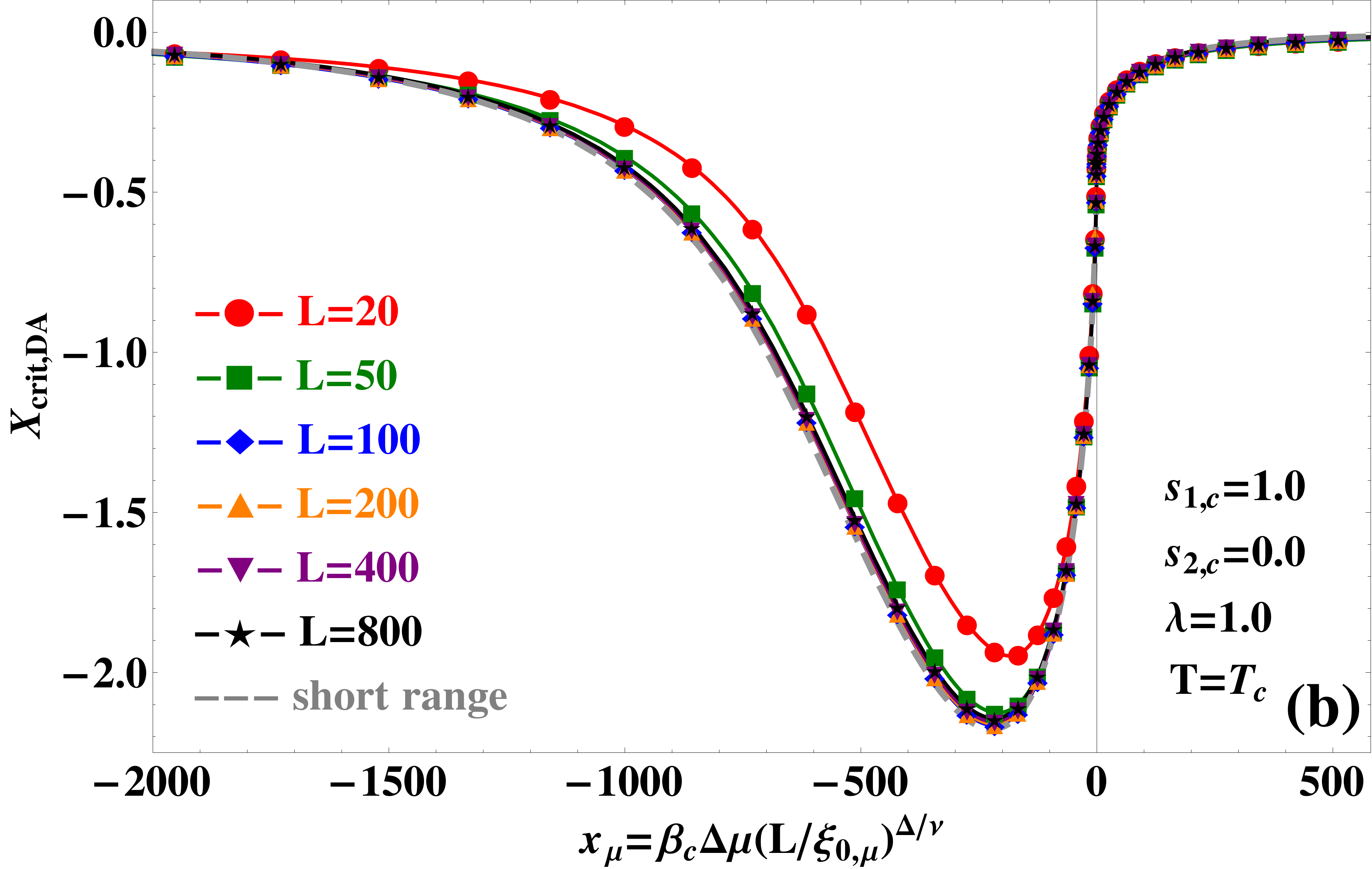}}}\\
\mbox{\subfigure{\includegraphics[width=8.7 cm]{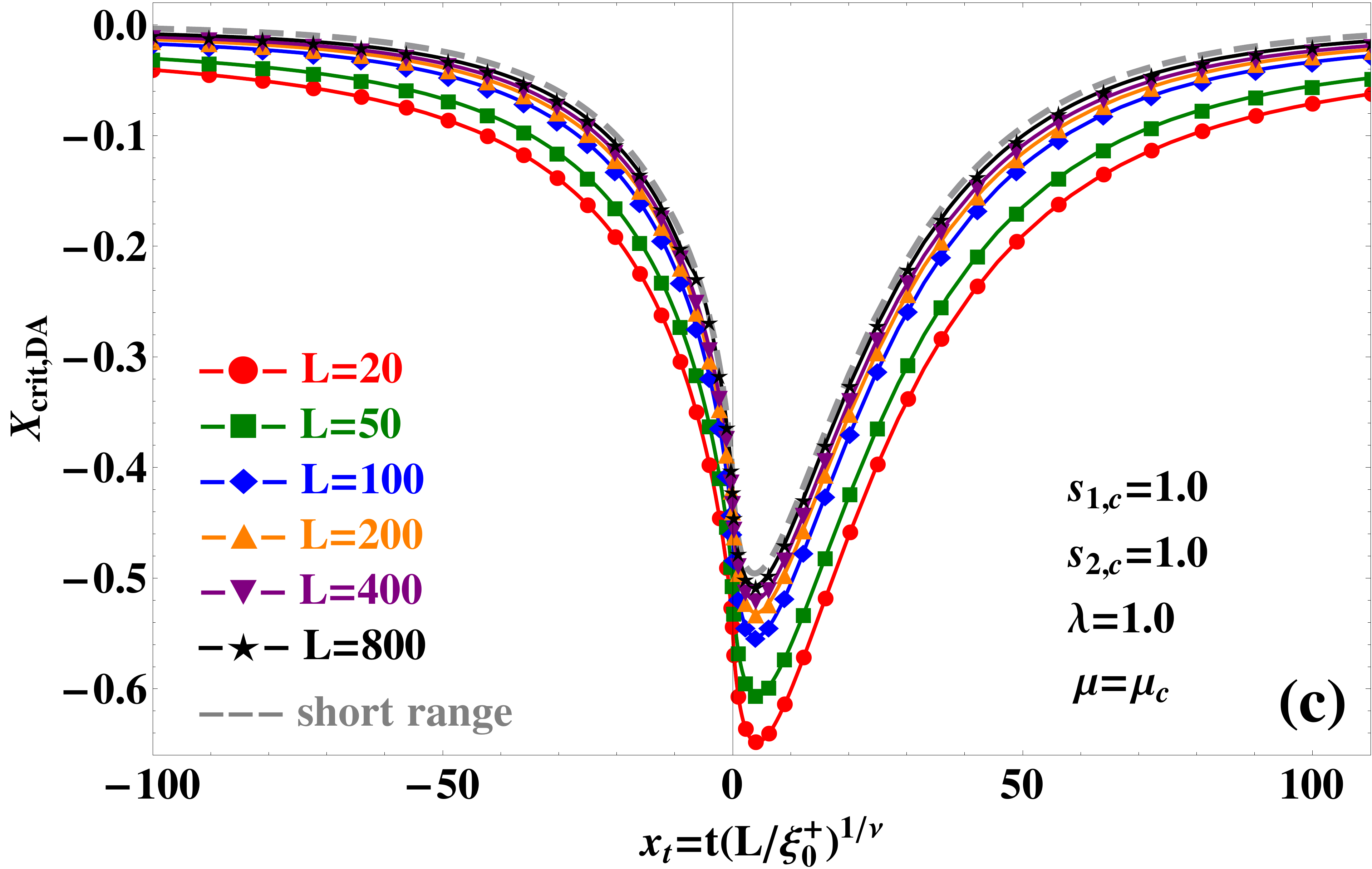}}\quad
      \subfigure{\includegraphics[width=8.7 cm]{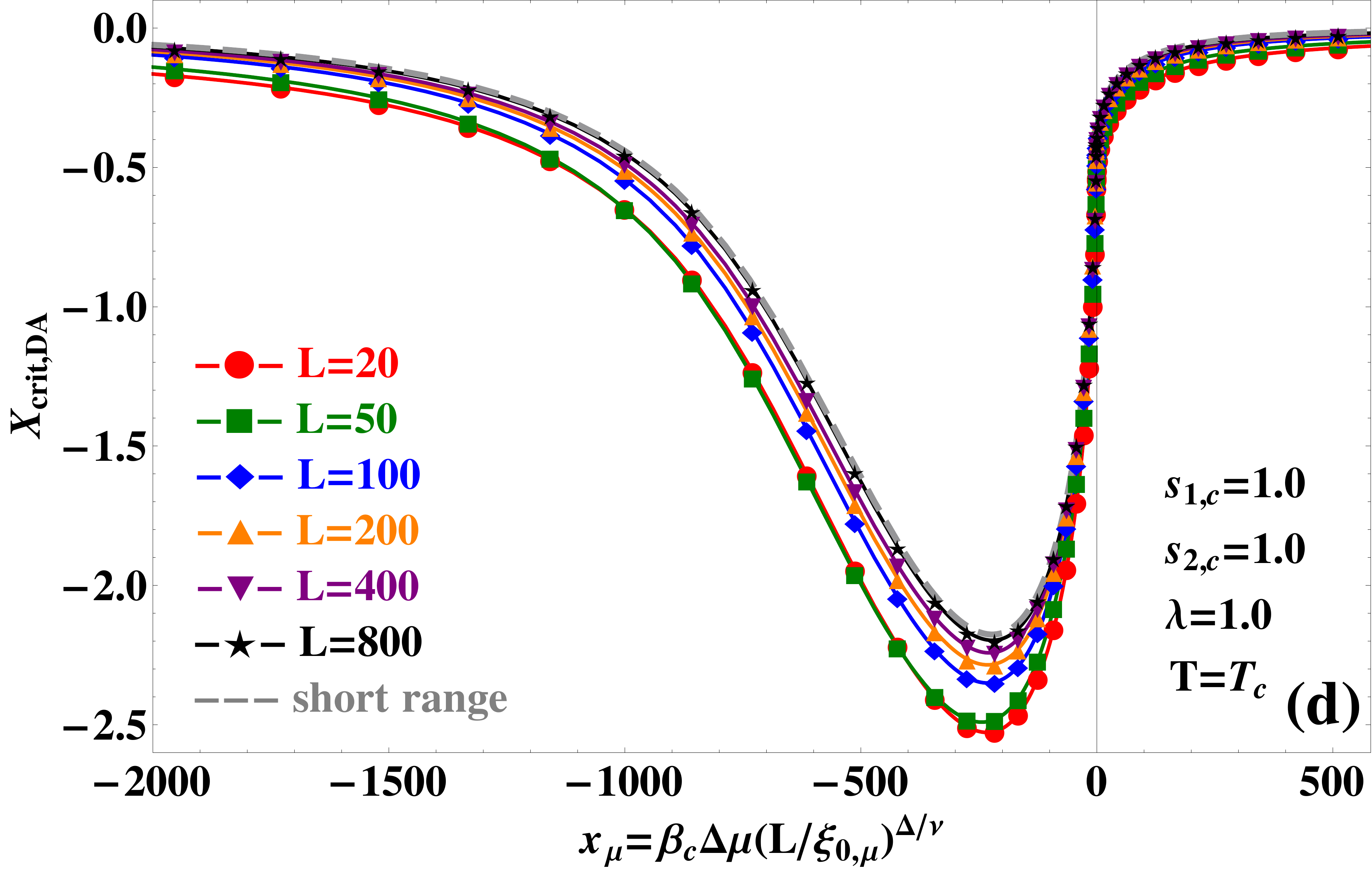}}}
  \caption{(Color online) Behavior of the scaling function $X_{\rm crit,DA}$ within the DA in a $d=3$ dimensional sphere-plate/sphere-sphere fluid system \cite{note2}. In all four sub-figures the parameters characterizing the interactions in the systems have values $s_{1,c}=1.0$ and $\lambda=1.0$, with $s_{2,c}=0.0$ in $\mathbf{(a)}$ and $\mathbf{(b)}$, while in $\mathbf{(c)}$ and $\mathbf{(d)}$ one has $s_{2,c}=1.0$ \cite{note3}. We observe that for $s_{2,c}=0.0$ all scaling functions of systems with $L>20$ are indistinguishable from the one for the short-range system [see $\mathbf{(a)}$ and $\mathbf{(b)}$]. When $s_{2,c}=1.0$ the scaling functions do trace separate curves for different separations $L$ and with increase of $L$ they approach the one for a purely short-ranged system. All shown scaling functions correspond to attractive force.}
  \label{fig:Xcrits11s2gz}
\end{figure*}
An improvement and generalization of the DA, called "surface integration approach" has  been proposed in Ref. \cite{DV2012}. It has been used there to study van der Waals interactions between objects of arbitrary shape and a plate of arbitrary thickness. It delivers {\it exact }results if the interactions involved can be described by pair potentials. The main advantage of this approach over the DA is that one is no longer bound by the restriction that the interacting objects must be much closer to each other than their characteristic sizes. The main result is that for the force acting between a $3\mathrm{d}$ object (say a colloid particle) $B\equiv\{(x,y,z),(x,y,z)\in B\}$ of general shape $S(x,y)=z$ and a flat surface bounded by the $(x,y)-$plane of a Cartesian coordinate system, one has
\begin{eqnarray}\label{SIAgeneralsimple}
F^{B,|}_{\rm SIA}(L) &=&\int_{A_{S}^{\rm to}}\int  f^{\parallel}_{{\cal A}}[{\rm S}(x,y)] \mathrm{d}x \mathrm{d}y\nonumber \\
&-&\int_{A_{S}^{\rm away}}\int  f^{\parallel}_{{\cal A}}[{\rm S}(x,y)] \mathrm{d}x \mathrm{d}y,
\end{eqnarray}
where $A_{S}$ is the projection of the surface $S$ of the particle over the $(x,y)-$ plane, $A_S=A_S^{\rm to}\bigcup A_S^{\rm away}$.
Eq. \eqref{SIAgeneralsimple} has a very simple intuitive meaning: in order to determine the force acting on the particle one has to subtract from the contributions stemming from surface regions $A_S^{\rm to}$ that "face towards" the projection plane  those from regions $A_S^{\rm away}$  that "face away" from it, where  $A_S^{\rm to}$ and $A_S^{\rm away}$  are the projections of the corresponding parts of the surface of the body on the $(x,y)-$plane. It is clear that if one takes into account only the contributions over $A_S^{\rm to}$  one obtains expression very similar to the DA. Both expressions in that case will differ only by the fact that while \eref{SIAgeneralsimple} takes into account that the force on a given point of the $S$ is {\it along the normal} to the surface at that point, the standard DA does not take this into account. Let us recall that \eref{SIAgeneralsimple} provides exact results for the interaction under the assumption that the constituents of the body interact via pair potentials. This is, of course, {\it not} the case of CCF. It is, however, clear that under mechanical equilibrium of the colloid in the fluid, the CCF is again along the normal to the surface at the point of the surface where it acts. Thus, one can get a reasonably good approximation to the effect of that force by keeping just the integration over part of the surface of the body that faces the plane. This leads to
\begin{eqnarray}\label{SIA_Casimir}
\beta F^{B,|}_{{\rm Cas},{\rm SIA}}(L) =\int_{A_{S}^{\rm to}}\int  f^{\parallel}_{\rm Cas}[{\rm S}(x,y)] \mathrm{d}x \mathrm{d}y.
\end{eqnarray}
In the simplest case of interaction between a spherical particle and a thick plate [see Fig. \ref{fig:SIAfig}${\bf(a)}$], induced by point-like sources, \eref{SIAgeneralsimple} takes the form
\begin{equation}\label{SpPlgeneralSIA1}
F_{\rm SIA}^{R,|}(L)=2\pi R\int_{L}^{L+2R}\left[1-\dfrac{z-L}{R}\right]f_{{\cal A}}^{\parallel}(z)\mathrm{d}z.
\end{equation}
The term $[\cdots]$ in the integrand  of \eref{SpPlgeneralSIA1} reflects how the projection to the normal to the surface of the sphere changes as a function of $z$. Substituting \eref{way} [see also the text below \eref{scaling_function_Casimir}] in the above expression for the van der Waals sphere-plate interaction we can write
\begin{eqnarray}\label{SpPlgeneralSIA}
F_{\rm vdW,SIA}^{R,|}(L)&&=\dfrac{2\pi H_{A}^{({\rm reg})}\vartheta^{\sigma-3}}{\sigma-2}\nonumber\\
&&\times \left[\frac{R(\sigma-2)-L}{L^{\sigma-1}}+\frac{L+R\sigma}{(L+2R)^{\sigma-1}}\right].
\end{eqnarray}
The corresponding expression for the CCF arising between a sphere and a plate, following \eref{SIA_Casimir}, can then be written as
\begin{eqnarray}\label{SpPlgeneralSIACas}
&&\beta F_{\rm Cas,SIA}^{R,|}(L)=2\pi R\int_{L}^{L+R}\left[1-\dfrac{z-L}{R}\right]f_{\rm Cas}^{\parallel}(z){\rm d}z\nonumber\\
&&=2\pi R\int_{L}^{L+R}\dfrac{1}{z^{3}}\left[1-\dfrac{z-L}{R}\right]X_{\mathrm{Cas}}^{\parallel}[x_t(z),x_\mu(z),\cdots]{\rm d}z.\ \ \ \ \ \ \ \
\end{eqnarray}

When one is interested in the interaction between two or more objects with nonplanar geometry a general expression like \eref{SIAgeneralsimple} is, as far as we are aware of, not known. An attempt in this direction was reported in \oncite{BEB1998}, but the equation offered there leads to energy of interaction between two spheres which differs from the classical one reported by Hamaker (see Fig. 3 in \oncite{BEB1998}). In the current article we will show that in the special case of pair of spherical particles [see Fig. \ref{fig:SIAfig}${\bf(b)}$] with radii $R_{1}$ and $R_{2}$ one can write the force in the form (for details see Appendix \ref{sec:AppSpSpSIA})
\begin{eqnarray}\label{sphere_sphere_force_SIA_intext}
F_{\rm SIA}^{R_{1},R_{2}}(L)=&&-R_{1}\int_{L}^{L+2R_{2}}\left[f_{{\cal A}}^{\parallel}(z_{2})
+f_{{\cal A}}^{\parallel}(z_{2}+2R_{1})\right]\zeta(z_{2})\mathrm{d}z_{2}\nonumber\\
&&+\int_{L}^{L+2R_{2}}\zeta(z_{2})\int_{z_{2}}^{z_{2}+2R_{1}}f_{{\cal A}}^{\parallel}(z_{1})\mathrm{d}z_{1}\mathrm{d}z_{2},
\end{eqnarray}
where the function $\zeta(z_{2})$ is
\begin{equation}\label{eq:dzeta_deriv_def}
\zeta(z_{2})=\dfrac{{\rm d}}{{\rm d}L}\left\{\dfrac{\pi\left[R_{2}^{2}-(L+R_{2}-z_{2})^{2}\right]}{(L+R_{1}+R_{2})}\right\}.
\end{equation}
From \eref{sphere_sphere_force_SIA_intext} with $\sigma=3$ and $\sigma=4$ one obtains
\begin{itemize}
	\item when $\sigma=3$
	\begin{eqnarray}\label{vdWSIAsigma3}
	F_{\rm vdW,SIA}^{R_{1},R_{2}}(L|3)&&=\dfrac{128H_{A}^{\rm (reg)}R_{1}^{3}R_{2}^{3}}{L^{2}(L+2R_{1})^{2}(L+2R_{2})^{2}}\nonumber\\&&\times
	\dfrac{(L+R_{1}+R_{2})}{(L+2R_{1}+2R_{2})^{2}}.
	\end{eqnarray}
	This result can be easily verified by simple differentiation with respect to the separation distance, of the potential obtained in \oncite{H37}.
	\item when $\sigma=4$
	\begin{eqnarray}\label{vdWSIAsigma4}
	&&F_{\rm vdW,SIA}^{R_{1},R_{2}}(L|4)=\dfrac{H_{A}^{\rm (reg)}\vartheta}{(L+R_{1}+R_{2})^{2}}\nonumber\\
	&&\times\left\{\dfrac{2R_{1}\mathfrak{P}_{1}(L)}{L^{3}(L+2R_{1})^{3}}-\dfrac{2R_{1}\mathfrak{P}_{2}(L)}{(L+2R_{2})^{3}(L+2R_{1}+2R_{2})^{3}}
	\right.\nonumber\\&&\left.-\ln\left[\dfrac{(L+2R_{1})(L+2R_{2})}{L(L+2R_{1}+2R_{2})}\right]\right\},
	\end{eqnarray}
	where the two polynomials which enter the above expression are
	\begin{widetext}
	\begin{subequations}\label{polynomialssigma4}
		\begin{eqnarray}\label{poly1s4}
		&&\mathfrak{P}_{1}(L)=L^5+5L^{4}R_{1}+10L^{3}R_{1}^{2}+2L^{2}R_{1}^{2}(5R_{1}-4R_{2})+4LR_{1}^{2}(R_{1}^{2}-4R_{1}R_{2}-2R_{1}^{2})
		-8R_{1}^{3}R_{2}(R_{1}+R_{2}),
		\end{eqnarray}
		and
		\begin{eqnarray}\label{poly2s4}
		&&\mathfrak{P}_{2}(L)=L^5 + 5L^{4}(R_{1}+2R_{2})+10L^{3}(R_{1}+2R_{2})^{2}+2L^{2}(5R_{1}^{3}+34R_{1}^{2}R_{2}+60R_{1}R_{2}^{2}+40R_{2}^{3})\nonumber\\
		&&+\, 4L(R_{1}^{4}+14R_{1}^{3}R_{2}+36R_{1}^{2}R_{2}^{2}+40R_{1}R_{2}^{3}+20R_{2}^{4})+16R_{2}(R_{1}+R_{2})(R_{1}+2R_{2})(R_{1}^{2}+R_{1}R_{2}+R_{2}^{2}).
		\end{eqnarray}
	\end{subequations}
\end{widetext}
\end{itemize}

\begin{figure*}[t!]
\centering
\mbox{\subfigure{\includegraphics[width=8.7 cm]{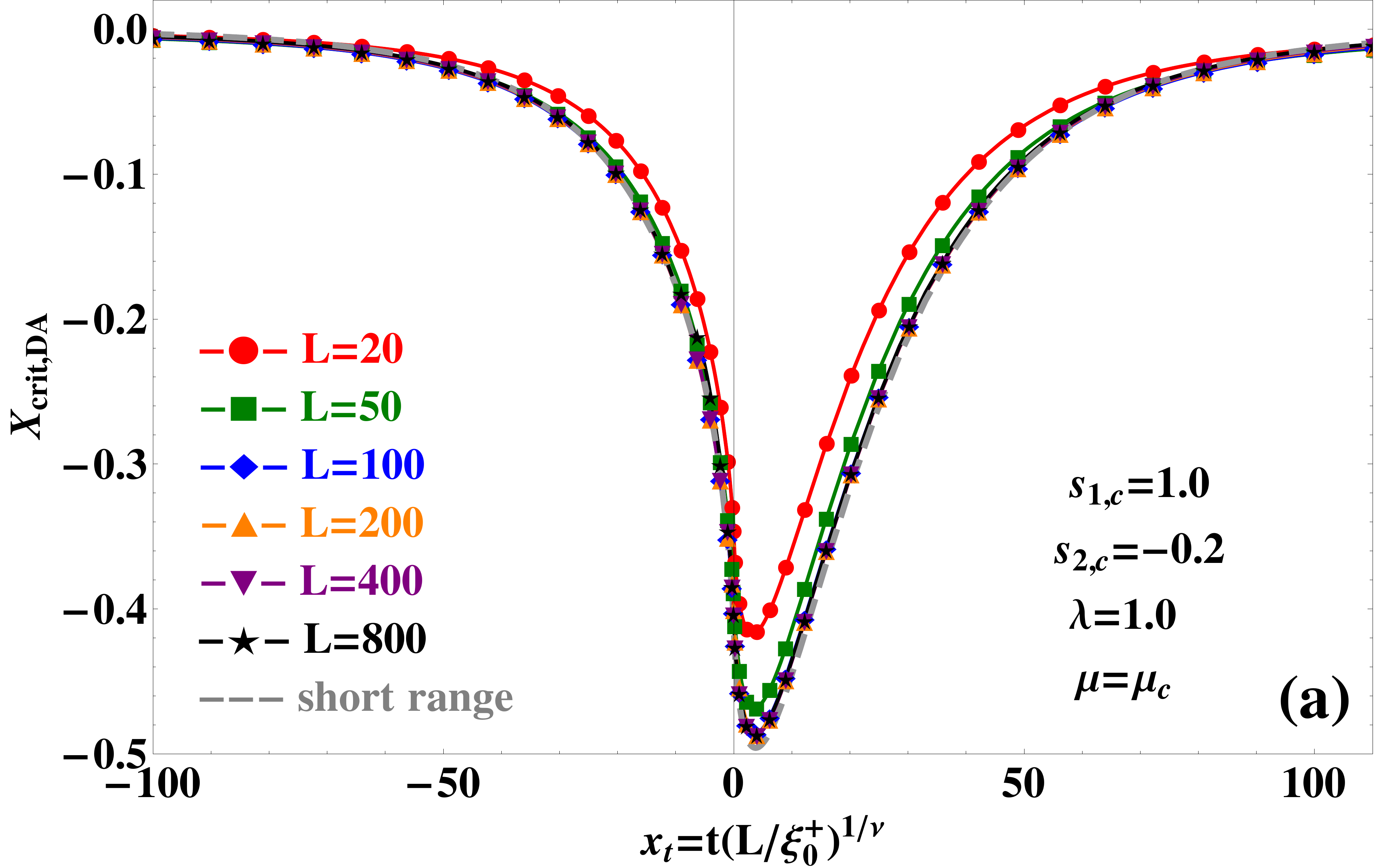}}\quad
      \subfigure{\includegraphics[width=8.7 cm]{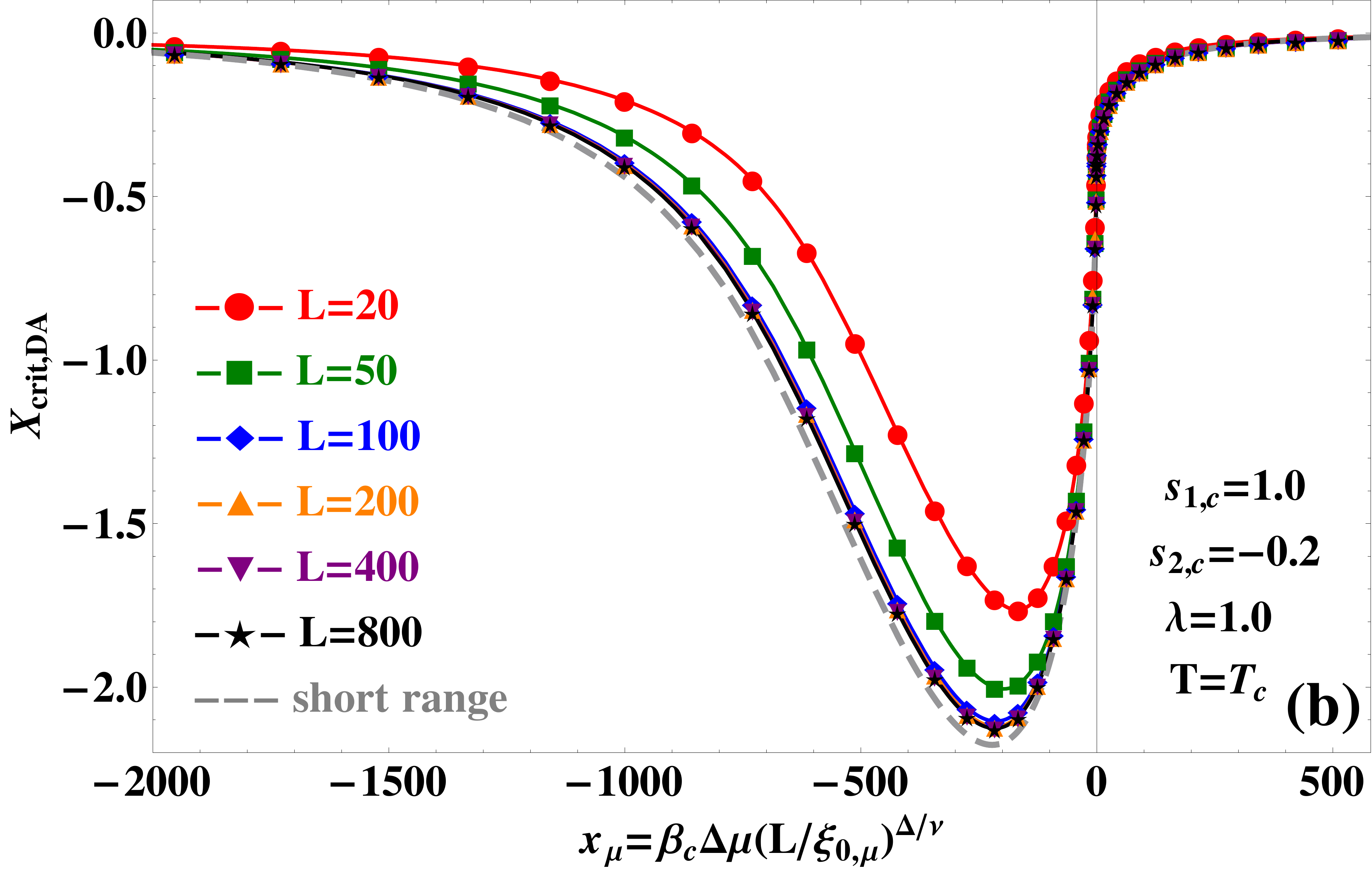}}}\\
\mbox{\subfigure{\includegraphics[width=8.7 cm]{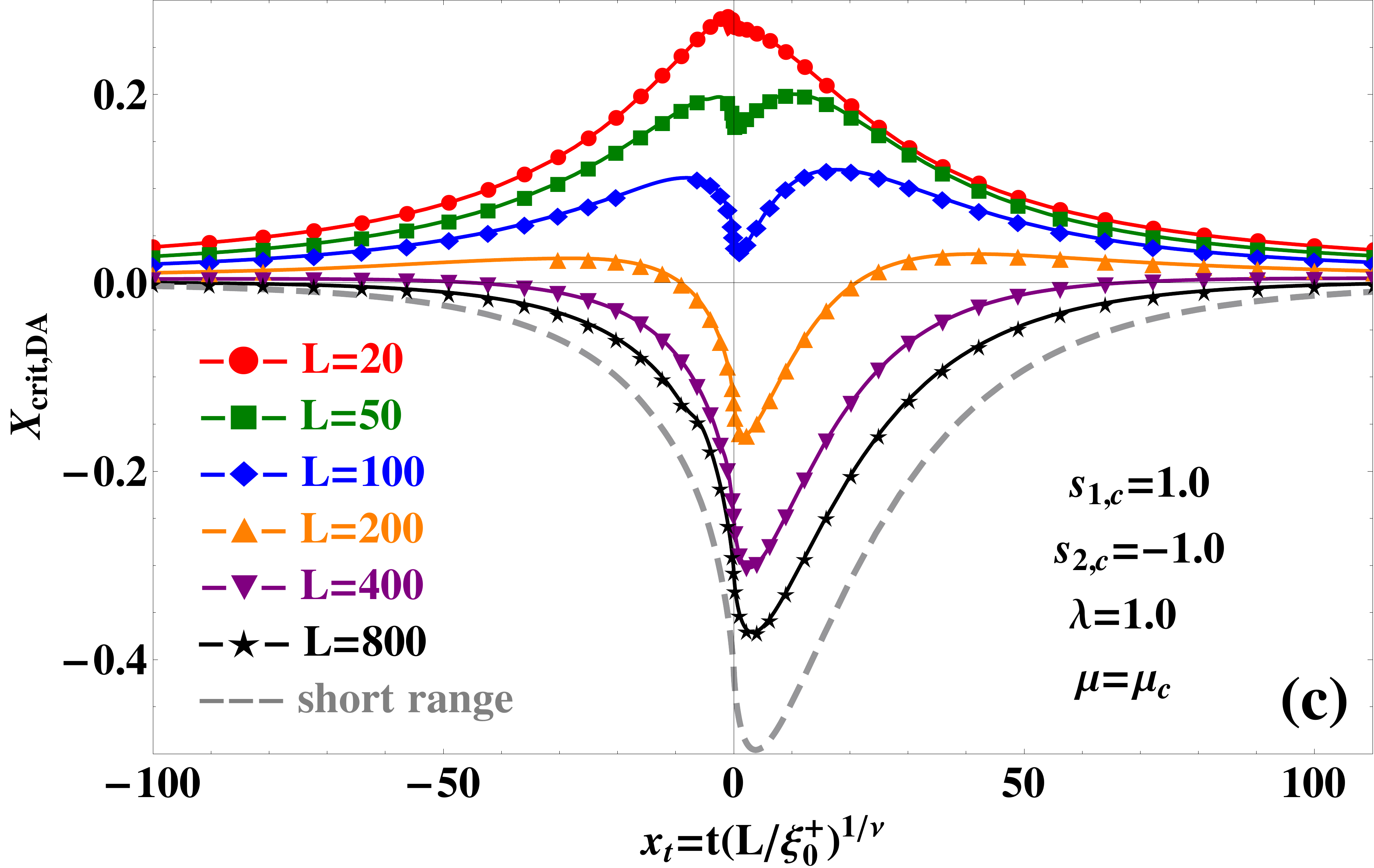}}\quad
      \subfigure{\includegraphics[width=8.7 cm]{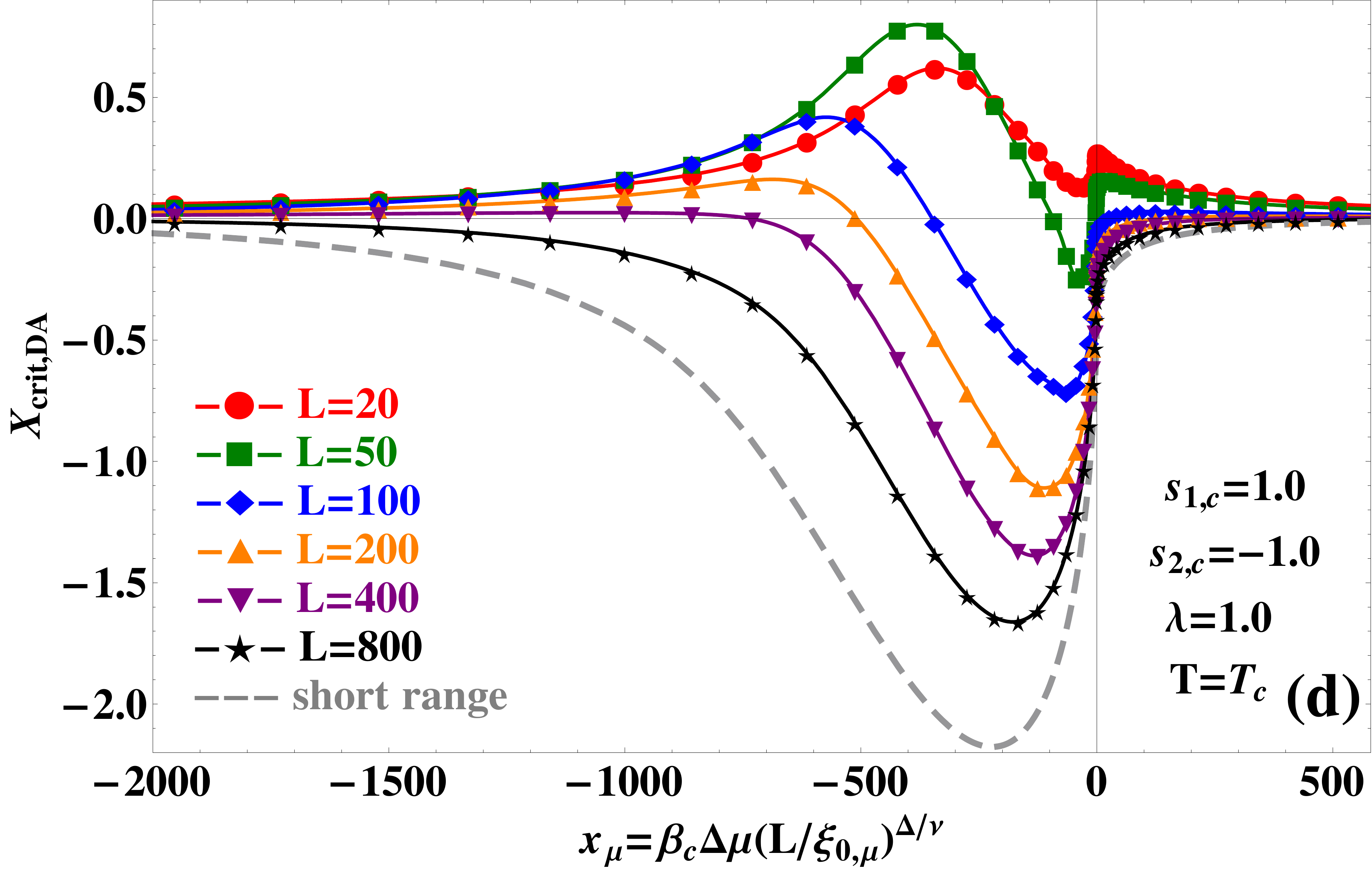}}}
  \caption{Behavior of the scaling function $X_{\rm crit,DA}$ within the DA in a $d=3$ dimensional sphere-plate/sphere-sphere fluid system \cite{note2}. In all four sub-figures the parameters characterizing the interactions in the systems have values $s_{1,c}=1.0$ and $\lambda=1.0$, with $s_{2,c}=-0.2$ in $\mathbf{(a)}$ and $\mathbf{(b)}$, while in $\mathbf{(c)}$ and $\mathbf{(d)}$ one has $s_{2,c}=-1.0$ \cite{note3}. Due to the low absolute value of $s_{2,c}$ with comparison to $s_{1,c}$ in $\mathbf{(a)}$ and $\mathbf{(b)}$ the scaling functions are not well dispersed and as it can be seen for $L>50$ they practically coincide with that for pure short-range system [compare with Figs. \ref{fig:Xcrits11s2gz}$\mathbf{(a)}$ and \ref{fig:Xcrits11s2gz}$\mathbf{(b)}$]. When $s_{2,c}=-1.0$ the scaling functions do trace separate curves for different film thicknesses $L$.  With increase of $L$ they slowly approach the one for a purely short-ranged system. For $\Delta\mu=0.0$ [see $\mathbf{(c)}$] one notices that for $L\leq 100$ the scaling functions correspond to {\it repulsive} forces. For $L=200$ the force changes sign twice. When $L$ is further increased the force becomes entirely attractive for $L>400$. At $T=T_{c}$ [see $\mathbf{(d)}$] some of the scaling functions change sign twice in the "gas" phase of the fluid medium (those for $L<100$), except the scaling function for $L=20$, which corresponds to entirely repulsive force. Upon increasing the separation the change occurs only once, and is not observed for $L>400$.}
  \label{fig:Xcrits11s2sz}
\end{figure*}
For the CCF between spherical objects after the substitution of \eref{scaling_function} in \eref{sphere_sphere_force_SIA_intext} and properly  taking the limits  of  integration leads to
\begin{widetext}
\begin{eqnarray}\label{sphere_sphere_force_SIA_Cas_intext2}
&&\beta F_{\rm Cas,SIA}^{R_{1},R_{2}}(L)=-R_{1}\int_{L}^{L+R_2}\dfrac{\zeta(z_{2})}{z_{2}^{3}}
X_{\mathrm{Cas}}^{\parallel}[x_t(z_{2}),x_\mu(z_{2}),\cdots]\mathrm{d}z_{2}+\int_{L}^{{L+R_2}}\zeta(z_{2})\int_{z_{2}}^{z_{2}+R_{1}}\dfrac{1}{z_{1}^{3}}
X_{\mathrm{Cas}}^{\parallel}[x_t(z_{1}),x_\mu(z_{1}),\cdots]\mathrm{d}z_{1}\mathrm{d}z_{2}.\ \ \ \ \ \ \ \ \ \
\end{eqnarray}
\end{widetext}

Since both DA and SIA utilize the knowledge of the behavior of the force per unit area arising between a pair of parallel plates, in the following section we present the corresponding model within which we describe such a system. Because the included expressions have already been presented in details in \oncite{VaDa2015}, here only key results, as well some notations, are going to be given which will be needed in the remainder of the article.
%
\section{The model}\label{sec:Model}
%
As explained in   \oncite{VaDa2015}, we consider a lattice-gas model of a fluid confined between two planar plates, separated at a distance $L$ from each other, with grand canonical potential $\Omega\left[\rho(\mathbf{r})\right]$ given by
\begin{eqnarray}\label{Garand_pot_l_lattice_model}
&&\Omega\left[\rho({\bf r})\right]= k_{B} T \sum_{{\bf r}\in \mathbb{M}}\left\{
\rho({\bf r}) \ln\left[\rho({\bf r})\right]\right.\nonumber\\ &&\left.
+\left[1-\rho({\bf r})\right] \ln\left[1-\rho({\bf r})\right]\right\}
+\dfrac{1}{2}\sum_{{\bf r}, {\bf r}'\in \mathbb{M}}\rho({\bf
r})w^{l}({\bf r}-{\bf r}')\rho({\bf r}') \nonumber \\ &&
+\sum_{{\bf r} \in \mathbb{M}}\left[V^{(s_{1}|l|s_{2})}(z)-\mu\right] \rho({\bf r}),
\end{eqnarray}
where $\mathbb{M}$ is a simple cubic lattice in the region occupied by the fluid medium -- $\infty^{d-1}\times[0,L]$  and $V^{(s_{1}|l|s_{2})}(z)$ is an external potential that reflects the interactions between the confining plates and the constituents of the fluid. In \eref{Garand_pot_l_lattice_model} $w^{l}({\bf r}-{\bf r}')=-4J^{l}({\bf r}-{\bf r}')$ is the nonlocal coupling (interaction potential) between the constituents of the confined medium and $\mu$ is the chemical potential.  All length scales here and in the remainder  are taken in units of the lattice constant $a_{0}$, so that the particle number density $\rho(\bf r)$ becomes a number which varies in the range $[0,1]$. We recall that in the framework of a mean-field treatment with respect to the critical behavior the effective spatial dimension is $d=4$, irrespective of the actual dimension of the model under consideration.

In \eref{Garand_pot_l_lattice_model}
\begin{eqnarray}\label{extrenalpot}
\mbox{i)} \; V^{(s_{1}|l|s_{2})}(z)&=&-\rho_{s_{1}}J_{\rm sr}^{s_{1},l}\delta(z)-\rho_{s_{2}}J_{\rm sr}^{s_{2},l}\delta(L-z)\nonumber\\&&
+v_{s_{1}}(z+1)^{-\sigma}+v_{s_{2}}(L+1-z)^{-\sigma},
\end{eqnarray}
where $v_{s_{i}}=-G(d,\sigma)\rho_{s_{i}}J^{s_{i},l},\ i=1,2$, with
\begin{equation}\label{v_s_l_def}
G(d,\sigma)=4\pi^{(d-1)/2}\dfrac{\Gamma\left(\frac{1+\sigma}{2}\right)}{\sigma\Gamma\left(\frac{d+\sigma}{2}\right)},
\end{equation}
and $\delta(x)$ is the discrete delta function;
\begin{eqnarray}
\mbox{ii)} \; J^{l}(\mathbf{r})=J_{\mathrm{sr}}^{l}\left\{\delta(|\mathbf{r}|)+\delta(|\mathbf{r}|-1)\right\}+
\dfrac{J^{l}\theta(|\mathbf{r}|-1)}{1+|\mathbf{r}|^{d+\sigma}}\label{Jldeftext},
\end{eqnarray}
is a proper lattice version of $-w^{l}(\mathbf{r})/4$ as the interaction energy between the fluid particles, and
\begin{eqnarray}
\mbox{iii)} \; J^{s_{i},l}(\mathbf{r})=J_{\mathrm{sr}}^{s_{i},l}\delta(|\mathbf{r}|-1)+
\dfrac{J^{s_{i},l}\theta(|\mathbf{r}|-1)}{|\mathbf{r}|^{d+\sigma}},\ i=1,2\ \ \ \ \ \ \ \ \label{Jsildeftext}
\end{eqnarray}
is the one between a fluid particle and a substrate particle, $\theta(x)$ is the Heaviside step function with the convention $\theta(0)=0$.

Taking into account the translational symmetry of the system along the bounding surfaces, the variation with respect to $\rho(\bf r)$ leads to an equation of state for the equilibrium density $\rho^{*}({\bf r})=[1+\phi^{*}({\bf r})]/2$, where  $\phi(\mathbf{r})\equiv\phi\left(\mathbf{r}_{\parallel},z\right)=\phi(z)$, with  $\mathbf{r}=\left\{\mathbf{r}_{\parallel},z\right\}$ is the local order parameter profile $\{\phi(z),\ 0\leq z\leq L\}$. In terms of $\phi(z)$  the equation of state can be written in the following form
\begin{eqnarray}\label{order_parameter_equation_d4}
&&\mathrm{arctanh}\left[\phi^{*}(z)\right]=\dfrac{\beta}{2}\left[\Delta\mu-\Delta V(z)\right]\nonumber\\
&&+K\left\{\text{\ {a}}_{d,\sigma}\left(\lambda\right)\phi^{*}(z)+\text{\ {a}}_{d,\sigma}^{nn}\left(\lambda\right)\left[\phi^{*}(z+1)+\phi^{*}(z-1)\right]\right.\nonumber\\&&\left.+\lambda
\sum_{z'=0}^{L}g_{d,\sigma}(|z-z'|)\theta(|z-z'|-1)\phi^{*}(z')\right\},
\end{eqnarray}
where $\Delta\mu=\mu-\mu_{c}$, $K=\beta J_{\mathrm{sr}}^{l}$, $\text{\ {a}}_{d,\sigma}\left(\lambda\right)=(2d-1)+\lambda(c_{d,\sigma}-d)$ and $\text{\ {a}}_{d,\sigma}^{nn}\left(\lambda\right)=1.0+\lambda(c_{d,\sigma}^{nn}-0.5)$ with $c_{d,\sigma}^{nn}=g_{d,\sigma}(1)+g_{d,\sigma}^{nn}(\pm 1)$. The functions $c_{d,\sigma}$, $g_{d,\sigma}(|z-z'|)$ and $g_{d,\sigma}^{nn}(|z-z'|)$ are determined in Eqs. (C10), (C11) and (C12) of \oncite{DSD2007}, respectively.

The excess grand canonical potential per unit area, $\omega_{\rm ex}\equiv\lim_{\cal{A}\to\infty}[\Omega/{\cal{A}}]-L\omega_{\rm bulk}$, has the form
\begin{eqnarray}\label{deltaomegaexz_equilibrium}
\beta\omega_{\rm ex}&&=\sum_{z=0}^{L} \left\{\dfrac{1}{2}\ln\left[1-\phi^{*}(z)^{2}\right]-\dfrac{1}{2}\ln\left[1-\phi_{b}^{2}\right]\right.\nonumber\\&&\left.
+\dfrac{1}{4}\phi^{*}(z)\ln\left[\dfrac{1+\phi^{*}(z)}{1-\phi^{*}(z)}\right]-\dfrac{1}{4}\phi_{b}\ln\left[\dfrac{1+\phi_{b}}{1-\phi_{b}}\right]\right.
\nonumber\\&&\left.+\dfrac{\beta}{2}\Delta V(z)\phi^{*}(z)-\dfrac{\beta\Delta\mu}{2}[\phi^{*}(z)-\phi_{b}]\right\}+\beta\omega_{\rm reg},\ \ \ \ \
\end{eqnarray}
where
\begin{eqnarray}\label{bethaomega_reg}
\beta(\sigma-1)\omega_{\rm reg}&&=\left[\dfrac{K}{K_{c}}(s_{1,c}+s_{2,c})-\dfrac{1}{4}G(d,\sigma)K\lambda\right]\nonumber\\&&
\times L^{-\sigma+1}\vartheta^{\sigma-d}.
\end{eqnarray}
Here $\phi_{b}$ is the bulk value of the order parameter (see Eq. (4.2) in \oncite{VaDa2015})$, K_{c}=\beta_{c}J_{\rm sr}^{l}$ and $s_{i,c},\ i=1,2$ are the values of the plates-fluid coupling parameters evaluated at the bulk critical point of the system $\{\beta=\beta_{c}=[\sum_{\bf r}J^{l}({\bf r})]^{-1}, \mu=\mu_{c}=-2\sum_{\bf r}J^{l}({\bf r})$, with the sum running over the whole lattice$\}$.

The {\it effective} surface potential $\beta\Delta V(z)/2$  in Eqs. (\ref{order_parameter_equation_d4}) and (\ref{deltaomegaexz_equilibrium}) is given by
\begin{eqnarray}\label{DeltaV_l_ab_thichlayers}
\dfrac{\beta}{2}\Delta V(z)=
\dfrac{s_{1}}{(z+1)^{\sigma}}+\dfrac{s_{2}}{(L+1-z)^{\sigma}},
\end{eqnarray}
where $1\le z\le L-1$ and
\begin{eqnarray}\label{s_def_l}
s_{i}=\dfrac{1}{2}\beta G(d,\sigma)
\left(\rho_{s_{i}}J^{s_{i},l}-\rho_c J^l\right),\ i=1,2
\end{eqnarray}
are the ($T$- and $\mu$-independent) dimensionless plates-fluid coupling parameters $\propto x_{s_{i}}$. In \eref{DeltaV_l_ab_thichlayers} the restriction $z\geq1$ holds because we consider the layers closest to the substrate to be completely occupied by the liquid phase of the fluid, i.e., $\phi(0) = \phi(L) = 1$, thus ensuring the $(+,+)$ boundary conditions. Physically this can be accomplished by choosing a proper coating of the surfaces of the plates.
The coupling parameter $\lambda\propto x_{l}$ probes the importance of the long-ranged parts of the interaction potential within the fluid medium
\begin{equation}\label{l_param_def_new}
\lambda=J^{l}/J_{\mathrm{sr}}^{l}.
\end{equation}
In \eref{s_def_l} $s_{i}>0$, i.e., $\rho_{s_{i}}J^{s_{i},l}>\rho_{c}J^{l}$ corresponds to plates "preferring" the liquid phase of the fluid, while $s_{i}<0$, or $\rho_{s_{i}}J^{s_{i},l}<\rho_{c}J^{l}$ mirrors the one with affinity to its gas phase. If the interactions
between the constituents of the fluid are of Lennard-Jones type one has $\lambda=2$, as commented in Refs.  \cite{DSD2007,VaDa2015}.
The marginal case $s_{i}=0$ together with $\lambda=0$ describes a pure short-range system (for details refer to Eqs. (4.10) and the text therein in \oncite{VaDa2015}).

Eqs. \eqref{Garand_pot_l_lattice_model} -- \eqref{l_param_def_new} provide the basis of the properties of the model that will be used to determine the sphere-plane and sphere-sphere interactions in the remainder.

Using Eqs. (\ref{deltaomegaexz_equilibrium}) and (\ref{eq:total_force}) for the {\it total} force $f_{\rm tot}^{\parallel}(L|T,\mu)$ (per unit area $\mathcal{A}$ and $k_{B}T$) acting between parallel plates bounding the fluid medium the following expression can be written
\begin{eqnarray}\label{totalforcegeneralexpression_new}
f_{\rm tot}^{\parallel}(L|T,\mu)&=&
-\dfrac{\beta}{2}\left[\omega_{\rm ex}(L+1|T,\mu)-\omega_{\rm ex}(L-1|T,\mu)\right]\nonumber\\
&&-\dfrac{4Ks_{1,c}s_{2,c}}{G(d,\sigma)K_{c}^{2}\lambda}L^{-\sigma}\vartheta^{\sigma-d},
\end{eqnarray}
where the last term represents the direct interaction between the plates (for details see the Appendix in \oncite{VaDa2015}). On the other hand, if one subtracts from the potential $\omega_{\rm ex}$ its regular part $\omega_{\rm reg}$, i.e., if we consider the quantity
\begin{equation}
\label{Delta_omega_def}
\Delta\omega\equiv\lim_{\cal{A}\to\infty}[(\Omega-\Omega_{\rm reg})/{\cal{A}}]-L\omega_{\rm bulk},
\end{equation}
then, in accord with Eqs. (\ref{eq:decomp}) -- (\ref{eq:def_Casimir}), the $L$ dependence of $\Delta\omega$ via \eref{eq:total_force} provides the singular part of the total force, i.e., $f_{\rm Cas}^{\parallel}(L|T,\mu)$. Explicitly, one has
\begin{eqnarray}\label{critCasforcegeneralexpression_new}
f_{{\rm Cas}}^{\parallel}(L|T,\mu)&=&
-\dfrac{\beta}{2}\left[\Delta\omega(L+1|T,\mu)\right.\nonumber\\&&\left.-\Delta\omega(L-1|T,\mu)\right].
\end{eqnarray}
Thus, near $T_c$ the TF and the CCF are related via the expression
\begin{eqnarray}\label{totCasrelation_new}
f_{\rm tot}^{\parallel}(L|T,\mu)=f_{{\rm Cas}}^{\parallel}(L|T,\mu)+(\sigma-1)\beta H_{A}^{({\rm reg})}L^{-\sigma}\vartheta^{\sigma-d},\ \ \ \ \ \ \ \ \ \
\end{eqnarray}
where the last term is the mathematical equivalent of \eref{eq:def_vdW} per unit area $\mathcal{A}$ and $k_{B}T$ for a film geometry with $H_{A}^{({\rm reg})}$ being the nonsingular (regular) part of the Hamaker term given by
\begin{eqnarray}\label{HA_ns}
(\sigma-1)\beta H_{A}^{({\rm reg})}=&&-\dfrac{4Ks_{1,c}s_{2,c}}{G(d,\sigma)K_{c}^{2}\lambda}\nonumber\\
&&+\left[\dfrac{K}{K_{c}}(s_{1,c}+s_{2,c})-\dfrac{1}{4}G(d,\sigma)K\lambda\right].\ \ \ \ \ \ \ \ \ \
\end{eqnarray}
For the singular part of $H_{A}$ within the presented model the following expression was derived (see Eq. (A6) in \oncite{VaDa2015})
\begin{eqnarray}\label{HA_sing}
(\sigma-1)\beta H_{A}^{({\rm sing})}(T,\mu)&&=\dfrac{K}{K_{c}}(s_{1,c}+s_{2,c})\phi_{b}\nonumber\\
 &&-\dfrac{1}{4}G(d,\sigma)K\lambda\phi_{b}^{2},
\end{eqnarray}
where the $T$ and $\mu$ dependance is carried by the bulk order parameter $\phi_{b}$ (see Eq. (4.2) in \oncite{VaDa2015}).

In the next section, based on the results reported in Sections \ref{sec:ThermalCas} and \ref{sec:Model}, we present numerical results for the behaviour of the above discussed forces in sphere-plate and sphere-sphere fluid systems for the cases $d=\sigma=3$. In Subsec. \ref{subsec:ExperimentalFeas} we also show that the values of the parameters used in the  numerical evaluations can be experimentally achieved for specific materials.
\begin{figure*}[t!]
\centering
\mbox{\subfigure{\includegraphics[width=8.7 cm]{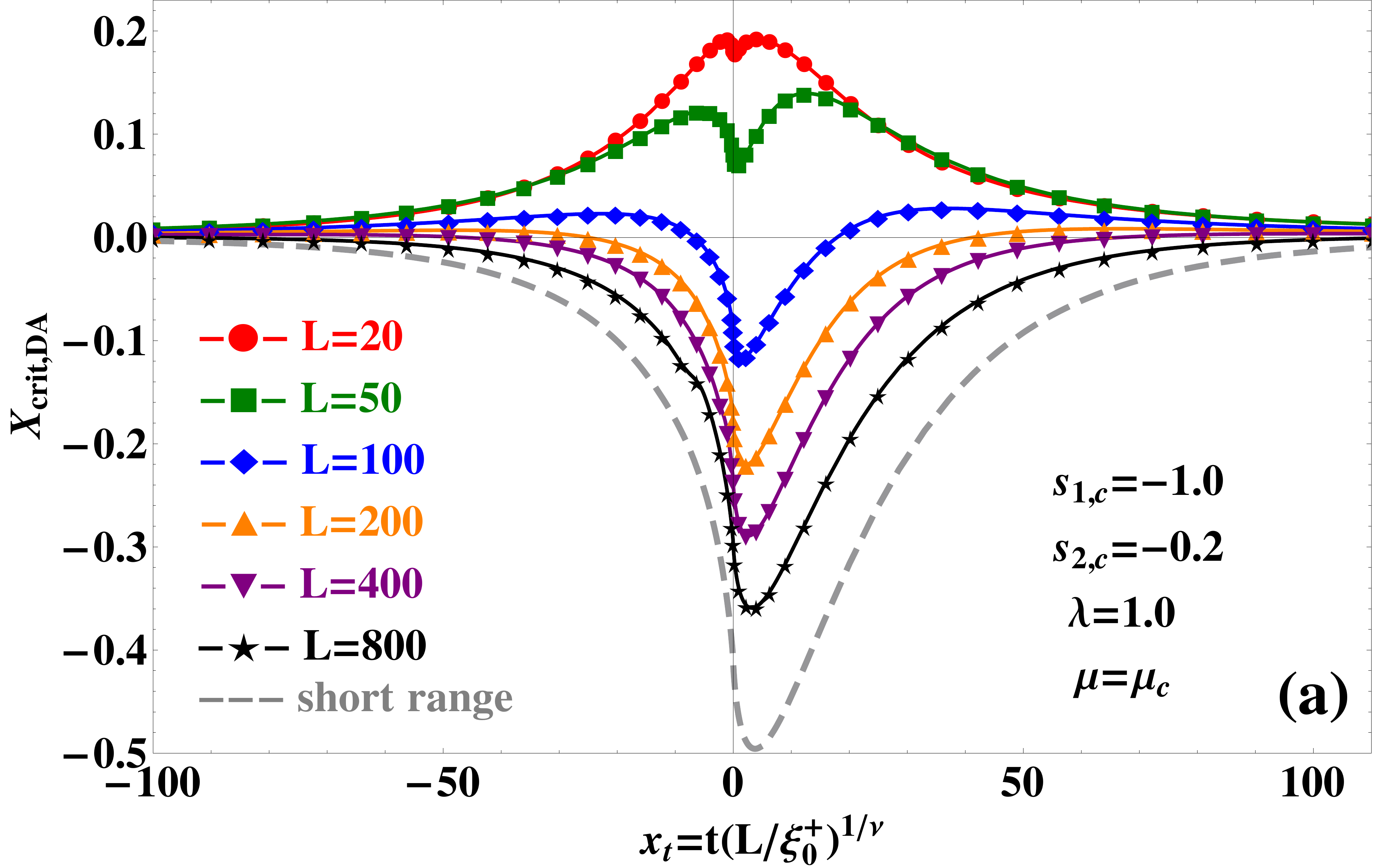}}\quad
      \subfigure{\includegraphics[width=8.7 cm]{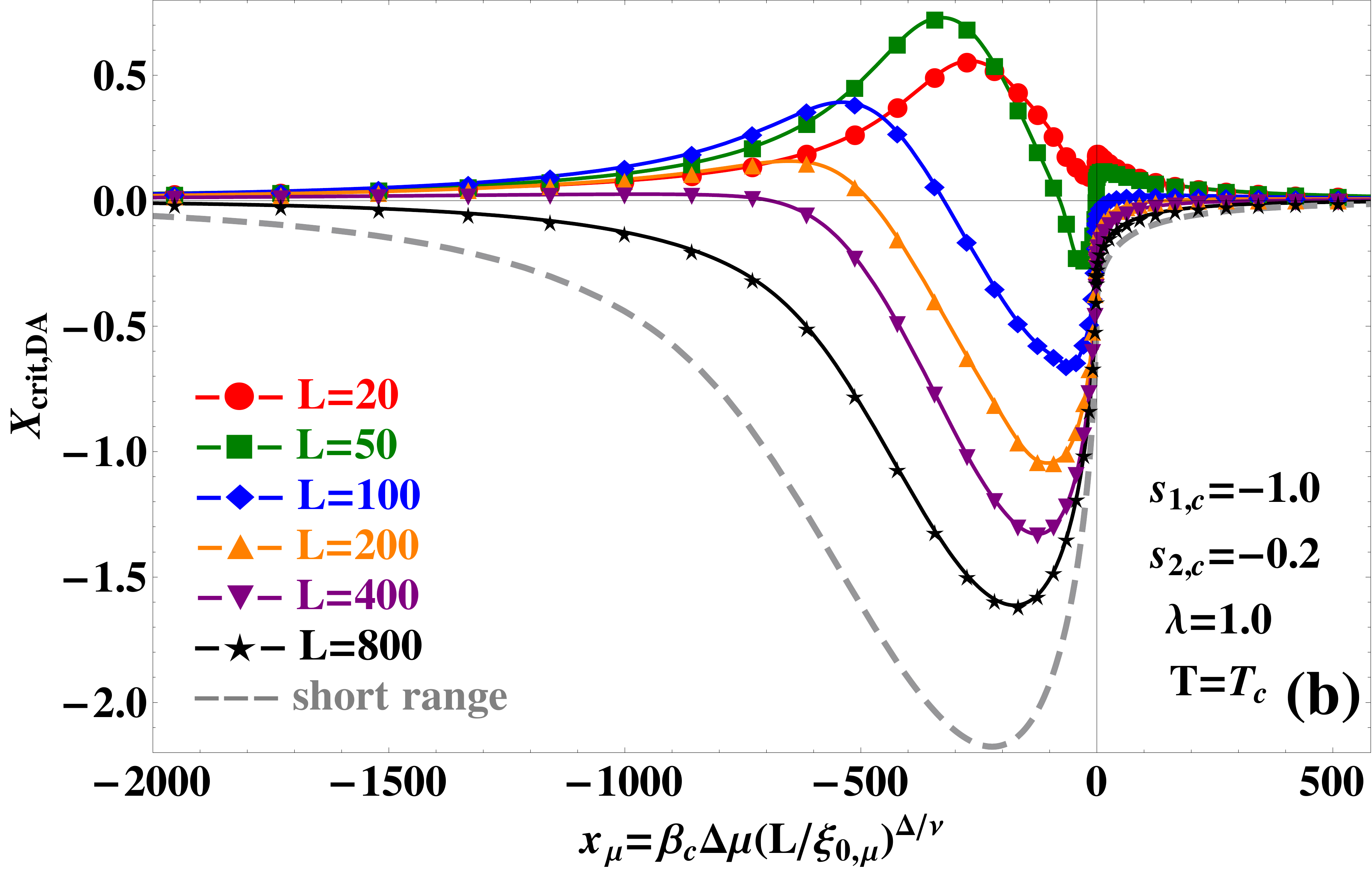}}}\\
\mbox{\subfigure{\includegraphics[width=8.7 cm]{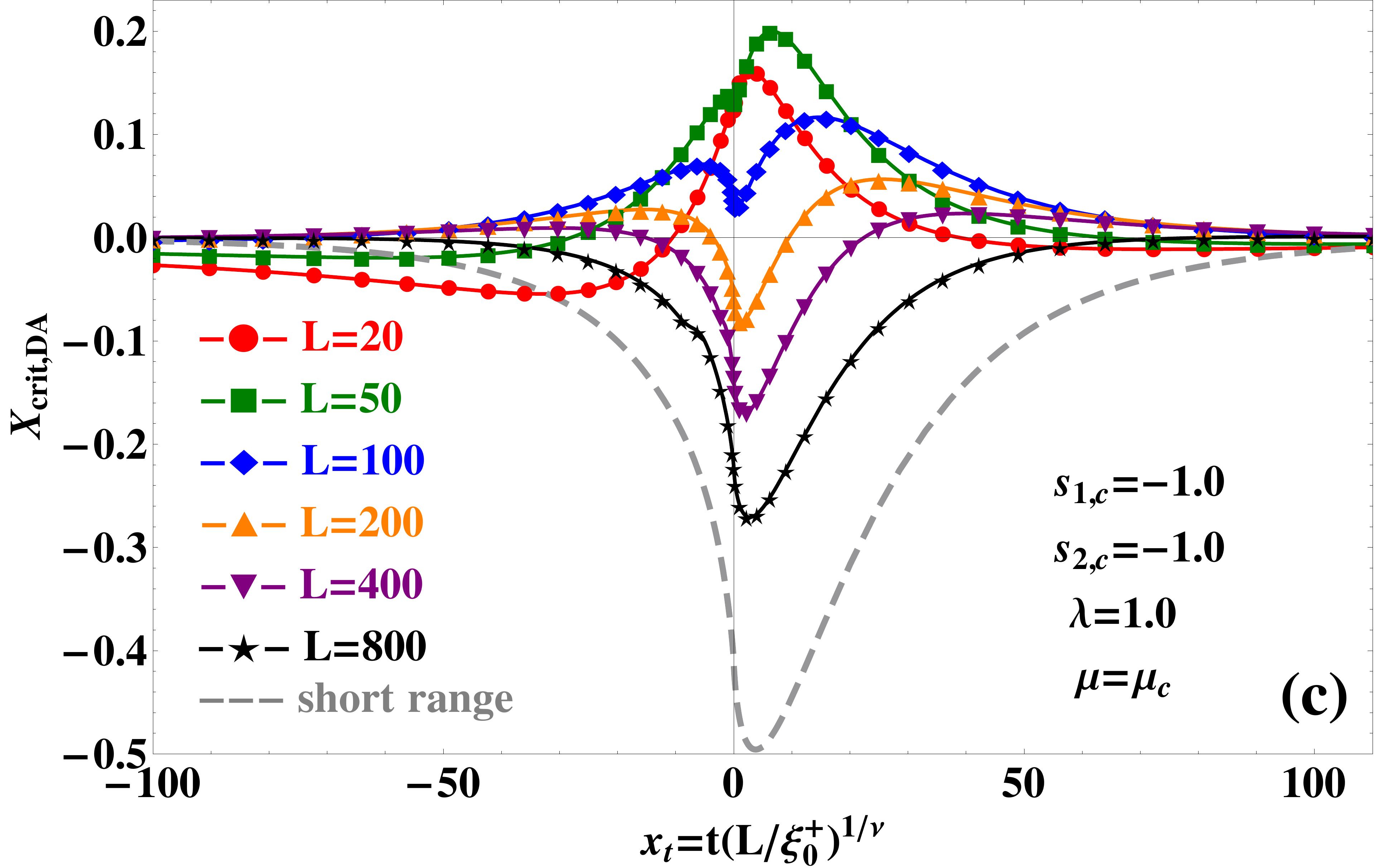}}\quad
      \subfigure{\includegraphics[width=8.7 cm]{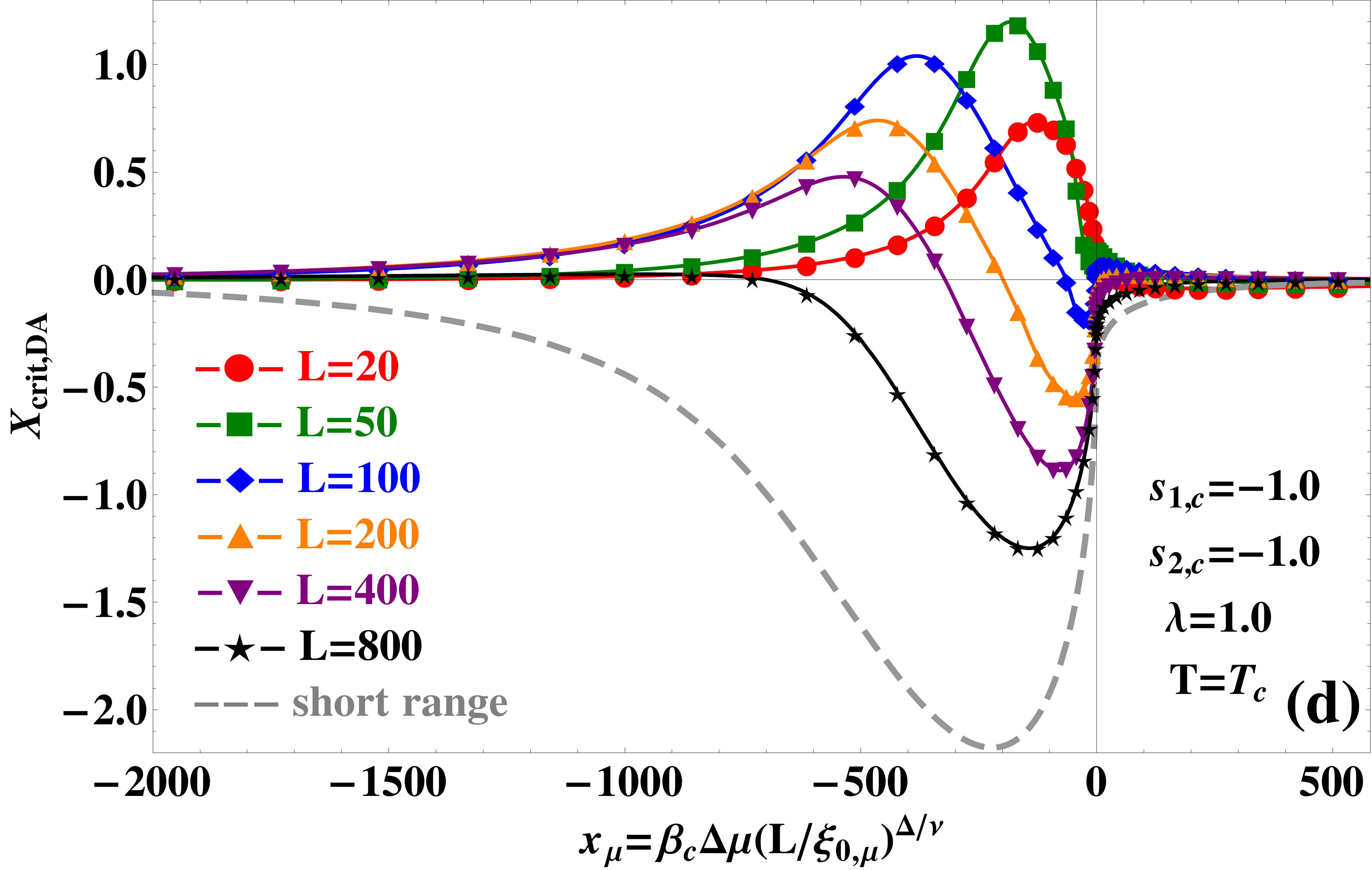}}}
  \caption{Behavior of the scaling function $X_{\rm crit,DA}$ within the DA in a $d=3$ dimensional sphere-plate/sphere-sphere fluid system \cite{note2}. In all four sub-figures the parameters characterizing the interactions in the systems have values $s_{1,c}=-1.0$ and $\lambda=1.0$, with $s_{2,c}=-0.2$ in $\mathbf{(a)}$ and $\mathbf{(b)}$, while in $\mathbf{(c)}$ and $\mathbf{(d)}$ one has $s_{2,c}=-1.0$ \cite{note3}. Although the temperature dependent scaling functions at $x_{\mu}=0.0$ resembles these shown in Fig. \ref{fig:Xcrits11s2sz}$\mathbf{(c)}$, the changed sign of $s_{1,c}$ here "pushes down" the curves toward the one for $L=400$. This results in fewer systems in which repulsion can occur. In contrast the behaviour of the field dependant scaling functions depicted on $\mathbf{(b)}$ is qualitatively and nearly quantitatively identical to these shown on Fig. \ref{fig:Xcrits11s2sz}$\mathbf{(d)}$. For $s_{2,c}=-1.0$ at $\Delta\mu=0.0$ [see $\mathbf{(c)}$] one observes that for $L=20$ ({\color{red_n}{{\Large-}\hspace{-0.1cm}{\Large-}\hspace{-0.1cm}{\large$\bullet$}\hspace{-0.09cm}{\Large-}\hspace{-0.1cm}{\Large-}}}) the scaling function changes sign twice, having two minima and a maximum. When the separation $L$ is increased the values of the minima decrease rapidly towards zero, while that of the maximum increases, being highest for $L=50$ ({\color{dark_green}{{\Large-}\hspace{-0.1cm}{\Large-}\hspace{-0.1cm}{\small$\blacksquare$}\hspace{-0.09cm}{\Large-}\hspace{-0.1cm}{\Large-}}}). When $L=100$ ({\color{blue_n}{{\Large-}\hspace{-0.1cm}{\Large-}\hspace{-0.1cm}$\blacklozenge$\hspace{-0.09cm}{\Large-}\hspace{-0.1cm}{\Large-}}}) the scaling function corresponds to repulsive force, now having two maxima and a single minimum. Reaching $L=200$ ({\color{orange_n}{{\Large-}\hspace{-0.1cm}{\Large-}\hspace{-0.1cm}$\blacktriangle$\hspace{-0.09cm}{\Large-}\hspace{-0.1cm}{\Large-}}}) the scaling function changes sign twice, and corresponds to attractive force for $L>400$ ({\color{purple_n}{{\Large-}\hspace{-0.1cm}{\Large-}\hspace{-0.1cm}$\blacktriangledown$\hspace{-0.09cm}{\Large-}\hspace{-0.1cm}{\Large-}}}). At $T=T_{c}$ [see $\mathbf{(d)}$] for $L=20$ and $50$ the scaling functions change sign once for $x_{\mu}>0$. Then for $L\in[100,400]$ one observes double sign change, with an increasing minimum occurring in the region $x_{\mu}<0$. The scaling function evaluated for $L=50$ exhibits a pronounced maximum which value is the highest in comparison to the other scaling functions calculated for various separations $L$. Upon increasing the separation sign change of the scaling function occurs only once, and is not observed for $L>800$. Note that for $L=800$ the scaling function still changes sign in the region $x_{\mu}<0$.}
  \label{fig:Xcrits1m1s2sz}
\end{figure*}
%
%
\section{Results and their experimental feasibility}\label{sec:ResultsAndExpFeas}
%
In the current section, using the results for $d=\sigma=4$ from the mean-field type numerical study discussed in Sections \ref{sec:ThermalCas} and \ref{sec:Model}, we will present some approximate results for the behavior of the CCF, vdWF and TF between two spherical particles as well as between a spherical particle and a plate in $d=3$, using both the DA and SIA approximations. As it became clear from the above shown equations, the key knowledge which is required for the desired calculations to be performed, is the force [scaling function(s)] per unit area between two parallel semi-infinite spaces (plates), for many different separations $L$, and at various values of $T$ and $\mu$ of the fluctuating fluid medium.

Within the mean-field theory the $T$ and $\mu$ dependance of the corresponding forces near the bulk critical point $(T=T_c,\Delta\mu=0)$ is given by \eref{eq:def_scaling_var}, with $\nu=1/2$ and $\Delta=3/2$. In our numerical treatment we take these variables to range in the intervals: $x_{t}\in\left[-24^{2};24^{2}\right]$ and $x_{\mu}\in\left[-24^{3};24^{3}\right]$. For the study of the scaling function of the CCF within the DA, the separation $L$ between the set of parallel plates is varied from 20 to 100 with step 10 and from 100 to 1000 with step 100. In addition, within the SIA approximation, $L$ was varied from 20 to 60 with a step of 2, from 60 to 100 with a step of 5, and from 100 to 200 with a step of 10. In order to demonstrate the general tendencies in the behaviour of the scaling functions, we consider one of the plates-fluid coupling parameters fixed (say $s_{1,c}$), having a value either $1.0$ or $-1.0$, while the other one, $s_{2,c}$, is varied from $1.0$ to $-1.0$ with step $-0.2$. The fluid-fluid coupling parameter $\lambda$ is supposed to be either $1.0$ or $2.0$.

\subsection{Calculation of the forces in sphere-plate and sphere-sphere systems within the DA and SIA in $d=3$ }\label{subsec:SpPlSpSpDASIAd3}
%
\begin{figure*}[t!]
	\centering
	\mbox{\subfigure{\includegraphics[width=8.7 cm]{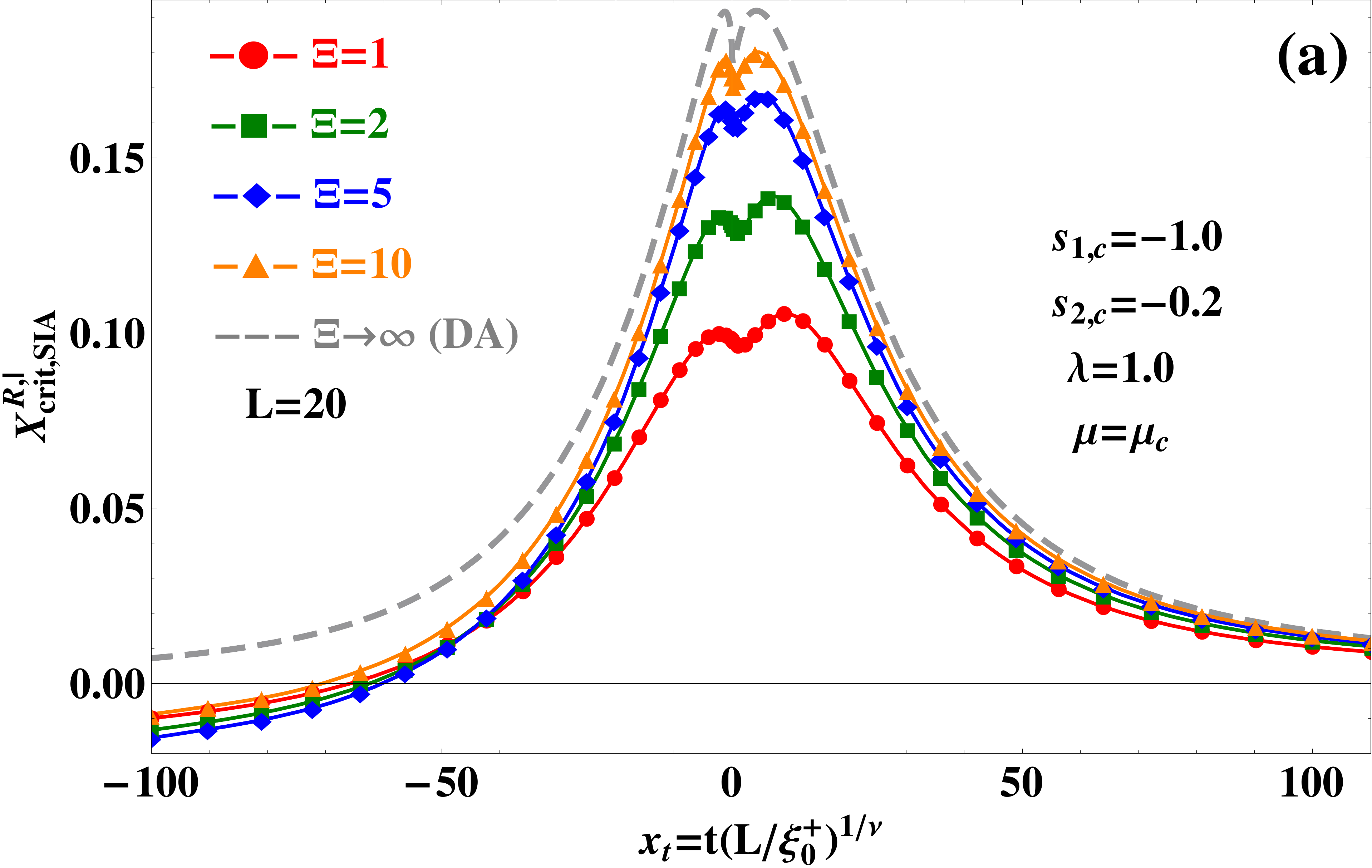}}\quad
		\subfigure{\includegraphics[width=8.6 cm]{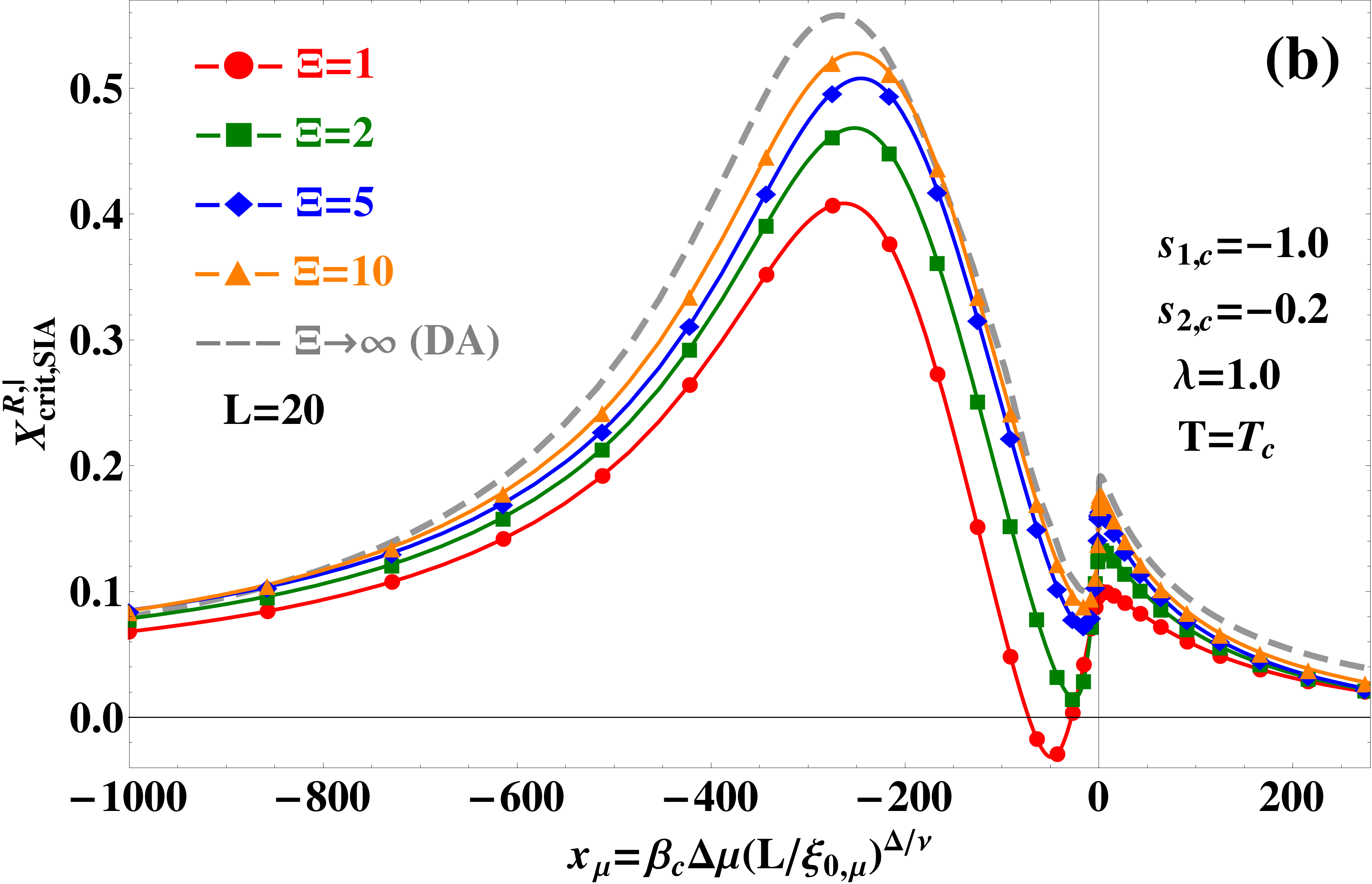}}}\\
	\mbox{\subfigure{\includegraphics[width=8.7 cm]{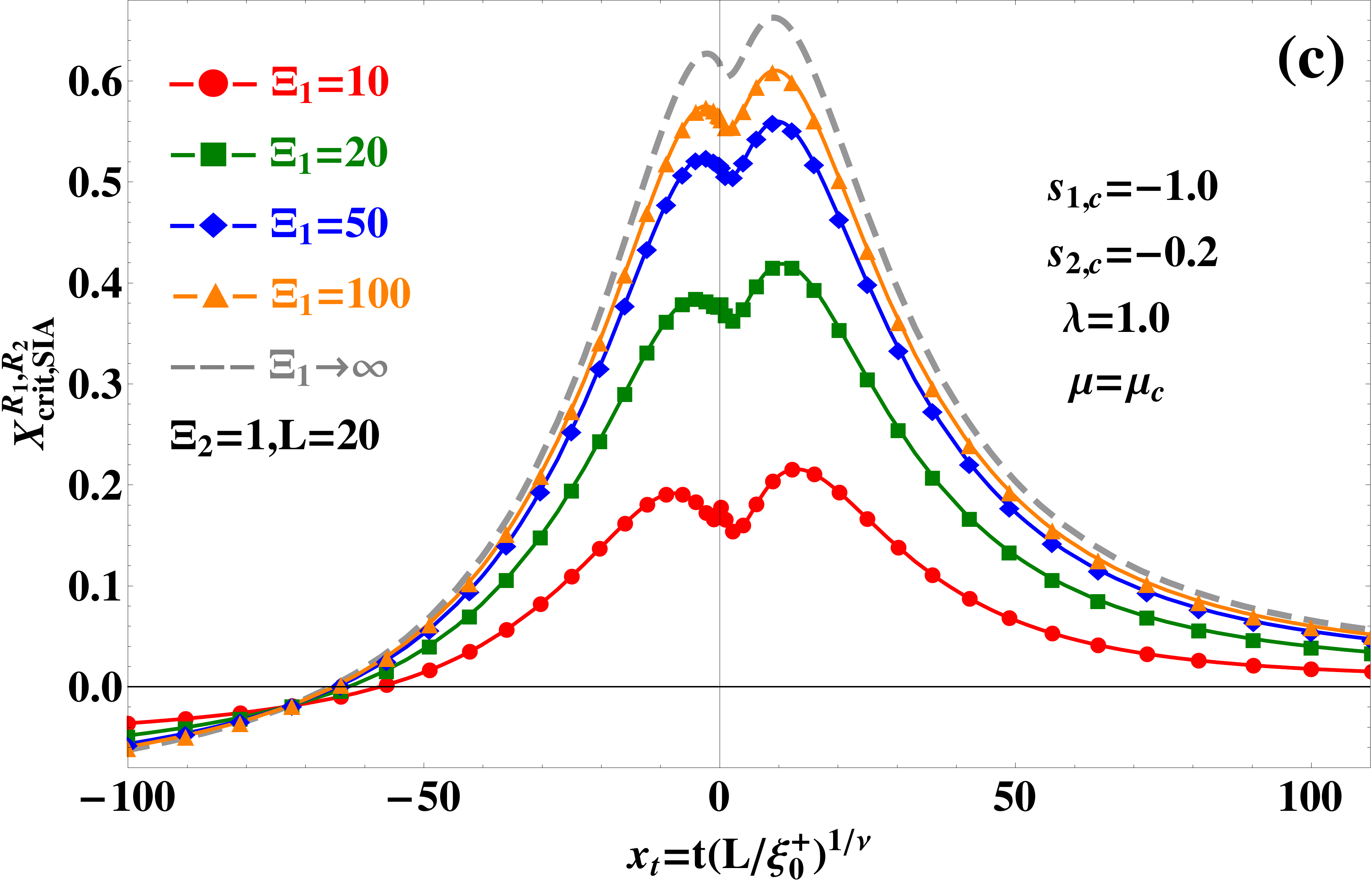}}\quad
		\subfigure{\includegraphics[width=8.8 cm]{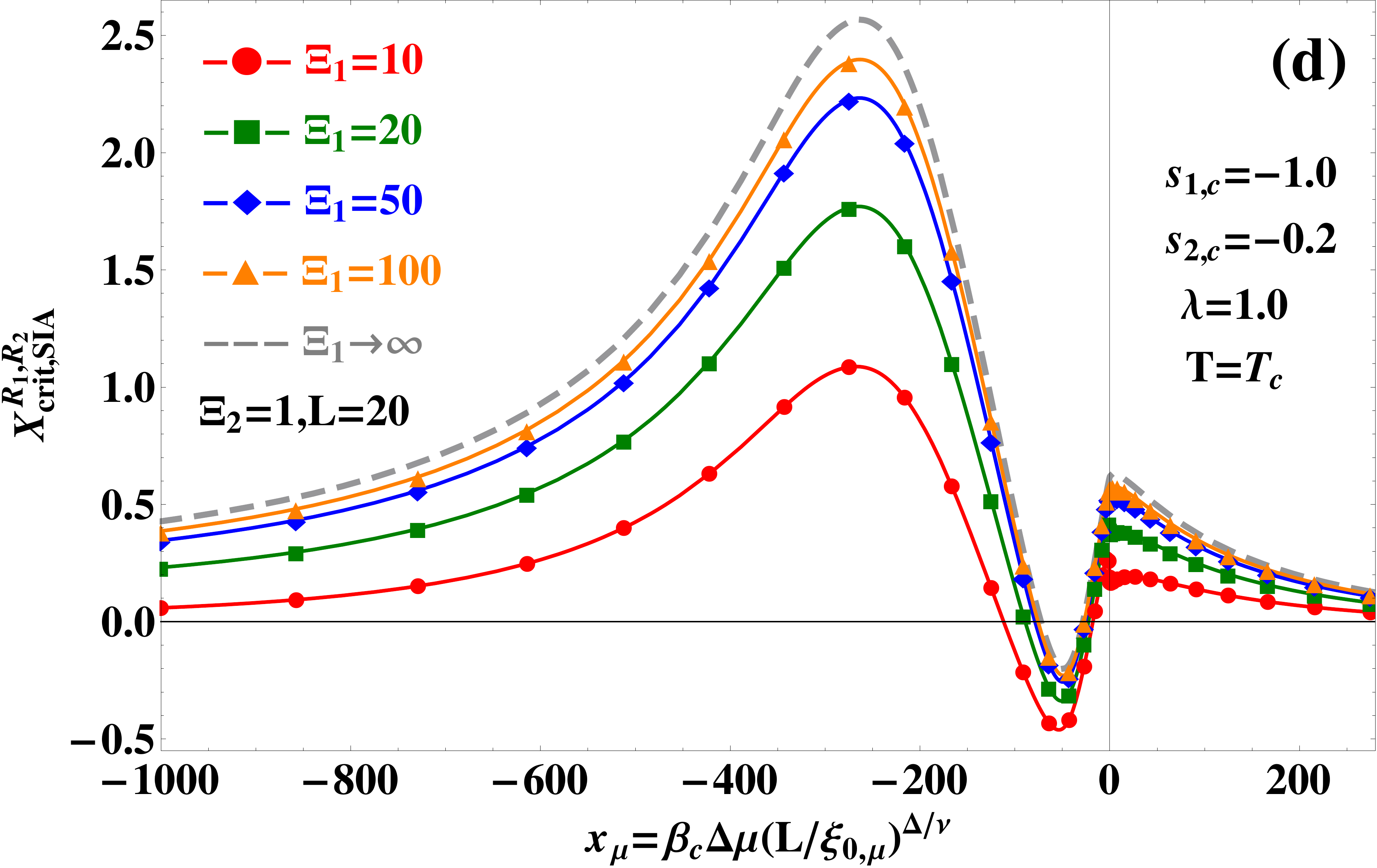}}}
	\caption{Behavior of the scaling functions $X_{\rm crit,SIA}^{R,|}$ and $X_{\rm crit,SIA}^{R_{1},R_{2}}$ within the SIA approximation in a $d=3$ dimensional sphere-plate and sphere-sphere fluid systems, respectively. The separation distance for every shown system is $L=20$ \cite{note2}. In all four sub-figures the parameters characterizing the interactions in the systems have values $s_{1,c}=-1.0$, $s_{2,c}=-0.2$ and $\lambda=1.0$. The plotted curves on sub-figures $\mathbf{(a)}$ and $\mathbf{(b)}$ are obtained with the use of \eref{ScalFunSIASpPl}, while these depicted in $\mathbf{(c)}$ and $\mathbf{(d)}$ using \eref{ScalFunSIASpSp}. As expected, when the ratio $\Xi\equiv R/L$ increases the overall behaviour of the scaling functions tends to that predicted by the DA [see sub-figures $\mathbf{(a)}$ and $\mathbf{(b)}$]. One notice that while the curve evaluated within the DA ({\color{gray_n}{$---$} $\Xi\rightarrow\infty$}) corresponds to repulsive force in the entire interval of values of $x_{t}$ and $x_{\mu}$, when the ration $\Xi$ is finite the scaling functions change sign once for $x_{t}<-50$ [see $\mathbf{(a)}$] and $x_{\mu}>500$, with a shallow attractive minimum, which gradually tends to zero with increase of $\Xi$. In contrast, as a function of the field scaling variable the curve corresponding to $\Xi=1$ ({\color{red_n}{{\Large-}\hspace{-0.1cm}{\Large-}\hspace{-0.1cm}{\large$\bullet$}\hspace{-0.09cm}{\Large-}\hspace{-0.1cm}{\Large-}}}) exhibits double sign change in the "gas phase" of the fluid medium. Such is no longer observed for $\Xi\geq2$. Also with increase of $\Xi$ the value of $x_{\mu}$ at which the minimum occurs goes to zero, while that of the global maximum changes slightly [see $\mathbf{(b)}$]. Sub-figures $\mathbf{(c)}$ and $\mathbf{(d)}$ depicts explicitly the validity of the limit $\lim_{\Xi_{1}\rightarrow\infty}X_{\rm crit,SIA}^{R_{1},R_{2}}=2\pi\Xi_{2}X_{\rm crit,SIA}^{R,|}$, both as a function of $x_{t}$ and $x_{\mu}$.}
	\label{fig:XcritSIA}
\end{figure*}
\begin{figure*}[t!]
	\centering
	\mbox{\subfigure{\includegraphics[width=8.7 cm]{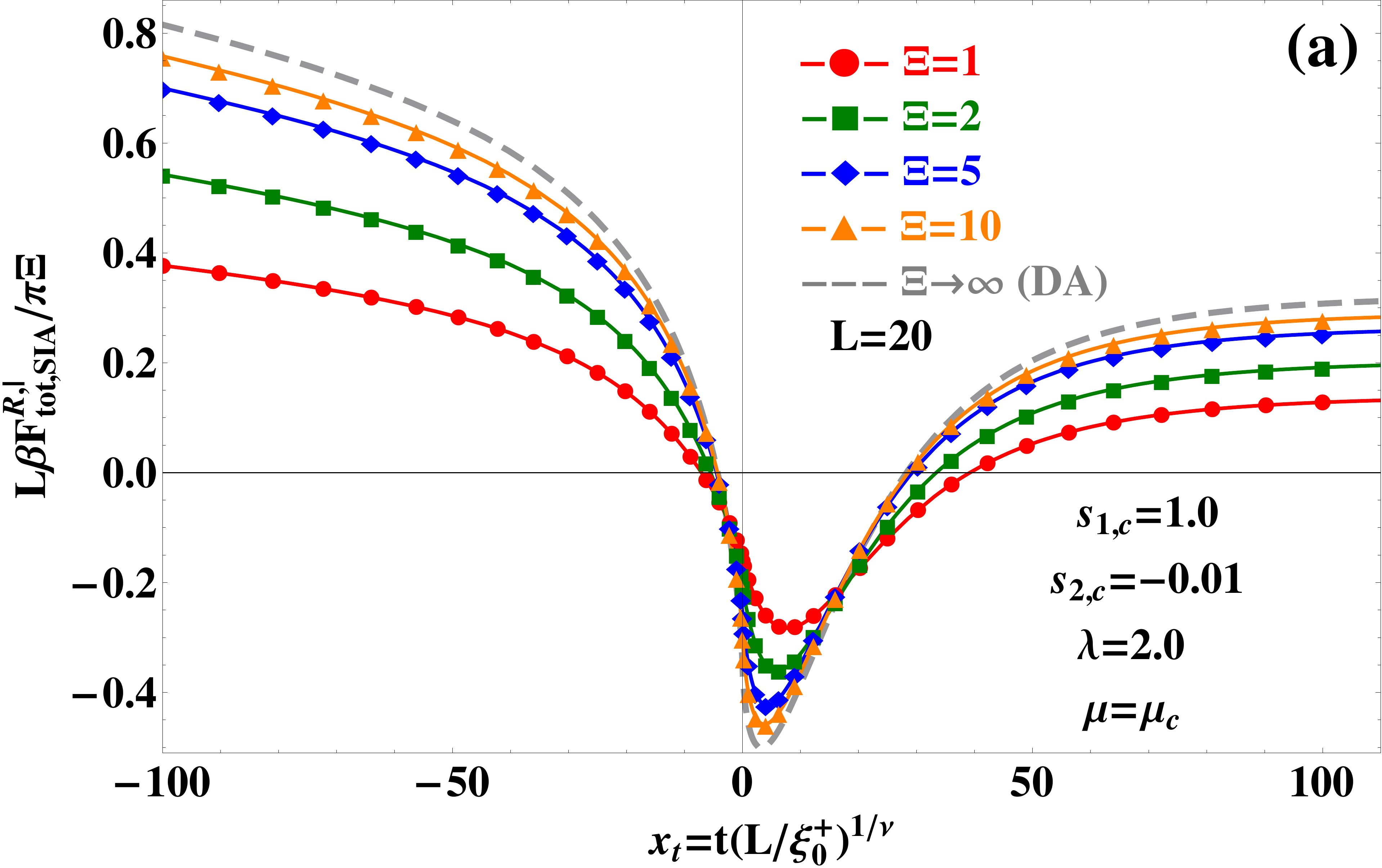}}\quad
		\subfigure{\includegraphics[width=8.5 cm]{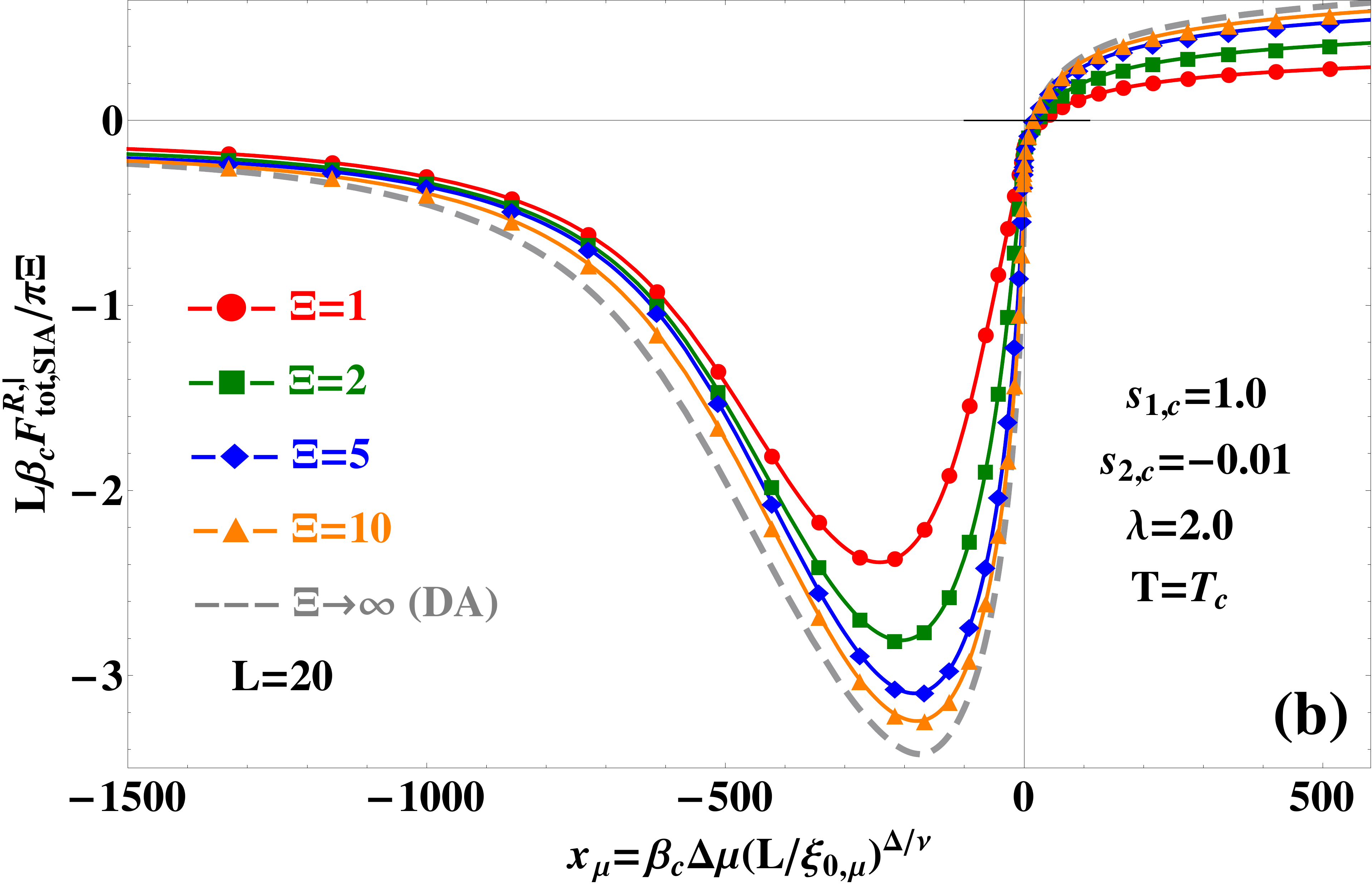}}}\\
	\mbox{\subfigure{\includegraphics[width=8.7 cm]{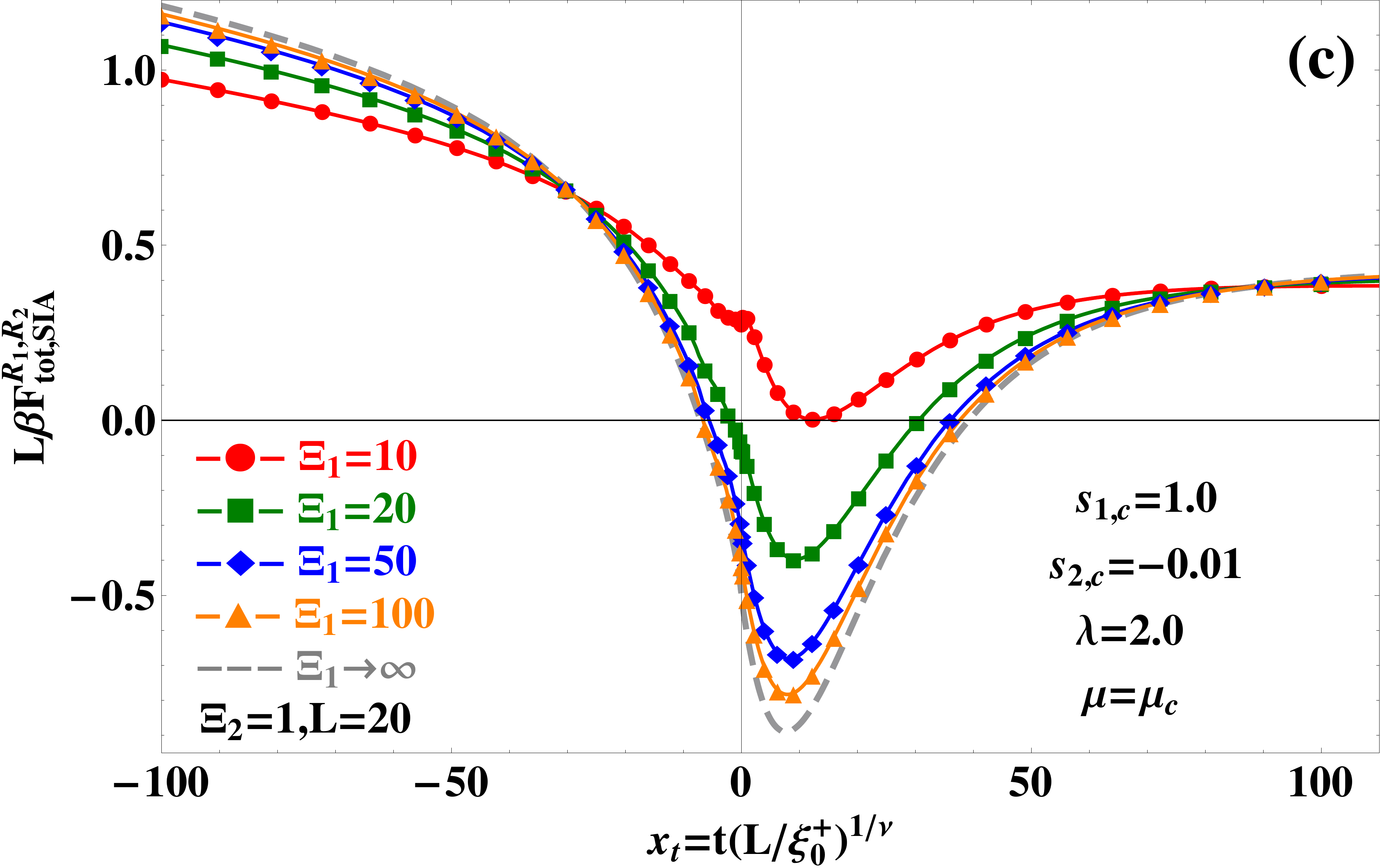}}\quad
		\subfigure{\includegraphics[width=8.5 cm]{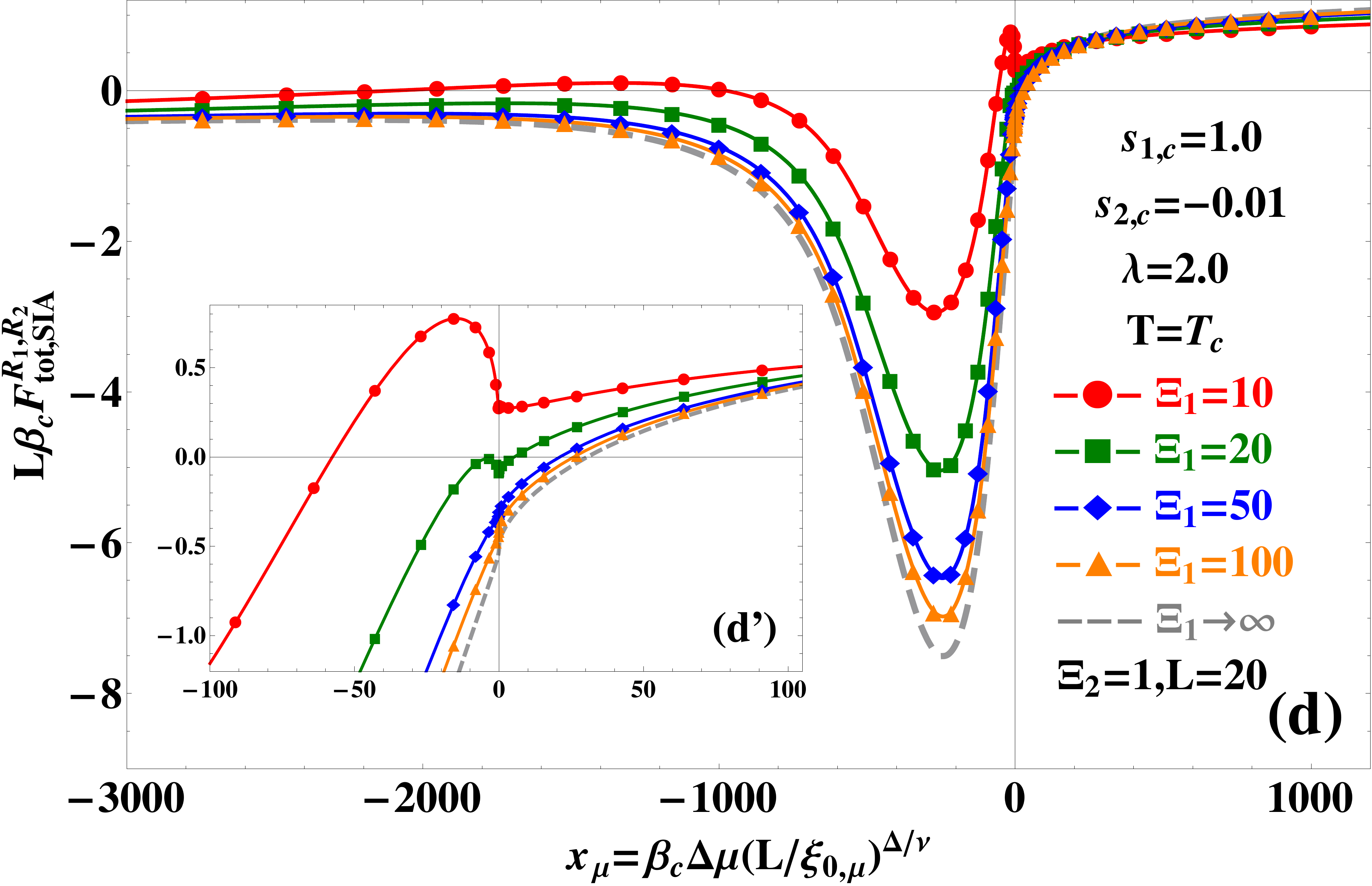}}}
	\caption{Interplay between the critical fluctuations of the fluid medium and the van der Waals interactions in it, resulting in the TF $L\beta F_{\rm tot,SIA}^{R,|}/\pi\Xi$ and $L\beta F_{\rm tot,SIA}^{R_{1},R_{2}}$ within the SIA in a $d=3$ dimensional sphere-plate and sphere-sphere fluid systems, respectively. The separation distance for every shown system is $L=20$ \cite{note2}. In all four sub-figures the parameters characterizing the interactions in the systems have values $s_{1,c}=1.0$, $s_{2,c}=-0.01$ and $\lambda=2.0$. The plotted curves on sub-figures $\mathbf{(a)}$ and $\mathbf{(b)}$ are obtained with the use of Eqs. (\ref{HA_ns}), (\ref{HA_sing}), (\ref{SIAvdWSpPl3d}), (\ref{SIACasSpPlate3d}) -- (\ref{IVfun}), while these depicted in $\mathbf{(c)}$ and $\mathbf{(d)}$ using Eqs. (\ref{HA_ns}), (\ref{HA_sing}), (\ref{vdWSIAsigma3}), (\ref{SIACasSpSpere3d}) -- (\ref{IR1R2}). Specifically as expected when the ratio $\Xi\equiv R/L$ increases the overall behaviour of the scaling functions tends to that predicted by the DA ({\color{gray_n}{$---$} $\Xi\rightarrow\infty$}) [see sub-figures $\mathbf{(a)}$ and $\mathbf{(b)}$]. On both sub-figures the curves show similar qualitative behaviour, as the force exhibits a double sign change [once in the "gas" phase ($x_{t}>0$) and once in the "liquid" one ($x_{t}<0$)] as a function of the temperature and a single one [only in the "liquid" phase ($x_{\mu}>0$)] as a function of the chemical potential. Sub-figures $\mathbf{(c)}$ and $\mathbf{(d)}$ depicts explicitly the validity of the limit $\lim_{\Xi_{1}\rightarrow\infty}(F_{\rm tot,SIA}^{R_{1},R_{2}})=F_{\rm tot,SIA}^{R,|}$, both as a function of $x_{t}$ and $x_{\mu}$. On $\mathbf{(c)}$ the model predicts a repulsive force ({\color{red}{{\Large-}\hspace{-0.1cm}{\Large-}\hspace{-0.1cm}{\large$\bullet$}\hspace{-0.09cm}{\Large-}\hspace{-0.1cm}{\Large-}}}), which becomes zero only at its minimum (achieved for $x_{t}>0$), when the interaction is between two spherical particles characterized by $\Xi_{1}=10$ and $\Xi_{2}=1$. Here one also notices that the behaviour of the force becomes slightly non-monotonic at the vicinity of the bulk critical point, which is due to the occurrence of a shallow minimum at $x_{t}<0$ of $X_{\rm crit,SIA}^{R_{1},R_{2}}$. One can speculate that for $10<\Xi_{1}<20$ a double sign change of the force will be observed in the "gas" phase of the fluid. As a function of the chemical potential [see sub-figures $\mathbf{(d)}$ and $\mathbf{(d')}$] a triple sign change ({\color{red}{{\Large-}\hspace{-0.1cm}{\Large-}\hspace{-0.1cm}{\large$\bullet$}\hspace{-0.09cm}{\Large-}\hspace{-0.1cm}{\Large-}}}) of the NTFs appears for $x_{\mu}<0$ when $\Xi_{1}=10$ and $\Xi_{2}=1$. Then, upon increasing the radius of the second sphere to and above  $\Xi_{1}=20$ ({\color{dark_green}{{\Large-}\hspace{-0.1cm}{\Large-}\hspace{-0.1cm}{\small$\blacksquare$}\hspace{-0.09cm}{\Large-}\hspace{-0.1cm}{\Large-}}}) the force changes sign once for $x_{\mu}>0$. This change is observed very near the critical point [see sub-figure $\mathbf{(d')}$].}
	\label{fig:TFSIA}
\end{figure*}
From Eqs. (\ref{DASpSp}) and (\ref{intDAvdW}), considering only genuine van der Waals interactions $(d=\sigma)$ in $d=3$, one has that within the DA the forces are calculated via the following expression
\begin{equation}\label{DAvdWSpSpSpPl3d}
F_{\rm vdW,DA}^{a,b}(L)=\dfrac{2\pi H_{A}^{\rm (reg)}}{L}\times
\begin{cases}
\Xi       \\
\Xi_{\rm eff}  \\
\end{cases},
\end{equation}
where in the case of sphere-plate interaction $(a=R,\ b=|)$ $\Xi\equiv R/L$ and $\Xi_{\rm eff}\equiv(\Xi_{1}\Xi_{2})/(\Xi_{1}+\Xi_{2})$, with $\Xi_{i}\equiv R_{i}/L, \ i=1,2$ when the interaction is between spheres $(a=R_{1},\ b=R_{2})$. Analogically for the CCF using Eqs. (\ref{DASpSp}) and (\ref{intDACas}) one can write
\begin{equation}\label{DACasSpSpSpPl3d}
\beta F_{\rm Cas,DA}^{a,b}(L)=\dfrac{2\pi X_{\rm Cas,DA}}{L}\times
\begin{cases}
\Xi       \\
\Xi_{\rm eff}  \\
\end{cases},
\end{equation}
where the scaling function $X_{\rm Cas,DA}$ is calculated as follows
\begin{eqnarray}\label{ScalFunDA}
X_{\rm Cas,DA}&&=\int_{1}^{\Lambda_{1}}\bar{z}^{-3}X_{\rm Cas}^{\parallel}[\mathfrak{a}_{1}(\bar{z})]{\rm d}\bar{z}\nonumber\\
&&+\sum_{i=j+1}^{j_{\max}-1}\int_{\Lambda_{i-1}}^{\Lambda_{i}}\bar{z}^{-3}X_{\rm Cas}^{\parallel}[\mathfrak{a}_{i}(\bar{z})]{\rm d}\bar{z}\nonumber\\
&&+\int_{\Lambda_{\max}}^{\infty}\bar{z}^{-3}X_{\rm Cas}^{\parallel}[\mathfrak{a}_{\max}(\bar{z})]{\rm d}\bar{z},
\end{eqnarray}
where $\Lambda_{j}\equiv L_{j}/L,\ j=1,j_{\max}$; $\bar{z}\equiv z/L$ is dimensionless variable; $L_{\max}\equiv L_{j_{\max}}$ is the largest system for which numerical data are available and the arguments of the scaling function are given by $[\mathfrak{a}_{j}(\bar{z})]\triangleq[x_{t}(L_{z;j}),x_{\mu}(L_{z;j}),...]$, with $L_{z;j}\equiv \bar{z}L_{j}$. In the calculations performed in the current article $L_{\max}=1000$. In order that the scaling function $X_{\rm crit}^{\parallel}$ calculated within the mean-field theory contributes properly to the critical Casimir and hence to the total force of interaction in $d<4$, one must normalize it accordingly. The need of such normalization is explained in details in Ref. \cite{DSD2007} (see there Sections IV.A.1 and IV.A.3). For boundary conditions $\tau$ one has
\begin{eqnarray}\label{Xbar}
X^{(\tau)}_{\rm crit}(\cdot)=\frac{2 \Delta^{(\tau)}(d=3)}{X_{\rm crit,sr}^{(\tau),\mathrm{MF}}(t=\Delta\mu=0)} \left[\frac{\xi_0^{+}(0)}{\xi_0^{+}\left(\lambda\right)}\right]^4 X^{(\tau),\mathrm{MF}}_{\rm crit}(\cdot), \ \ \ \ \ \
\end{eqnarray}
where $X_{\rm crit,sr}^{(\tau),\mathrm{MF}}$ is the value of the scaling function for a system within mean-field treatment governed by short-range interactions at its corresponding bulk critical point, and $X^{(\tau),\mathrm{MF}}_{\rm crit}(\cdot)$ is the scaling function of the CCF, calculated for $d=\sigma=4$, with $\lambda \neq 0,\ s_{1,c}\neq 0,\ s_{2,c}\neq 0$.  Here $\Delta^{(\tau)}(d=3)$ is the Casimir amplitude for the $d=3$ Ising universality class with boundary conditions $\tau$, while $\xi_0^{+}(0)$ and $\xi_0^{+}(\lambda)$ are the amplitudes of the bulk correlation length in mean-field systems with, correspondingly, short-ranged ($\lambda=0$) and long-ranged ($\lambda\ne 0$) fluid-fluid interactions, as $\xi_0^{+}\left(\lambda\right)=\sqrt{v_{2}}$ (see Eqs. (4.15) and (4.17) in \oncite{DSD2007}). Therefore for $\lambda=0,\ 1\ \text{and}\ 2$ one has $v_{2}=1/9,\ 0.1640,\ \text{and}\ 0.1998$, and hence $\xi_{0}^{+}=1/3,\ 0.4050,\ \text{and}\ 0.4470$ respectively. Taking into account that the value of the Casimir amplitude for the $d=3$ Ising universality class with $(+,+)$ boundary conditions \cite{Has2010} is
\begin{equation}\label{Casamppp}
\Delta^{(+,+)}=-0.410(8)
\end{equation}
and that $X_{\mathrm{Cas},\mathrm{sr}}^{(\tau),\mathrm{MF}}(t=\Delta\mu=0)=-1.7315$, for the normalizing coefficients one obtains: $0.217$ for $\lambda=1$ and  $ 0.147$ when $\lambda=2$.

Within the SIA the corresponding genuine vdWF in a sphere-plate and sphere-sphere systems is estimated with the use of Eqs. (\ref{SpPlgeneralSIA}) with $\sigma=3$
\begin{equation}
\label{SIAvdWSpPl3d}
F_{\rm vdW,SIA}^{R,|}(L)=\dfrac{8\pi H_{A}^{\rm (reg)}\Xi^{3} }{L(1+2\Xi)^{2}},
\end{equation}
and \eref{vdWSIAsigma3},  respectively.

For the CCF in the sphere-plate case, using \eref{SpPlgeneralSIACas}, one has
\begin{equation}\label{SIACasSpPlate3d}
\beta F_{\rm Cas,SIA}^{R,|}(L)=\dfrac{2\pi}{L}\Xi X_{\rm Cas,SIA}^{R,|},
\end{equation}
where the scaling function $X_{\rm Cas,SIA}^{R,|}$ for the sphere-plate CCF is calculated as follows
\begin{widetext}
\begin{eqnarray}\label{ScalFunSIASpPl}
&& X_{\rm Cas,SIA}^{R,|}=\int_{1}^{\Lambda_{1}}I^{R,|}(\bar{z}|\Lambda_{1})X_{\rm Cas}^{\parallel}[\mathfrak{a}_{1}(\bar{z})]{\rm d}\bar{z}+\sum_{i=j+1}^{j_{\max}-1}\int_{\Lambda_{i-1}}^{\Lambda_{i}}I^{R,|}(\bar{z}|\Lambda_{i})X_{\rm Cas}^{\parallel}[\mathfrak{a}_{i}(\bar{z})]{\rm d}\bar{z}+\int_{\Lambda_{\max}}^{1+\Xi}I^{R,|}(\bar{z}|\Lambda_{\max})X_{\rm Cas}^{\parallel}[\mathfrak{a}_{\max}(\bar{z})]{\rm d}\bar{z},
\end{eqnarray}
\end{widetext}
where
\begin{equation}\label{IVfun}
I^{R,|}(\bar{z}|\Lambda_{j})=\dfrac{1}{\bar{z}^{3}}\left[1-\dfrac{\bar{z}-1}{\Xi}\right]\theta(\Xi-\Lambda_{j}),
\end{equation}
with $\theta$ being the Heaviside step function with the convention $\theta(0)=0$.

In the case of sphere-sphere geometry the expression for calculating the CCF within the SIA, using \eref{sphere_sphere_force_SIA_Cas_intext2}, is
\begin{equation}\label{SIACasSpSpere3d}
\beta F_{\rm Cas,SIA}^{R_{1},R_{2}}(L)=\dfrac{X_{\rm Cas,SIA}^{R_{1},R_{2}}}{L},
\end{equation}
where
\begin{widetext}
\begin{eqnarray}\label{ScalFunSIASpSp}
&&X_{\rm Cas,SIA}^{R_{1},R_{2}}=\int_{1}^{\Lambda_{1}}I_{1}^{R_{1},R_{2}}(\bar{z}_{2}|\Lambda_{1})X_{\rm Cas}^{\parallel}[\mathfrak{a}_{1}(\bar{z}_{2})]{\rm d}\bar{z}_{2} +\sum_{i=j+1}^{j_{\max}-1}\int_{\Lambda_{i-1}}^{\Lambda_{i}}I_{1}^{R_{1},R_{2}}(\bar{z}_{2}|\Lambda_{i})X_{\rm Cas}^{\parallel}[\mathfrak{a}_{i}(\bar{z}_{2})]{\rm d}\bar{z}_{2}\nonumber\\
&&+\int_{\Lambda_{\max}}^{1+\Xi_{2}}I_{1}^{R_{1},R_{2}}(\bar{z}_{2}|\Lambda_{\max})X_{\rm Cas}^{\parallel}[\mathfrak{a}_{\max}(\bar{z}_{2})]{\rm d}\bar{z}_{2}+\int_{1}^{1+\Xi_{2}}\left\{\int_{1}^{\Lambda_{j}/\bar{z}_{2}}I_{2}^{R_{1},R_{2}}(\tilde{z},\bar{z}_{2}|\Lambda_{j})X_{\rm Cas}^{\parallel}[\mathfrak{a}_{j}(\tilde{z})]{\rm d}\tilde{z}\right. \nonumber\\
&&\left.+\sum_{i=j+1}^{j_{\max}-1}\int_{\Lambda_{i-1}/\bar{z}_{2}}^{\Lambda_{i}/\bar{z}_{2}}
I_{2}^{R_{1},R_{2}}(\tilde{z},\bar{z}_{2}|\Lambda_{i})
X_{\rm Cas}^{\parallel}[\mathfrak{a}_{i}(\tilde{z})]{\rm d}\tilde{z}\right.\left.+\int_{\Lambda_{\max}/\bar{z}_{2}}^{1+(\Xi_{1}/\bar{z}_{2})}I_{2}^{R_{1},R_{2}}(\tilde{z},\bar{z}_{2}|\Lambda_{\max})X_{\rm Cas}^{\parallel}[\mathfrak{a}_{\max}(\tilde{z})]{\rm d}\tilde{z}\right\}{\rm d}\bar{z}_{2},\ \ \ \ \ \ \ \
\end{eqnarray}
\end{widetext}
with $\tilde{z}\equiv \bar{z}_{1}/\bar{z}_{2}$,
\begin{subequations}\label{IR1R2}
	\begin{eqnarray}\label{IR1R2z2}
	I_{1}^{R_{1},R_{2}}(\bar{z}_{2}|\Lambda_{j})=-\Xi_{1}\dfrac{\zeta(\bar{z}_{2})}{\bar{z}_{2}^{3}}\theta(\Xi_{2}-\Lambda_{j}),
	\end{eqnarray}
	\begin{eqnarray}\label{IR1R2z1z2}
	I_{2}^{R_{1},R_{2}}(\tilde{z},\bar{z}_{2}|\Lambda_{j})=\dfrac{\zeta(\bar{z}_{2})}{\bar{z}_{2}^{2}\tilde{z}^{3}}\theta(\Xi_{1}-\Lambda_{j}),
	\end{eqnarray}
	\begin{eqnarray}\label{dzetadim}
	\zeta(\bar{z}_{2})=\dfrac{\pi\left[\bar{z}_{2}^{2}+2\Xi_{1}\bar{z}_{2}-2(1+\Xi_{2})(\Xi_{1}+\Xi_{2})-1\right]}{(1+\Xi_{1}+\Xi_{2})^{2}},\ \ \ \ \ \ \ \
	\end{eqnarray}
\end{subequations}
and the arguments of the scaling function defined via $[\mathfrak{a}_{j}(\bar{z}_{k})]\triangleq[x_{t}(L_{z_{k};j}),x_{\mu}(L_{z_{k};j}),...],\ k=1,2;\ j=1,j_{\max}$.

After presenting the mathematical means to calculate the CCF and vdWF, we now pass to the detailed discussion of the results and argumentation of the experimental feasibility of the parameters used in the model calculations utilizing DA and SIA.
%
%
%
%
\subsection{Discussion of the results}\label{subsec:Discussion}
%
We start with the simplest case of systems (any of the depicted in Fig. \ref{fig:fluid_systems}) governed by purely short-range interactions, i.e. $s_{i,c}=\lambda=0,\ i=1,2$.  In this case it is clear the CCF is simply $\propto X_{\rm Cas}^{\rm sr}$. It results only from the correlated fluctuations and the size-dependent spatial order parameter $\phi({\bf r})$. Following Eqs. (\ref{CasimirF_l}), (\ref{HA_ns}) and (\ref{HA_sing}) we see that $X_{\rm Cas}^{\rm sr}$ coincides with $X_{\rm crit}^{\rm sr}$. Irrespective of the geometry of the interacting objects $X_{\rm Cas}^{\rm sr}$ is \textit{negative}, which corresponds to {\it attractive} force, at any separation $L$ and for any value of the scaling variables $x_t$ and $x_\mu$ under $(+,+)$ boundary conditions. For sphere-plate and sphere-sphere systems the study of $X_{\rm crit}$ within the DA showed that when $L>50$ the curves fall on that obtained via \eref{ScalFunDA}, using:{ \it i) } the exact analytical results based on Eqs. (4.10) reported in \oncite{VaDa2015} when $\mu=\mu_{c}$; {\it ii)} the data for the plate-plate interaction calculated within the presented mean-field theory for $T=T_{c}$. Note that the scaling function of the CCF (and $X_{\rm crit}$ respectively), as well as the vdWF, is {\it one and the same} for sphere-plate and sphere-sphere systems, irrespective of the interaction type which takes place between the constituents of a considered system, when the behaviour is studied within the DA [see Eqs. (\ref{DAvdWSpSpSpPl3d}) -- (\ref{ScalFunDA})]. As it is clear from these equations, the corresponding forces differ {\it only} by a multiplication factor which depends on some basic geometrical characteristics of the interacting objects.

Staying within the DA, we study the behaviour of $X_{\rm crit, DA}$ for a system with long-range interactions characterized by $s_{1,c}=1.0$, $s_{2,c}=0.0$ and $\lambda=1.0$. We see that practically all curves except that for $L=20$ coincide with $X_{\rm crit,DA}^{\rm sr}$ as depicted in Fig. \ref{fig:Xcrits11s2gz}$\mathbf{(a)}$ and $\mathbf{(b)}$. Upon increasing $s_{2,c}$ the curves disperse and all appear below $X_{\rm crit,DA}^{\rm sr}$. For $s_{2,c}=1.0$ all scaling functions are clearly distinguishable from one another [see Fig. \ref{fig:Xcrits11s2gz}$\mathbf{(c)}$ and $\mathbf{(d)}$]. This is easy to understand, given that the nonnegative values of the coupling parameters $s$ and $\lambda$ are associated with an additional enhancement of the ordering in the system, both near the surfaces of the interacting bodies and in the bulk of the system. Considering Eq. (2.12) in \oncite{VaDa2015} we see that at small distances the nonuniversal behaviour dominates and adds to the universal one (described by $X_{\rm crit}^{\rm sr}$) with a positive sign, resulting in net scaling functions with minima deeper than that of a short-range system. As expected, when the distance is increased the influence of the long-range interactions decreases in the bulk of the system, and as a result all scaling functions tend to the {\it universal one}.

In this line of thinking for systems characterized by $s_{1,c}>0$ and $-s_{1,c}\ll s_{2,c}\leq0$ we expect  $X_{\rm crit,DA}$ to remain again negative for any separation $L$, $x_{t}$ and $x_{\mu}$, irrespective of the value of $\lambda$ (which is always nonnegative). Indeed, this turns out to be true and is depicted in Fig. \ref{fig:Xcrits11s2sz}(\textbf{a}) and (\textbf{b}). However when $s_{2,c}\approx-s_{1,c}$ and the separation between the walls is relatively small a significant part of the system is disordered which results in nonnegative or sign-changing scaling function [see Fig. \ref{fig:Xcrits11s2sz}(\textbf{c}) and (\textbf{d})].  As the distance $L$ is increased the influence of $\Delta V(z)$ [see \eref{DeltaV_l_ab_thichlayers}] quickly decreases and only the additional ordering effect of the fluid-fluid interactions influences the behavior of the order parameter and hence of $X_{\rm crit,DA}$.

When both wall-fluid coupling parameters are negative, $X_{\rm crit,DA}$ is negative for any $x_{t}$ and $x_{\mu}$ only for very large separations $L$ where the effect of the long-ranged interactions on the behavior of the system is negligible. Naturally, since the short-ranged surface potentials do support $(+,+)$ boundary conditions, the role of the negative substrate potentials, which oppose the order near the boundary, will be stronger than that of the positive substrate potentials which try to reinforce the phase preferred near the boundary. For example, we observe that the behavior of $X_{\rm crit,DA}$ in a system with $s_{1,c}=1.0,\ s_{2,c}=-1.0,\ \lambda=1.0$ and in such with $s_{1,c}=-1.0,\ s_{2,c}=-0.2,\ \lambda=1.0$ is almost identical for any $L$ as a function of $x_{\mu}$ [compare Fig. \ref{fig:Xcrits11s2sz}(\textbf{d}) and Fig. \ref{fig:Xcrits1m1s2sz}(\textbf{b})]. Thus, for a fixed $\lambda$ the behavior of the scaling function is mainly determined by the interplay between the short-ranged surface fields and the strong negative wall-fluid coupling $s_{2,c}$. If $s_{1,c}=s_{2,c}=-1.0$ and $\lambda<2.0$ the scaling function  $X_{\rm crit,DA}$ exhibits an unexpected behavior as a function of $L$: for moderate values of $L$ in the range $20$ to $100$, the maximum of the {\it repulsive} part of the force increases with increasing $L$ both as a function of $x_{t}$ and $x_{\mu}$ [see Fig. \ref{fig:Xcrits1m1s2sz}(\textbf{c}) and (\textbf{d})]; for larger values of $L$ the maximum decreases, as expected, and the overall behavior of the scaling function approaches that one of the system with completely short-ranged interactions.

So far the discussion was focused on the behavior of $X_{\rm crit}$ within the DA, i.e. for separations $L$ much smaller than the characteristic geometrical extend of the interacting objects [in the context of the current article, the sphere(s) radius(ii)]. The quantitative and qualitative comparison in the behavior of the scaling functions within the DA and SIA is presented on Fig. \ref{fig:XcritSIA}. Let us emphasize again that in contrast to the DA, the use of SIA {\it does not} put any restrictions on the sizes and separation lengths that can be considered. In order to illustrate only the main idea, here we restrain ourselves to the  choice of parameters $s_{1,c}=-1.0$, $s_{2,c}=-0.2$ and $\lambda=1.0$ in a system with $L=20$ fluid layers. The first distinction one notices is that in comparison to the DA, within the SIA the mathematical expressions for calculating the CCF and vdWF are different for the sphere-plate and sphere-sphere systems, respectively -- see the corresponding equations in Subsecs. \ref{sec:SpPlandSpSpGeomForces} and \ref{subsec:SpPlSpSpDASIAd3}: when studying $X_{\rm crit}$, for the interaction between a plate and a spherical particle \eref{ScalFunSIASpPl} is used, while  between a pair of spherical particles one uses \eref{ScalFunSIASpSp}. The results presented  on Fig. \ref{fig:XcritSIA}(\textbf{a}) and (\textbf{b}) demonstrate that with the increase of the ratio $\Xi$ the overall behaviour of the scaling functions tends, as expected, to that predicted by the DA. One also notices that while the curve evaluated within the DA corresponds to repulsive force in the entire interval of values of $x_{t}$ and $x_{\mu}$, when the ratio $\Xi$ is finite $X_{\rm crit,SIA}$ changes sign once for $x_{t}<-50$ [Fig. \ref{fig:XcritSIA}(\textbf{a})] and $x_{\mu}>500$, with a shallow attractive minimum, which gradually tends to zero with increase of $\Xi$. More specifically, studying the temperature dependance of the scaling functions we notice that while those evaluated within the DA exhibit a single minimum and two maxima of equal height, the  maxima of the scaling function evaluated for a system with $\Xi=1$ differ with about 10 \%. Another comparison shows that the difference between the global maximum of $X_{\rm crit,SIA}$ evaluated for $\Xi=1$ and any of the two of $X_{\rm crit,DA}$ is nearly 2 times. As a function of the field scaling variable [Fig. \ref{fig:XcritSIA}(\textbf{b})] the curve corresponding to $\Xi=1$ exhibits double {\it sign change} in the "gas" phase of the fluid medium. For $\Xi$ just under $2$ the minimum of the scaling function is zero. For $x_{\mu}<0$ all curves obtained for $\Xi\geq2$ describe repulsive force. Also with increase of $\Xi$ the value of $x_{\mu}$ at which the minimum occurs goes to zero, while that of the global maximum changes slightly. For the field dependance of $X_{\rm crit}$ one finds that the difference between the global maxima of $X_{\rm crit,SIA}(\Xi=1)$ and $X_{\rm crit,DA}$ is 73 \%.

For the study of the CCF between two spherical particles within the SIA we choose to vary the ratio $\Xi_{1}$ and fix that of $\Xi_{2}$ to 1 -- see Fig. \ref{fig:XcritSIA}(\textbf{c}) and (\textbf{d}). The depicted on these two subfigures justifies the mathematically predicted limit $\lim_{\Xi_{1}\rightarrow\infty}X_{\rm crit,SIA}^{R_{1},R_{2}}=2\pi\Xi_{2}X_{\rm crit,SIA}^{R,|}$, both as a function of $x_{t}$ and $x_{\mu}$. On Fig. \ref{fig:XcritSIA}(\textbf{c}) one sees that for moderate values of $\Xi_{1}$, the scaling function exhibits 3 maxima (one in each phase and at the critical point) and a couple of minima. With the increase of $\Xi_{1}$ the maximum at $x_{t}=0$ and the minimum which appears in the "liquid" phase both disappear, as the scaling function of a sphere-sphere system with $\Xi_{1}=100\Xi_{2})$ approaches that for the sphere-plate system with $\Xi_{2}=1$. On a quantitative level, as a function of $x_{t}$ the difference between the values at the global maxima of $X_{\rm crit,SIA}^{R_{1},R_{2}}(\Xi_{1}=10)$ and $X_{\rm crit,SIA}^{R_{1},R_{2}}(\Xi_{1}\rightarrow\infty)$ is about 3 times. The scaling functions evaluated for various values of $\Xi_{1}$ and $x_{\mu}$ differ only quantitatively. In analogy with the comparison made for $X_{\rm crit,SIA}^{R_{1},R_{2}}(x_{t},x_{\mu}=0)$, here we have that the difference between the values at the global maxima as well as minima of $X_{\rm crit,SIA}^{R_{1},R_{2}}(\Xi_{1}=10)$ and $X_{\rm crit,SIA}^{R_{1},R_{2}}(\Xi_{1}\rightarrow\infty)$ is about 2.4 times.

We will close the discussion with some comments on the behavior of the TF, in a sphere-palate and sphere-sphere systems studied within the SIA. The obtained results are depicted on Fig. \ref{fig:TFSIA}. In scope of arguing the experimental feasibility of the presented theory, which will be done in the following section, here the comment will be put on a particular system with $L=20$ characterized by the following coupling parameters: $s_{1,c}=1.0$, $s_{2,c}=-0.01$ and $\lambda=2.0$. The behavior of the Hamaker term (associated with the vdWF) both as a function of $x_{t}$ and $x_{\mu}$ is rather trivial, when one considers a lattice-gas model. An example of it for a system with $s_{1,c}s_{2,c}=0$ is depicted on Fig. 5 in \oncite{VaDa2015}. Unlike the case presented there, when $s_{1,c}s_{2,c}\neq0$, as considered here, the direct substrate-substrate interaction [the first term in \eref{HA_ns}] adds to overall behaviour of the interaction. For $s_{1,c}s_{2,c}<0$ at $x_{\mu}=0$ $H_{A}$ is nonzero and corresponds to repulsion for any value of $x_{t}$. On the other hand, as a function of $x_{\mu}$, at $T=T_{c}$ and for $s_{1,c}s_{2,c}<0$, $H_{A}$ changes sign once in the "gas" phase of the fluid, being attractive for $x_{\mu}<900$ and repulsive otherwise, with an infinite slope at $x_{\mu}=0$. With regard to the behaviour of $X_{\rm crit,SIA}^{R,|}$ both as a function of $x_{t}$ and $x_{\mu}$ for the system specified above, one observes a single sign change with a low repulsive maximum, appearing in the "liquid" phase of the fluid, which gradually tends to zero with increase of $\Xi$. After superimposing both quantities, namely $2\beta H_{A}$ and $X_{\rm crit,SIA}^{R,|}$, using Eqs. (\ref{HA_ns}), (\ref{HA_sing}), (\ref{SIAvdWSpPl3d}), (\ref{SIACasSpPlate3d}) -- (\ref{IVfun}), one ends up with the temperature [see Fig. \ref{fig:TFSIA}(\textbf{a})] and field [see Fig. \ref{fig:TFSIA}(\textbf{b})] dependencies of the TF between a spherical particle of arbitrary radius and a planar substrate within the SIA.

%
\subsection{Experimental feasibility of the predicted effects}\label{subsec:ExperimentalFeas}
%
In the previous section it has been argued that if the coupling parameters are tuned in a certain way, one can realise not only repulsive or sign changing CCF, but even the {\it total} force can exhibit such a behaviour. Particular values of the model parameters at which these phenomena are observed are, e.g., $s_{1,c}=1.0$, $s_{2,c}=-0.01$ and $\lambda=2.0$. In the current section we will show that there are materials characterized by the above mentioned values of the parameters in question, thus demonstrating the feasibility of the theoretical predictions presented.  Let us consider a sphere-plate and sphere-sphere systems immersed in xenon (Xe), which exhibits critical fluctuations near its bulk critical point. This is obviously a nonpolar simple fluid. Its physical characteristics are presented in Table I in \oncite{VaDa2015}. We will make the assumption that all interactions between the constituents of a given system are of Lennard-Jones type, i.e., $J^{l,s}=2J_{\rm sr}^{l,s}$ and $\lambda\equiv J^{l}/J_{\rm sr}^{l}=2$ (see Eq. (7.1) in \oncite{VaDa2015} and the text therein). For the calculation of the substrate-fluid coupling constant $s_{c}$ within the mean-field treatment of the problem we use the following expression: $s_{c}\equiv \, 0.5\,G(4,4)\,[\rho_{\rm nd}\beta_{c}J^{l,s}-\rho_{c}\beta_{c}J^{l}]$, where $J^{l,s}$ (see column 3 of Table \ref{table_fluid_sub}) and $J^{l}$ (see column 7 of Table I in \oncite{VaDa2015}) are the long-range inter-particle interaction energies and $\rho_{\rm nd}$ (see column 7 of Table \ref{table_fluid_sub}) is the number density of the bulk substrate relative to the critical one $\rho_{c}$ of the fluid medium. The quantitative assessment of the interatomic interactions between the xenon and some concrete substances listed in column 1 of Table \ref{table_fluid_sub}, manifests itself with values of $s_{c}$ (see column 8 of Table \ref{table_fluid_sub}) very near to the above mentioned model ones. With regard to the experimentally realizable spherical particles studied here, the following composites are considered: ruthenium \cite{LVCVML2016} (Ru) or platinum \cite{BE2016} (Pt) core, encapsulated by thin (about 4 unit cells or $\sim 1\ \rm{nm}$) $\rm{ZrO}_{2}$ or $\rm{CeO}_{2}$ film \cite{SS2011}; lithium (Li) core, encapsulated in a carbon shell \cite{YLLXHLZCC2016}; carbon \cite{ZQD2015} (C) or silica \cite{KLRKMJJCLS2016} (SiO$_{2}$) aerogels. The contact surface of any of the interacting geometries (spherical or planar) is assumed coated by monolayer of lead \cite{LZWYJ2016} (Pb) or thallium nitride \cite{PLJD2012} (TlN) to ensure the (+,+) boundary conditions. In addition, from the data reported in \oncite{RCCGS92} we have that $\beta_{c}J_{\rm sr}^{\rm Xe,Pb}=0.956$ and  $\beta_{c}J_{\rm sr}^{\rm Xe,TlN}=0.968$.
\begin{table}[th!]
	\renewcommand\thetable{1}
	\centering
	\caption{\label{table_fluid_sub} Physical characteristics and interaction parameters of the considered core materials: ruthenium (Ru), platinum (Pt), lithium (Li), carbon (C) and silica (SiO$_{2}$) aerogels with the fluid medium (Xe). The columns show: the distances $r_{0}^{s}$ and $r_{0}^{l,s}$ in {\AA} (columns 2 and 4) at which the inter-particle potential within the substrate and between it and the fluid is zero, the corresponding potential well depths $J_{\rm sr}^{s}$ and $J_{\rm sr}^{l,s}$ in units $k_{B}T_{c}$ (columns 3 and 5), the density $\rho$ of the substrates in $\rm{g/cm^{3}}$ (column 6), the number density $\rho_{\rm nd}$ (column 7) and the substrate-fluid coupling parameter evaluated at the critical temperature -- $s_{c}$ (column 8). The values of $r_{0}^{s}$ and $J_{\rm sr}^{s}$ are taken from \oncite{RCCGS92}, while these of $r_{0}^{l,s}$ and $J_{\rm sr}^{l,s}$ are calculated via Kong's mixing rules (see Eqs. (7.2) in \oncite{VaDa2015}). The densities of the aerogel substrates \cite{ZZZMFL2015} are taken in such a way as to render the desired value of $s_{c}$, namely $-0.01$.}
	\centering
	\begin{tabular}{ccccccccccccccc}
		\hline \hline
		1  && 2 && 3 && 4 && 5   && 6   && 7 && 8\\
		
		$\text{core}$  && $r_{0}^{s}$ && $\beta_{c}J_{\rm sr}^{s}$ && $r_{0}^{l,s}$ && $\beta_{c}J_{\rm sr}^{l,s}$ && $\rho$ && $\rho_{\rm nd}$ && $s_{c}$\\
		\hline
		$\rm{Ru}$           &&   2.963    && 0.097 && 3.964 && 0.171  &&  12.370   && 7.219 && 1.03\\
		$\rm{Pt}$           &&   2.754    && 0.139 && 3.868 && 0.190  &&  21.450   && 6.485 && 1.03\\
		$\rm{Li}$           &&   2.451    && 0.043 && 3.916 && 0.069  &&  0.535    && 4.546 && -0.01\\
		$\rm{C}$            &&   3.851    && 0.182 && 4.245 && 0.341  &&  0.232    && 1.139 && -0.01\\
		$\rm{SiO}_{2}$      &&   1.233    && 0.698 && 3.425 && 0.079  &&  0.250    && 0.525 && -0.01\\
		\hline \hline
	\end{tabular}
\label{table}
\end{table}

Although the mean-field theory gives poor quantitative estimation of the behavior of the studied forces, it is tempting to evaluate them, nevertheless. As it has been done in the calculations leading to the results reported in Fig. \ref{fig:TFSIA}, we fix the separation between the interacting objects to $L=20 a_{0}\simeq12\ {\rm nm}$, where $a_{0}$ is the suitable distance between the xenon atoms  at the bulk critical point (see column 5 of Table I in \oncite{VaDa2015}). Then, studying the temperature dependance of the TF occurring in a sphere-plate system, see Fig. \ref{fig:TFSIA}(\textbf{a}) with $R=12\ {\rm nm}$ at $\mu=\mu_{c}$, one shell observe a double sign change at $T_{1}\simeq0.9965T_{c}$ and $T_{2}=1.0195T_{c}$, with an attractive global minimum at $T_{\min}=1.0039T_{c}$, which magnitude is $F_{\rm tot,min}^{R,|}\simeq-0.31\ {\rm pN}$. If now the interaction takes place between a plate and a sphere with radius $120\ {\rm nm}$, the force sign change will be at $T_{1}\simeq0.9978T_{c}$ and $T_{2}=1.01435T_{c}$, the minimum will remain attractive, observed at $T_{\min}=1.0018T_{c}$, and with magnitude of $F_{\rm tot,min}^{R,|}\simeq-4.97\ {\rm pN}$.

As a function of the chemical potential difference at $T=T_{c}$, see Fig. \ref{fig:TFSIA}(\textbf{b}), the force changes sign only once irrespectively on the sphere's radius, with the occurrence of a single attractive minimum. Thus, when $R=12\ {\rm nm}$ the force is repulsive for $\beta_{c}\Delta\mu\geq1.11\times 10^{-4}$ and attractive otherwise. The minimum is reached at $(\beta_{c}\Delta\mu)_{\min}\simeq-8.868\times 10^{-4}$, with a magnitude of $F_{\rm tot,min}^{R,|}\simeq-2.58\ {\rm pN}$. The increase of the sphere's radius 10 times results in insignificant change in the value of $\beta_{c}\Delta\mu$ at which the sign change occurs, but the magnitude of the force at its minimum increases substantially to $F_{\rm tot,min}^{R,|}\simeq-35.12\ {\rm pN}$, observed at $(\beta_{c}\Delta\mu)_{\min}\simeq-6.534\times 10^{-4}$. For even larger spherical particles $\Xi\gg10$ the value of the force at its minimum is a linear function of the ratio $\Xi$ [see \eref{DACasSpSpSpPl3d}] and for the particular case considered here can be estimated with the use of the expressions: $F_{\rm tot,min}^{R,|}(x_{\tau,\min}=3.24)\simeq-0.54\times\Xi\ {\rm pN}$ and $F_{\rm tot,min}^{R,|}(x_{\mu,\min}=-175.616)\simeq-3.705\times\Xi\ {\rm pN}$.

Now we focus our attention on the interaction between spherical particles in critical xenon. Let the radius of one of the particles we consider fixed is, say $R_{2}=12\ {\rm nm}$, and vary the other one $R_{1}$. Starting with the temperature dependance Fig. \ref{fig:TFSIA}(\textbf{c}), we see that when $R_{1}=120\ {\rm nm}$ the TF is repulsive at any temperature, only becoming zero at $T=1.006T_{c}$. With the increase of $R_{1}$ a double sign change  and a single attractive minimum are observed. For $R_{1}=1.2\ \mu{\rm m}$ the magnitude of the force's minimum is $F_{\rm tot,min}^{R_{1},R_{2}}\simeq-0.271\ {\rm pN}$. At $T=T_{c}$ and $R_{1}=120\ {\rm nm}$ one observes not so trivial behaviour of the TF as a function of the field scaling variable [see the last paragraph of Subsec. \ref{subsec:Discussion} as well as Fig. \ref{fig:TFSIA}(\textbf{d}) and (\textbf{d'})]. For such a system, the global minimum has a value of $F_{\rm tot,min}^{R_{1},R_{2}}\simeq-1.01\ {\rm pN}$ and a maximum of $F_{\rm tot,max}^{R_{1},R_{2}}\simeq0.26\ {\rm pN}$. Upon increasing the value of $R_{1}$ to $1.2\ \mu{\rm m}$, the force changes sign only once, having a single attractive minimum with magnitude of $F_{\rm tot,min}^{R_{1},R_{2}}\simeq-2.40\ {\rm pN}$.

For the sake of completeness, we give the critical temperature of xenon $T_{c}=289.765\ {\rm K}$ and the value of the critical chemical potential per $k_{B}T_{c}$: $\beta_{c}\mu_{c}=-16.213$, calculated using Eqs. (2.2b), (5.37b) and (8.8) from \oncite{DCJ2014}.

\section{Discussion and Concluding Remarks}\label{sec:DisandConcRem}

In the current article we have studied the interactions between objects governed by dispersion forces immersed in a nonpolar critical medium also governed by such forces.  We envisage the case in which the critical medium is either a one-component fluid or a binary liquid mixture, i.e., its critical bulk behavior belongs to the Ising surface universality class.  Because of the modifications of the order parameter of the fluid, as well as of its fluctuation spectrum, in addition to the dispersion force one also has an additional effective force known as the CCF, acting between these objects. Since the dispersion forces influence the critical medium, they change the order parameter profile and the fluctuations of the system. Thus, studying the CCF in such systems one, unavoidably, studies the interplay between these two long-ranged and fluctuation induced forces. In addition to the contribution of any of these forces to the overall interaction between the immersed objects, we also studied the TF between them.  In order to achieve this we have used general scaling arguments and mean-field type calculations utilizing the Derjaguin and the surface integration approaches -- see Subsec. \ref{sec:SpPlandSpSpGeomForces}.  Any of these two approximations uses  data for the forces between two parallel slabs separated at a distance $L$ from each other, made of the same materials as the objects and confining the same fluctuating fluid medium characterized by its temperature $T$ and chemical potential $\mu$.  The corresponding model, which we have used to produce the data needed for the current study, is presented in Sec. \ref{sec:Model}.

In the article we concentrated on a system that involves either a sphere and a thick planar slab, or two spheres with, in general, different radii -- see Fig. \ref{fig:fluid_systems}. The surface of any of the objects immersed in the fluid is supposed coated by thin layers exerting strong preference to the liquid phase of the fluid, or one of the components of the mixture, modeled by strong adsorbing local surface potentials ensuring the so-called  $(+,+)$ boundary conditions.  We suppose that the core region of the slab and the particles,  on the other hand, influence the fluid by long-ranged {\it competing} dispersion potentials.

Figures \ref{fig:Xcrits11s2gz} -- \ref{fig:XcritSIA} show that for a suitable set of colloid-fluid, slabs-fluid, and fluid-fluid coupling parameters the competition between the effects due to the coatings and the core regions of the objects involved result in a {\it sign change} of the critical component of the CF between the objects considered. Figures  \ref{fig:Xcrits11s2gz} -- \ref{fig:Xcrits1m1s2sz} present these results obtained within the DA, while Fig. \ref{fig:XcritSIA} shows the ones obtained within the SIA. As it is clear from the data presented, the sign change happens  when one changes $T$, $\mu$ or $L$ between the  colloid and the slab or between the colloids.

Figure \ref{fig:TFSIA} shows, on its turn, that a {\it sign change}  can also be achieved for the {\it total} force between these objects. It is worthwhile to emphasize that this  can be used for governing behavior of objects, say colloidal particles, at small distances, say in colloid suspensions for segregation of colloids. It can also provide a strategy for solving problems with handling, feeding, trapping and fixing of micro parts in nanotechnology resolving the issues related to sticking of the particles on the surface of the mechanical manipulator utilizing, e.g., the {\it reversible dependence} on the forces under minute changes of the temperature of the critical medium. One can perform grabbing of particles for small values of $x_t$, where the force is attractive and release them at a given spacial position after slightly increasing or decreasing of temperature achieving in that way a value of $x_t$ with a repulsive TF. Data given in Table \ref{table} for specific substances demonstrate that the values of the parameters used in the theoretical calculations are experimentally feasible.

Finally let us make some remarks on the SIA which has been suggested  in Ref. \cite{DV2012} as an improvement and generalization of the DA. It delivers an {\it exact} expression between a body with a general shape and a slab, expressing the interaction in terms of the plate-plate interaction, provided the interactions involved can be described by pair potentials, say van der Waals type potentials. As \eref{SIAgeneralsimple} shows, the final result is expressed in terms of integral over projection $A_S^{\rm to}$ of the surface of the colloid particle on the plane which faces towards this plane, minus the contribution over the surface projection $A_S^{\rm away}$ that faces away from the plane. The main advantage of this approach over the DA, which involves only integration over $A_S^{\rm to}$, is that it is not bound by the restriction that the interacting objects must be much closer to each other than their characteristic sizes. The integral solely over $A_S^{\rm to}$ can also be used, as it is customary within the DA, to evaluate the interaction between the surface of the colloid and a plane also for the cases when the interaction can not be prescribed to some point like sources. The last includes, e.g., the way it is used for calculating the CCF. If the colloid is in a mechanical equilibrium in a fluid it is clear that the force at a given point on the surface of the colloid is orthogonal to the surface at that point. Then one can suggest an improvement over DA also for the case of CCF type interactions since the standard DA does not respect this fact.  The corresponding improvement is suggested in \eref{SIA_Casimir} for a colloid particle of any shape and a plane. Unfortunately, we are not aware for an expression for the interaction of two bodies of general shape expressed in terms of the corresponding plate-plate interactions. We have been able, however, to derive such an expression for the case of two spheres -- see \eref{sphere_sphere_force_SIA_intext}. It has been shown that it delivers correct results for interactions decaying with the distance  $\propto r^{-d-\sigma}$, with $d=3$ and $2<\sigma\le 4$ reproducing, for $\sigma=3$ the classical result of Hamaker \cite{H37}, see \eref{vdWSIAsigma3}, and delivering a new analytical result for the retarded van der Waals interaction with $\sigma=4$ -- see \eref{vdWSIAsigma4}. Furthermore, this expression delivers in the limit $R_1/L, R_2/L\to\infty$ the standard Derjaguin result \eref{DASpSp} for the sphere-sphere interactions -- for proof see Appandix \ref{sec:AppSpSpSIA}. The corresponding expression for the CCF between two spheres within the SIA approximation that takes into account only the interactions between surfaces facing each other, similar to the approach of DA, is given in \eref{sphere_sphere_force_SIA_Cas_intext2}. A comparison of the results obtained via DA and SIA is presented in Figs.  \ref{fig:XcritSIA} and \ref{fig:TFSIA}.

\acknowledgments
The authors gratefully acknowledge the support of the Program for Career Development of Young Scientists, BAS via contract No. DFNP-193. All numerical data used for the evaluation of the sphere-plate and sphere-sphere interactions were obtained with the use of Avitohol -- Cluster Platform of the Institute of Information and Communication Technologies at BAS. G.V. expresses its sincere thanks to Dr. K. Shterev for the helpful advices on the numerical calculations procedures.

\appendix
\section{Derivation of the sphere-sphere interaction force within the SIA approximation}\label{sec:AppSpSpSIA}

Within the {\it pairwise} summation hypothesis \cite{H37} proposed by H. Hamaker in 1937, the van der Waals interaction energy $\omega^{B_{1},B_{2}}$ between two {\it macroscopic} bodies in $d=3$ can be written as an integral over the individual interactions $\omega^{p,p}$ between all particles constituting the two objects
\begin{equation}\label{body_body_general_1}
\omega^{B_{1},B_{2}}(L)=\int_{V_{1}}{\rm d}v_{1}\int_{V_{2}}{\rm d}v_{2}\rho_{s_{1}}\rho_{s_{2}}\omega^{p,p}(r)
\end{equation}
with
\begin{equation}\label{body_body_general_2}
 \omega^{p,p}(r)=-\dfrac{J}{r^{3+\sigma}}\vartheta^{\sigma-3},
\end{equation}
where $L$ is the minimal distance between the objects; $V_{i},\ i=1,2$ are the volumes occupied by them; $\rho_{s_{i}},\ i=1,2$ are their number densities; $r$ denotes the distance between the elementary volume elements ${\rm d}v_{i},\ i=1,2$ and $J$ is any of the van der Waals coupling parameters.

Therefore within the Hamaker approach, for the interaction energy between a sphere of radius $R_{1}$ and a point-like object $p$ at a distance $R\equiv R_{1}+z_{2}$ to the sphere's center one can write (see Eq. (5) in \oncite{H37})
\begin{equation}\label{point_sphere_Ham}
\omega_{\rm Ham}^{p,R_{1}}(R)=\int_{R-R_{1}}^{R+R_{1}}\omega^{p,p}(r)\rho_{s_{1}}\pi\dfrac{r}{R}\left[R_{1}^{2}-(R-r)^{2}\right]{\rm d}r.
\end{equation}
Now by making use of the expression for the interaction energy $\omega^{p,|}$ between a point-like object and a half-infinite space (plate) (for details see Eqs. (7), (9) and (17) in Ref. \cite{DV2012}), \eref{point_sphere_Ham} can be written equivalently as
\begin{eqnarray}\label{point_sphere_SIA}
\omega_{\rm SIA}^{p,R_{1}}(R)&&=\int_{R-R_{1}}^{R+R_{1}}\left(\dfrac{z_{1}}{R}-1\right){\rm d}\omega^{p,|}(z_{1})\nonumber\\
&&=\dfrac{R_{1}}{R}\left[\omega^{p,|}(R+R_{1})+\omega^{p,|}(R-R_{1})\right]\nonumber\\&&-\dfrac{1}{R}
\int_{R-R_{1}}^{R+R_{1}}\omega^{p,|}(z_{1}){\rm d}z_{1},
\end{eqnarray}
where
\begin{equation}\label{point_plane}
\omega^{p,|}(z_{1})=-\dfrac{2\pi J \rho_{s_{1}}}{\sigma(\sigma+1)}\dfrac{1}{z_{1}^{\sigma}}\vartheta^{\sigma-3}.
\end{equation}
Using the same considerations as in the derivation of \eref{point_sphere_Ham}, Hamaker managed to obtain for the interaction energy $\omega^{R_{1},R_{2}}$ between two spheres of radii $R_{1}$ and $R_{2}$, separated at a distance $C\equiv R_{1}+R_{2}+L$ apart, the following expression
\begin{eqnarray}\label{phere_sphere_Ham}
&&\omega_{\rm Ham}^{R_{1},R_{2}}(C)=\nonumber\\
&&=\int_{C-R_{2}}^{C+R_{2}}\omega^{p,R_{1}}(R)\rho_{s_{2}}\pi\dfrac{R}{C}\left[R_{2}^{2}-(C-R)^{2}\right]{\rm d}R.
\end{eqnarray}
Substituting \eref{point_sphere_SIA} into \eref{phere_sphere_Ham}, taking into account that $f_{\cal A}^{\parallel}\equiv \omega^{p,|}\rho_{s}$ and performing differentiation with respect to the separation $L$, one ends up with an expression relating the interaction force $F^{R_{1},R_{2}}$ between two spherical particles with the force per unit area between two parallel plates
\begin{widetext}
\begin{eqnarray}\label{sphere_sphere_force_SIA}
F_{\rm SIA}^{R_{1},R_{2}}(L)=-R_{1}\int_{L}^{L+2R_{2}}\left[f_{\cal A}^{\parallel}(z_{2})
+f_{\cal A}^{\parallel}(z_{2}+2R_{1})\right]\zeta(z_{2})\mathrm{d}z_{2}+\int_{L}^{L+2R_{2}}\int_{z_{2}}^{z_{2}+2R_{1}}\zeta(z_{2})f_{\cal A}^{\parallel}(z_{1})\mathrm{d}z_{1}\mathrm{d}z_{2},
\end{eqnarray}
\end{widetext}
where the function $\zeta(z_{2})$ is given in \eref{eq:dzeta_deriv_def} in the main text.
Because in the derivation of \eref{sphere_sphere_force_SIA} no restrictions on the allowed sizes and separations between the particles are considered, one can say that this expression was derived within the SIA.

Now if one performs the integration with \eref{way} [see also the text below \eref{scaling_function_Casimir}] in $d=3$, the general expression for the vdWF between two spheres reads
\begin{widetext}
\begin{eqnarray}\label{vdWSpSpgensigmaApp}
F_{\rm vdW,SIA}^{R_{1},R_{2}}(L|\sigma)&&=-H_{A}^{\rm (reg)}\vartheta^{\sigma-3}\int_{L}^{L+2R_{2}}\left\{\dfrac{[(\sigma-1)R_{1}-z_{2}]}{z_{2}^{\sigma}}
+\dfrac{[(\sigma+1)R_{1}+z_{2}]}{(z_{2}+2R_{1})^{\sigma}}\right\}\zeta(z_{2})\mathrm{d}z_{2}\nonumber\\&&=
\dfrac{2H_{A}^{\rm (reg)}\vartheta^{\sigma-3}}{(\sigma-2)(\sigma-3)(L+R_{1}+R_{2})}\left[\dfrac{{\cal E}(L|\sigma)}{(\sigma-4)(L+R_{1}+R_{2})}+{\cal E}(L|\sigma+1)\right],
\end{eqnarray}
which is applicable for any $\sigma\in(2;4)$ with
\begin{subequations}\label{polynomial}
\begin{equation}\label{ExpresionLsigma}
{\cal E}(L|\sigma)=\dfrac{\mathcal{P}_{1}(L|\sigma)}{L^{\sigma-2}}-\dfrac{\mathcal{P}_{2}(L|\sigma)}{(L+2R_{1})^{\sigma-2}}
-\dfrac{\mathcal{P}_{3}(L|\sigma)}{(L+2R_{2})^{\sigma-2}}+\dfrac{\mathcal{P}_{4}(L|\sigma)}{(L+2R_{1}+2R_{2})^{\sigma-2}}
\end{equation}
\begin{equation}\label{poly1}
\mathcal{P}_{1}(L|\sigma)=L^2-L(R_{1}+R_{2})(\sigma-4)+R_{1}R_{2}(\sigma-4)(\sigma-3),
\end{equation}
\begin{equation}\label{poly2}
\mathcal{P}_{2}(L|\sigma)=L^{2}+L[R_{2}\sigma-R_{1}(\sigma-4)]+R_{2}[2R_{2}(\sigma-2)-R_{1}(\sigma-4)(\sigma-1)],
\end{equation}
\begin{equation}\label{poly3}
\mathcal{P}_{3}(L|\sigma)=L^{2}+L[R_{1}\sigma-R_{2}(\sigma-4)]+R_{1}[2R_{1}(\sigma-2)-R_{2}(\sigma-4)(\sigma-1)],
\end{equation}
\begin{equation}\label{poly4}
\mathcal{P}_{4}(L|\sigma)=L^{2}+L(R_{1}+R_{2})\sigma+2(R_{1}^{2}+R_{2}^{2})(\sigma-2)+R_{1}R_{2}[4+(\sigma-3)\sigma].
\end{equation}
\end{subequations}
\end{widetext}
For the two most commonly considered cases $\sigma=3$ and $\sigma=4$, \eref{sphere_sphere_force_SIA} delivers the results reported in \eref{vdWSIAsigma3} and \eref{vdWSIAsigma4}, correspondingly, in the main text.

From \eref{sphere_sphere_force_SIA} the general expression describing the $L$ dependance of the sphere-plate interaction follows directly. Indeed by taking, say $R_{1}\rightarrow\infty$ [i.e. one of the spheres becomes a half-space (plate)], one has that for any inverse distance force law $F_{A}^{\parallel}(x)\xrightarrow{x\rightarrow\infty}0$ and $\zeta(z_{2})\rightarrow-2\pi(L+R_{2}-z_{2})$. Thus
\begin{equation}\label{sphere_plane_force_SIA}
F_{\rm SIA}^{R_{2},|}(L)=2\pi\int_{L}^{L+2R_{2}}(L+R_{2}-z_{2})f_{\cal A}^{\parallel}(z_{2}){\rm d}z_{2},
\end{equation}
which is exactly the result Eq. (20) reported in \oncite{DV2012}. It is easy to show that from \eref{sphere_sphere_force_SIA} follows the expression obtained by Derjaguin (see for example Eq. (1) in \oncite{DV2012}) describing the interaction between spherical particles at close proximity to one another [i.e. $R_{1},R_{2}\gg L$, or $L\rightarrow 0$].  Introducing the notations $\bar{z}_{i}\equiv z_{i}/L;\ \Xi_{i}\equiv R_{i}/L$, \eref{sphere_sphere_force_SIA} takes the form
\begin{eqnarray}\label{sphere_sphere_force_SIA_dim}
F_{\rm SIA}^{R_{1},R_{2}}(L)=&&-\dfrac{\Xi_{1}}{L^{\sigma-2}} \int_{1}^{1+2\Xi_{2}}\left[f_{\cal A}^{\parallel}(\bar{z}_{2})+f_{\cal A}^{\parallel}(\bar{z}_{2}+2\Xi_{1})\right]\zeta(\bar{z}_{2}){\rm d}\bar{z}_{2}\nonumber\\
&&+\dfrac{1}{L^{\sigma-2}}\int_{1}^{1+2\Xi_{2}}\int_{\bar{z}_{2}}^{\bar{z}_{2}+2\Xi_{1}}\zeta(\bar{z}_{2})f_{\cal A}^{\parallel}(\bar{z}_{1}){\rm d}\bar{z}_{1}{\rm d}\bar{z}_{2}.
\end{eqnarray}
Now we notice that within the DA $\Xi_{i}\rightarrow\infty,\ i=1,2;$  then $f_{\cal A}^{\parallel}(\bar{z}_{2}+2\Xi_{1})\rightarrow 0$, $\lim_{L\rightarrow 0}\int_{\bar{z}_{2}}^{\bar{z}_{2}+2\Xi_{1}}...=\int_{\infty}^{\infty}...=0$ and $\lim_{L\rightarrow 0}\zeta(\bar{z}_{2})=2\Xi_{2}/(\Xi_{1}+\Xi_{2})$. Therefore
\begin{eqnarray}\label{sphere_sphere_force_DA_dim}
F_{\rm DA}^{R_{1},R_{2}}(L)&&=2\pi \dfrac{1}{L^{\sigma-2}}\dfrac{\Xi_{1}\Xi_{2}}{\Xi_{1}+\Xi_{2}}\int_{1}^{\infty}f_{\cal A}^{\parallel}(\bar{z}_{2}){\rm d}\bar{z}_{2}\nonumber\\&&=2\pi \dfrac{R_{1}R_{2}}{R_{1}+R_{2}}\int_{L}^{\infty}f_{\cal A}^{\parallel}(z_{2}){\rm d}z_{2}=2\pi R_{\rm eff}\omega_{A}^{\parallel}(L).\ \ \ \ \ \ \ \
\end{eqnarray}

In a way analogous to the use of the DA for determination of the CCF between objects of complicated geometry, one can make use of the SIA technique to do the same. Then, as usual, the restriction appears that the integration must be carried on only over these parts of the objects surfaces which are facing each other (see for details the text above). Taking this into account we can write the following expressions for the CCF between two spherical particles of arbitrary radii as well as between a spherical particle and a planar substrate respectively
\begin{subequations}\label{sp_sp_and_sp_plate_Cas}
\begin{eqnarray}\label{sphere_sphere_force_SIA_Cas}
&&F_{\rm Cas,SIA}^{R_{1},R_{2}}(L)=-\int_{L}^{L+R_{2}}R_{1}\zeta(z_{2})f_{{\cal A},{\rm Cas}}^{\parallel}(z_{2}){\rm d}z_{2}\nonumber\\
&&+\int_{L}^{L+R_{2}}\zeta(z_{2})\int_{z_{2}}^{z_{2}+R_{1}}f_{{\cal A},{\rm Cas}}^{\parallel}(z_{1}){\rm d}z_{1}{\rm d}z_{2},
\end{eqnarray}
\begin{equation}\label{sphere_plane_force_SIA_Cas}
F_{\rm Cas,SIA}^{R,|}(L)=2\pi\int_{L}^{L+R}(L+R-z)f_{{\cal A},{\rm Cas}}^{\parallel}(z){\rm d}z.
\end{equation}
\end{subequations}


\begin{thebibliography}{110}
\bibitem{note} The excess potential characterizes the difference between that one of the fluid system with and without the objects immersed in it.
\bibitem[{not()}]{note2}
\bibinfo{note}{ On $\mathbf{(a)}$ and $\mathbf{(c)}$ the temperature dependence is shown at $\Delta\mu=0.0$, (i.e., $x_{\mu}=0.0$), while on $\mathbf{(b)}$ and $\mathbf{(d)}$ -- the one on the chemical potential at $T=T_{c}$, (i.e., at $x_{t}=0.0$).}
\end{thebibliography}

\end{document}